\documentclass[12pt]{article}

\usepackage{amsmath,amssymb,epsfig,epsf}
\usepackage{graphicx}

\newcommand{\BK}{{\ensuremath{\bf{k}}}}

\begin{document}

\title{\bf Generalized dynamical mean-field theory in physics of strongly correlated
systems}

\author{E.\ Z.\ Kuchinskii$^1$, I.\ A.\ Nekrasov$^1$, M.\ V.~Sadovskii$^{1,2}$\\
{\sl $^1$Institute for Electrophysics, Russian Academy of Sciences, Ural Branch,}\\ {\sl Amundsen str. 106, Ekaterinburg 620016, Russia}\\
{\sl $^2$Institute for Metal Physics, Russian Academy of Sciences, Ural Branch,}\\ {\sl S. Kovalevskaya str. 18, Ekaterinburg 620219, Russia}}

\date{}

\maketitle

\begin{abstract}

This review is devoted to generalization of dynamical mean-field theory (DMFT)
for strongly correlated electronic systems towards the account of
different types of additional interactions, necessary for
correct physical description of many experimentally observed phenomena in such 
systems.
As additional interactions we consider: 
(1) interaction of electrons with antiferromagnetic (or charge) fluctuations
of order parameter in high-Tc superconductors leading to the formation of pseudogap state,
(2) scattering of electrons on static disorder and its role in general picture of
Anderson-Hubbard metal-insulator transition,
(3) electron-phonon interaction and corresponding anomalies of electronic spectra
in strongly correlated systems. 
Proposed DMFT+$\Sigma$ approach is based on taking into account above mentioned
interactions by introducing additional self-energy $\Sigma$ (in general momentum dependent)
into conventional DMFT scheme and
calculated in a self-consistent way within the standard set of DMFT equations
Here we formulate general scheme of calculation of both one-particle
(spectral functions and densities of states) and two-particle
(optical conductivity) properties.
We examine the problem of pseudogap formation, including the 
Fermi arc formation and partial destruction of the Fermi surface,
metal-insulator transition in disordered Anderson-Hubbard model,
and general picture of kink formation within electronic spectra
in strongly correlated systems.

DMFT+$\Sigma$ approach is generalized to describe realistic materials
with strong electron-electron correlations based on LDA+DMFT method.
General scheme of LDA+DMFT method is presented together with some of its
applications to real systems. The LDA+DMFT+$\Sigma$ approach is employed
to modelling of pseudogap state of electron and hole doped high-T$_c$ cuprates.
Comparison with variety of ARPES experiments is given.\\ 


\end{abstract}

\newpage

\tableofcontents

\newpage

\section{Introduction}

\label{intro}

Strongly correlated electronic systems (SCS), which are mainly realized in a 
number compounds of transition or rare-earth elements with partially filled 
$3d$, $4f$ and $5f$ shells, for more than half a century attract 
a lot of interest of scientists because of their unusual physical properties 
and difficulties in their theoretical description. 
Problem of metal-insulator phase transition, observed in many transition
metal oxides, heavy fermions systems, with great variety of different phase 
transitions and related phenomena, manganites with giant magnetoresistance -- 
all these systems become a subject of great attention for both experimentalists 
and theorists. Perhaps the most significant development in this area was the 
discovery of high temperature superconductivity in copper oxides, which 
provoked a new wave of interest in the synthesis and description of such systems.

As already stressed above, the diversity of physical phenomena in all these 
compounds is due to partially filled $3d$, $4f$ and $5f$ shells. Strong 
interaction of electrons within narrow bands belonging to these orbitals
shells with each other or with itinerant electrons of outer shells is basically 
responsible for unique properties of these systems. Early qualitative ideas 
formulated by Mott \cite{NFM} were further developed in many theoretical works 
to follow, forming a new area of modern theory of condensed matter. 
There are now thousands of such papers and many new reviews and books 
are regularly published \cite{IzAn}.

Of course, nowadays, a single review can not cover all this area of 
research and the aim of authors is rather modest. The object of this work is 
description of a number of theoretical approaches formulated during recent years 
to account for some additional interactions which are important for the coherent
description of strongly correlated systems. Here we speak not only about 
``external'' perturbations due to interaction of correlated electrons with 
bosonic excitations such as phonons, spin fluctuations or scattering of 
electrons by disorder, but also about attempts to improve most developed and 
widely used theoretical approaches such as dynamical mean-field theory
(DMFT) \cite{vollha93,pruschke,georges96,Vollh10}. 

These tasks are closely related with variety of topical problems under active 
study at present, such as the nature of the pseudogap state of high-T$_c$ 
cuprates, the evolution of their Fermi surfaces upon doping with formation of 
so called ``Fermi arcs'', observed in ARPES experiments, the problem of 
formation of kinks in electronic spectrum, the general problem of
metal-insulator transition in disordered systems, due to mutual interference of 
strong correlations and Anderson localization. In this review to some extent 
we deal with all of these problems.

During last years the general theory of strongly correlated systems based on 
DMFT practically merged with the so called ``first-principle'' approaches to
calculations of electronic spectra of real solids (LDA+DMFT approach), with
significant successes already achieved \cite{IzAn,PT}.
To this end in our review we devote some attention to first attempts of 
generalizing these approaches towards the account of the above mentioned physical 
effects.

\section{Strongly correlated systems and dynamical mean-field theory
(DMFT).}

\subsection{Hubbard model and basics of DMFT.}

\label{secDMFT}

Starting with pioneering works of Hubbard \cite{Hubbard} in the early 60th
the simplest model to describe strongly correlated systems is the
so called Hubbard model.
One band Hubbard model Hamiltonian is:
\begin{equation}
H=-t\sum_{\langle ij\rangle \sigma }c_{i\sigma }^{\dagger }c_{j\sigma }+
U\sum_{i}n_{i\uparrow }n_{i\downarrow },  
\label{Hubbard}
\end{equation}
where $t>0$ -- nearest neighbors hopping amplitude, $U$ --
single site repulsion,
$n_{i\sigma }=c_{i\sigma }^{\dagger }c_{i\sigma }^{{\phantom{\dagger}}}$ --
particle number operator on site $i$,
$c_{i\sigma }$ ($c_{i\sigma }^{\dagger}$) -- annihilation (creation)
electron operators for spin $\sigma$.
The model has only two competing energy parameters.
Parameter  $t$ defines kinetic energy of electron and facilitates intersite hoppings
(delocalization), while parameter $U$ defines potential energy
and favors localization of an electron on a lattice cite.
Energy bands formed by $3d$, $4f$ and $5f$ orbitals
are rather narrow, thus quite often kinetic and potential energy
are of the same order of magnitude ($t\sim U$).
In this case there is no small parameter in the model
and it is impossible to build any kind of perturbation theory.
This fact alone leads to all the  difficulties in description of SCS
even for such oversimplified model.

Almost for 30 years there were no satisfactory approaches to analyze SCS.
It seemed that theory of these systems will forever remain fragmentary and semiquantitative.
The breakthrough came in 1989 in the work by Metzner and Vollhardt \cite{MetzVoll89}.
They suggested formal consideration of the system of interacting electrons in large space
dimensions $d\to \infty$ (or in a lattice with large coordination number$z\to \infty$ 
\footnote{For hypercubic lattice $z=2d$ and these two limits practically coincide.
However, even for three dimensional lattices $z$ could be quite large, for example, in body centered
lattice $z=8$ and for face centered $z=12$.
To this end it is more correct to speak about limit of large $z$.}). 

Employing this limit it is possible to neglect spatial fluctuations in the systems
while full local dynamics is preserved.
In Ref. \cite{MetzVoll89} it was shown that in the limit of infinite
spatial dimensions (or more precise infinite coordination number)
the main role is played only by local contributions to self-energy part of
full interacting Green's function. All non local contributions are
proportional to $1/\sqrt z\sim 1/\sqrt d$ and can be dropped.
In this limit electron self-energy does not depend on momentum ${\bf k}$, and is a 
function of frequency only (real ($\omega $) or Matsubara ($\omega _n$) one) 
\footnote{Large coordination numbers allow one to apply this approximation
rather successfully even for rather small $d$}:
\begin{equation}
\Sigma _{\sigma}({\bf k},\omega)=\Sigma _{\sigma}(\omega).
\label{loc_sigma}
\end{equation}
This statement is the main simplification appearing in the limit of
$d\to \infty$ \cite{vollha93,georges96,Vollh10,PT}.

\begin{figure}
\includegraphics[clip=true,width=0.75\textwidth]{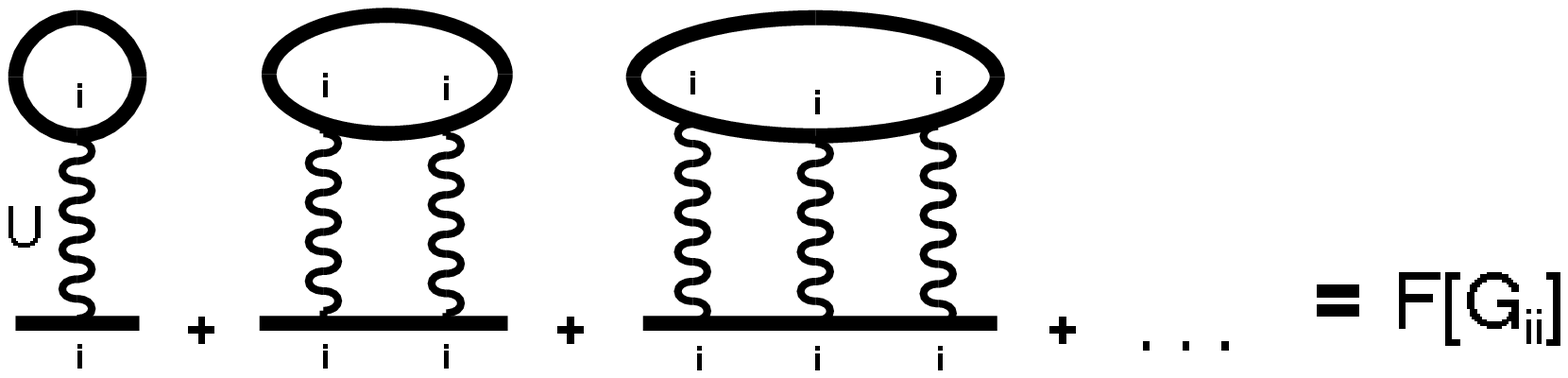}
\caption{\small ``Skeleton diagrams of local self-energy $\Sigma$ in DMFT.
}
\label{sigmDMFT}
\end{figure}

In Fig. \ref{sigmDMFT} we show the ``skeleton'' diagrams of DMFT self-energy $\Sigma$.
Wavy lines represent local (Hubbard) interaction $U$,
full lines represent local Green functions $G_{ii}$.
Strictly speaking in the limit of $d\to \infty$ the self-energy is not only a local one,
but also in each vertex of ``skeleton'' diagram, only one particular site enters, e.g. the $i$-th ones,
as shown in Fig. \ref{sigmDMFT}. Thus this self-energy is a functional
$\Sigma_{ii , \sigma}=F[G_{ii , \sigma}]$ of interacting local Green's function $G_{ii , \sigma}$.
Generally it is not sufficient to make our problem the fully local one, since
interacting Green function $G_{ij , \sigma}$ is still nonlocal.
Then a question arises, whether it is possible to choose purely local
non interacting (in the absence of $U$) problem with completely equivalent self-energy?
Surely it can be done! 
Let $\mathcal{G}_{\sigma}(\omega)$ to be ``bare'' dynamical Green function of
such local problem without Coulomb interaction $U$, 
while $G_{d\sigma}(\omega)$ and $\Sigma_{d\sigma}(\omega)$ are
corresponding interacting Green's function and self-energy.
If one can guarantee the equality  $G_{d\sigma}(\omega)=G_{ii , \sigma}(\omega)$, 
then corresponding self-energies are also equivalent,
because the structure of diagrams of weak coupling $U$ perturbation theory
is totally preserved, which means that self-energy of the local problem
is defined by the same functional $\Sigma_{d\sigma}=F[G_{d\sigma}]$.
But $\mathcal{G}_{\sigma}$, $G_{d\sigma}$ and $\Sigma_{d\sigma}$
are connected through the Dyson equation, which immediately gives us
the ``bare'' dynamical Green function of the local problem.
\begin{equation}
\mathcal{G}^{-1}_{\sigma}(\omega ) = \Sigma_\sigma (\omega)+G_{ii , \sigma}^{-1}(\omega );
\label{calG_DMFT}
\end{equation}

\begin{figure}
\begin{center}
\includegraphics[clip=true,width=0.3\textwidth]{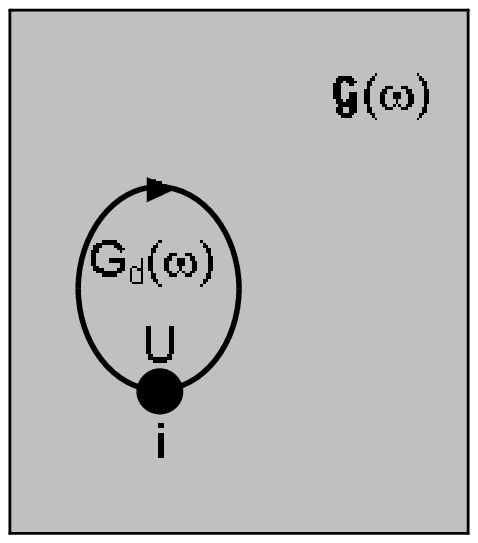}
\end{center}
\caption{\small Within the DMFT lattice Hubbard model maps to 
interacting electrons on a single site (``impurity''), surrounded
by the fermionic bath defining dynamical (Weiss) mean-field $\mathcal{G}(\omega)$.}
\label{fig_impDMFT}
\end{figure}

Thus the lattice Hubbard model in the limit of $d\to \infty$ exactly maps onto
purely local dynamical problem. Physically it corresponds
(as shown in Fig. \ref{fig_impDMFT}) to the problem of interacting
electrons on a single ``Anderson impurity'' in a ``bath'' and interaction
with the bath is contained in dynamical mean-field $\mathcal{G}(\omega )$.
Quite often, in analogy with molecular field theory in magnetism,
this field is called ``Weiss field''.
This explains the name of such an approach - dynamical mean-field theory (DMFT). 

This purely dynamic problem is still quite complicated. However, the problem
is equivalent to a single impurity Anderson model (SIAM)  \cite{Anderson61}.
This model can be studied in detail by a number of different methods
and its physics is now well understood. For this model there are well developed
approximate analytical methods such as iterative perturbation theory (IPT) \cite{georges96}
and non-crossing approximation (NCA) \cite{pruschke89,pruschke93}, but most remarkable is
the possibility to solve this model by exact numerical methods like quantum Monte-Carlo (QMC) \cite{QMC} 
or numerical renormalization group (NRG) \cite{NRG,BPH}.
Solution of an effective SIAM employing any of these methods, usually 
called an ``impurity solver'',  completes the general scheme of DMFT approach. 

\begin{figure}
\includegraphics[clip=true,width=0.6\textwidth]{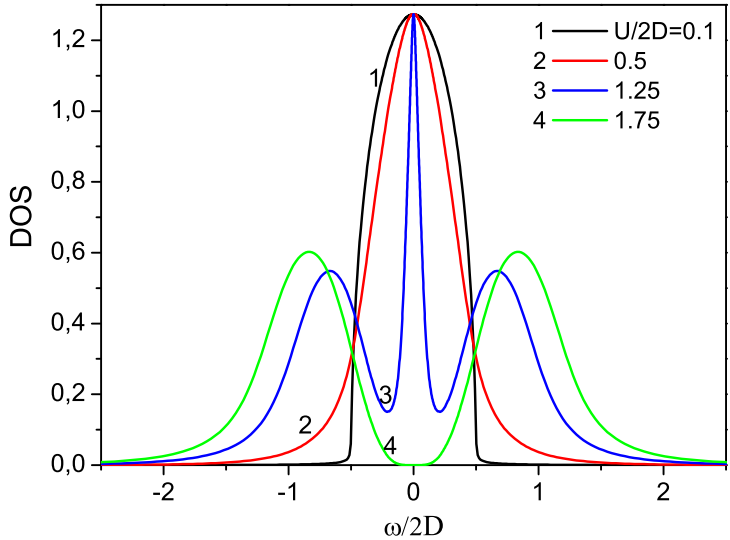}
\caption{\small DMFT(NRG) densities of states at half-filling obtained for
semielliptic ``bare'' density of states for different values of $U$.}
\label{fig_DOS_DMFT}
\end{figure}

Apparently, today DMFT is the most elaborate and reliable theoretical method to describe
SCS. In its framework the so called three-peak structure of the density of states of SCS 
was obtained for the first time \cite{georges96}, consisting of the 
central (quasiparticle) peak on the Fermi level and two wide maxima, corresponding to upper 
and lower Hubbard bands. Also the reliable theoretical description of Mott-Hubbard metal-insulator 
transition was obtained. 
In Fig. \ref{fig_DOS_DMFT} we show DMFT(NRG) densities of states
of the half-filled Hubbard model with semielliptic ``bare'' density of states with bandwidth $2D$.
As correlation strength $U$ grows the density of states demonstrates the formation of characteristic
three-peak structure and further increase of $U$ leads to a collapse of the quasiparticle
peak at $U/2D\approx 1.5$, leading to metal-insulator transition.

It turns out that within DMFT it is also possible to investigate some
two-particle properties. In particular, it is quite easy to obtain
dynamic optical conductivity  \cite{georges96,pruschke}.
During the recent years DMFT approach was generalized to describe realistic SCS
merging it  with ``$ab~initio$'' one-electron density functional theory
in local density approximation (DFT/LDA), leading to the
combined computational scheme of LDA+DMFT
\cite{poter97,LDADMFT1,Nekrasov00, psik, LDADMFT}, 
which will be described later.

Despite all the obvious advantages of DMFT this approach has a number of shortcomings.
Namely, as we stressed above, all non-local correlations are completely neglected.
A number of cluster generalizations of DMFT \cite{TMrmp,KSPB} were proposed recently 
to overcome this drawback. However,  all these methods are quite
computer time consuming and are rather restricted in a cluster size and with respect
to their generalization to multi-orbital case. Also in these approaches it is quite difficult
to investigate two-particle properties. To overcome these difficulties we proposed recently 
\cite{cm05,JTL05,FNT06} the new generalization of the conventional DMFT, allowing to consider
non-local correlations or additional (with respect to the Hubbard one) interactions (in principle 
of any kind), remaining within a single impurity DFMT picture and preserving self-consistent set 
of DMFT equations.

\subsection{Generalized DMFT+$\Sigma$ approach.}
\label{dmft_sk_t}

The main idea of the new approach is to use the exact in the limit of
$d\to \infty$ DMFT solution as a ``high energy'' zeroth order approximation,
describing electronic spectra on a large energy scale of the order of
bandwidth or $U$ value, while low energy scale details caused by
non-local effects or by effects of interaction of correlated electrons with 
different collective modes (e.g. phonons or spin fluctuations) are to be taken 
into account within some kind of perturbation theory, conserving, as far as 
possible, the general structure of DMFT equations. Actually, such a scheme can 
be realized in rather simple way \cite{cm05,JTL05,FNT06}.

To be more specific, in the following we consider the standard one band Hubbard 
model. Generalizations towards multi orbital case are also possible.
Main assumption of our approach is to choose Matsubara lattice Green's function 
as:
\begin{equation}
G_{\bf k}(i\omega)=\frac{1}{i\omega+\mu-\varepsilon({\bf k})-\Sigma(i\omega)
-\Sigma_{\bf k}(i\omega)},\qquad \omega=\pi T(2n+1),
\label{Gk9}
\end{equation}
where $\mu$ is the chemical potential, $\Sigma(i\omega)$ --  {\em local} 
DMFT self-energy due to Hubbard interaction and $\Sigma_{\bf k}(i\omega)$ --
some ``external'' (generally non-local, momentum dependent) self-energy.
This last contribution can arise from interaction of correlated electrons
with some ``additional'' collective modes or order parameter fluctuations
appearing in the Hubbard model itself, or from any other interactions 
(fluctuations) external with respect to the standard Hubbard model.
For example these can be phonons or scattering by impurities, when it is 
actually local (momentum independent).

One should emphasize that $\Sigma_{\bf k}(i\omega)$
can contain local (momentum independent) contribution even if the self-energy
is considered in the framework of the Hubbard model. However this contribution
{\em disappears} in the infinite spatial dimensions limit of $d\to\infty$
and is not accounted within the conventional DMFT, so that within our approach 
we not encounter any double counting problem even in this case.
This question does not come out at all for self-energy  $\Sigma_{\bf k}(i\omega)$
caused by ``external'' interactions.

\begin{figure}
\includegraphics[clip=true,width=1.\textwidth]{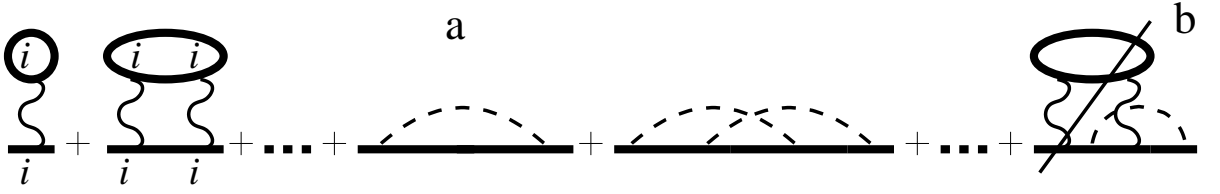}
\caption{\small Typical ``skeleton'' self-energy diagrams of 
DMFT+$\Sigma$ approach.}
\label{dDMFT_PG}
\end{figure}

More important is that our assumption of the additive form of a self-energy
 $\Sigma(i\omega)+\Sigma_{\bf k}(i\omega)$ implicitly corresponds to the neglect
of interference of the local (DMFT) and non-local contributions.
In Fig. \ref{dDMFT_PG} typical ``skeleton'' diagrams for self-energy
of DMFT+$\Sigma$ approach are given.
First two terms are local DFMT self-energy diagrams,
two diagrams in the middle show contributions to non-local
part of self-energy from ``additional'' interactions with collective modes
or order parameter fluctuations, and the last diagram (b) is an 
example of diagram with interference between local and non-local parts
which is neglected. Indeed, once we neglect such interference
(i.e. diagram shown in Fig. \ref{dDMFT_PG}(b)) the
total self-energy is defined as a simple sum of these two contributions
shown in Fig.  \ref{dDMFT_PG}.
Two last diagrams in Fig. \ref{dDMFT_PG}(a) are an example of ``skeleton'' 
diagrams for non-local self-energy, where full line is the Green's function  
$G_{\bf k}$ (\ref{Gk9}) and dashed line corresponds to an ``additional'' 
interaction with collective modes or order parameter fluctuations.

Finally, diagrammatic structure of the local self-energy remains identical
to that of the standard DMFT and we obtain the following self-consistent 
equations of the generalized DMFT+$\Sigma$ approach  \cite{cm05,JTL05,FNT06}:

\begin{enumerate}
\item{Start from some initial guess for the {\em local} self-energy
$\Sigma(i\omega)$, e.g., $\Sigma(i\omega)=0$}.  
\item{Calculate self-energy $\Sigma_{\bf k}(i\omega)$ in the framework of some
(approximate) scheme, taking into account interaction of correlated electrons 
with collective modes or order parameter fluctuations,
which, in general, can depend on $\Sigma(i\omega)$ and $\mu$.} 
\item{Calculates local Green's function:
\begin{equation}
G_{ii}(i\omega)=\frac{1}{N}\sum_{\bf k}\frac{1}{i\omega+\mu
-\varepsilon({\bf k})-\Sigma(i\omega)-\Sigma_{\bf k}(i\omega)}.
\label{Gloc}
\end{equation}
}
\item{Define the ``Weiss field'' as:
\begin{equation}
{\cal G}^{-1}_0(i\omega)=\Sigma(i\omega)+G^{-1}_{ii}(i\omega).
\label{Wss}
\end{equation}
}
\item{Using some ``impurity solver'' calculate the
single-particle Green's function of an effective single impurity Anderson 
model, i.e. compute the following integral over Grassmann variables
$ c^+_{i\sigma}$ and $c_{i\sigma}$:
\begin{equation}
G_{d}(\tau-\tau')=\frac{1}{Z_{\text{eff}}}
\int Dc^+_{i\sigma}Dc_{i\sigma}
c_{i\sigma}(\tau)c^+_{i\sigma}(\tau')\exp(-S_{\text{eff}}),
\label{AndImp}
\end{equation}
where an effective action for the fixed site (``Anderson impurity'') $i$ is:
\begin{equation}
S_{\text{eff}}=-\int_{0}^{\beta}d\tau_1\int_{0}^{\beta}
d\tau_2c_{i\sigma}(\tau_1){\cal G}^{-1}_0(\tau_1-\tau_2)c^+_{i\sigma}(\tau_2)
+\int_{0}^{\beta}d\tau Un_{i\uparrow}(\tau)n_{i\downarrow}(\tau)\;\;,
\label{Seff}
\end{equation}
with ``partition function''
$Z_{\text{eff}}=\int Dc^+_{i\sigma}Dc_{i\sigma}\exp(-S_{\text{eff}})$, and
$\beta=T^{-1}$. 
} 
\item{Define {\em new} local self-energy as:
\begin{equation}
\Sigma(i\omega)={\cal G}^{-1}_0(i\omega)-
G^{-1}_{d}(i\omega).
\label{StS}
\end{equation}
}
\item{Using this self-energy as an ``initial'' on step 1, continue the loop 
procedure until it converges to 
\begin{equation}
G_{ii}(i\omega)=G_{d}(i\omega).
\label{G009}
\end{equation}
with a given accuracy.
}
\end{enumerate}

At the end we obtain the final Green's function in the form of Eq. (\ref{Gk9}),
where $\Sigma(i\omega)$ and $\Sigma_{\bf k}(i\omega)$ are self-energies coming 
out of our iterative procedure.

Success of such approach (as well as its main drawback) is connected with an
additive form of total self-energy (neglect of interference between different 
contributions) in Eq. (\ref{Gk9}).
This allows one to preserve self-consistent set of equations of the standard 
DMFT. However there are two significant distinctions from conventional DMFT.
First of all, the local Green's function of an effective single impurity problem
has the form of Eq. (\ref{Gloc}) on each step of DMFT procedure.
Secondly, during each DMFT iteration ``external'' self-energy 
$\Sigma_{\bf k}(i\omega)$ is recalculated within some (approximate) scheme,
taking into account interaction with collective modes (phonons, magnons 
etc.) or with fluctuations of some order parameter.
To define non-local contribution $\Sigma_{\bf k}(i\omega )$ it is convenient to
introduce
\begin{equation} 
{\cal G}_{0\bf k}(i\omega)=\frac{1}{G_{\bf k}^{-1}(i\omega)+
\Sigma_{\bf k}(i\omega)}
=\frac{1}{i\omega+\mu-\varepsilon({\bf k})-\Sigma(i\omega)},
\label{G0pg}
\end{equation}
which plays the role of ``bare'' Green's function to build perturbation theory
over ``external'' interaction. The choice of the ``bare'' Green's function in 
the form of Eq. (\ref{G0pg}) guarantees Green's function ``dressed'' by such 
interaction $G_{\bf k}^{-1}(i\omega)={\cal G}_{0\bf k}^{-1}(i\omega)-\Sigma_{\bf k}(i\omega)$, 
entering ``skeleton'' diagrams for  $\Sigma_{\bf k}(i\omega)$, 
coincides exactly with full Green's function $G_{\bf k}(i\omega)$.

\begin{figure}
\includegraphics[clip=true,width=0.5\textwidth]{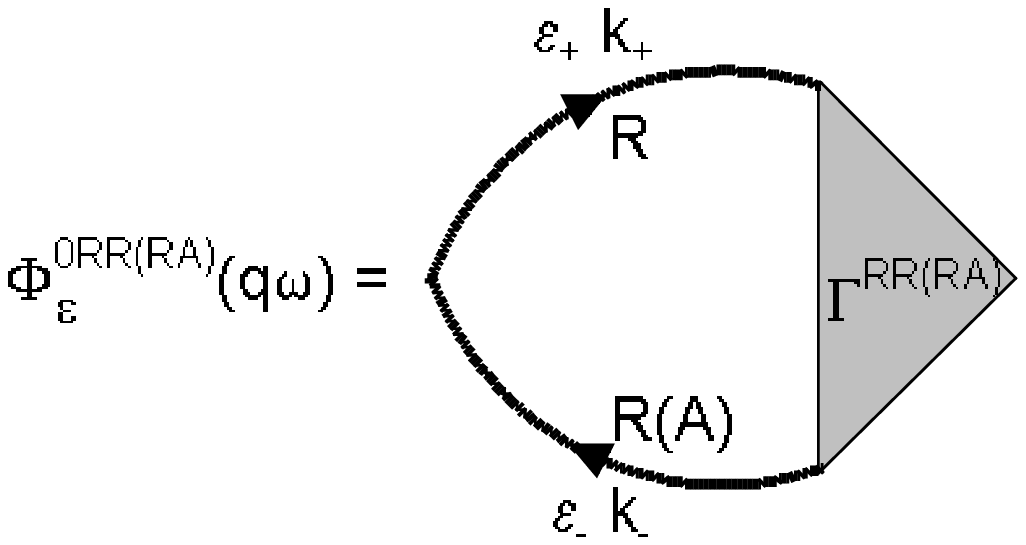}
\caption{\small Diagrammatic representation of $\Phi^{0RA}_{\varepsilon}(\omega,{\bf q})$ 
and $\Phi^{0RR}_{\varepsilon}(\omega,{\bf q})$.} 
\label{loop_fi} 
\end{figure}

Remarkable feature of our approach is the possibility of its generalization
to calculate two-particle properties e.g. optical conductivity \cite{PRB07,HubDis}.
Conductivity of a system is expressed via retarded density--density response function
$\chi^R(\omega,{\bf q})$ \cite{VW,Diagr}:
\begin{equation}
\sigma(\omega)=-\lim_{q\to 0}\frac{ie^2\omega}{q^2}\chi^R(\omega,{\bf q}),
\label{cond_op}
\end{equation}
where $e$ is electron charge. This response function is defined by
analytical continuation to real frequencies of the full polarization loop
in Matsubara representation \cite{VW}.
Note that conductivity is completely defined by first derivative
of this response function with respect to  $q^2$ in the limit of $q\to 0$.
This circumstance, as well as the neglect of interference between Hubbard
and ``external'' interactions in DMFT+$\Sigma$ approach and locality
of irreducible vertices of Hubbard interaction allows one to perform a
partial resummation of diagrams relevant for conductivity, making the use of
an exact (in the limit of $q\to 0$) Ward identity.
At the end the real part of optical conductivity in the DMFT+$\Sigma$ approach 
is \cite{PRB07,HubDis}:
\begin{eqnarray}
{\rm{Re}}\sigma(\omega)=\frac{e^2\omega}{2\pi}
\int_{-\infty}^{\infty}d\varepsilon\left[f(\varepsilon_-)
-f(\varepsilon_+)\right]{\rm{Re}}\left\{\phi^{0RA}_{\varepsilon}(\omega)\left[1-
\frac{\Sigma^R(\varepsilon_+)-\Sigma^A(\varepsilon_-)}{\omega}\right]^2-
\right.\nonumber\\
\left.-\phi^{0RR}_{\varepsilon}(\omega)\left[1-
\frac{\Sigma^R(\varepsilon_+)-\Sigma^R(\varepsilon_-)}{\omega}\right]^2
\right\}.
\label{cond_final}
\end{eqnarray}
where
\begin{equation}
\phi^{0RR(RA)}_{\varepsilon}(\omega)=\lim_{q\to 0}
\frac{\Phi^{0RR(RA)}_{\varepsilon}(\omega,{\bf q})-
\Phi^{0RR(RA)}_{\varepsilon}(\omega,0)}{q^2},
\label{fi0RA_func}
\end{equation}
and we introduced the two-particle Green functions of the following form:
\begin{equation}
\Phi^{0RR(RA)}_{\varepsilon}(\omega,{\bf q})=\sum_{\bf k}
G^R(\varepsilon_+,{\bf k_+})G^{R(A)}(\varepsilon_-,{\bf k_-})
\Gamma^{RR(RA)}(\varepsilon_-,{\bf k}_-;\varepsilon_+,{\bf k}_+),
\label{PhiRA}
\end{equation}
which are diagrammatically represented by Fig. \ref{loop_fi}
( ${\bf k}_{\pm}={\bf k}\pm\frac{\bf q}{2}$, $\varepsilon_{\pm}=\varepsilon\pm\frac{\omega}{2}$). 
Vertices
$\Gamma^{RR(RA)}(\varepsilon_-,{\bf k}_-;\varepsilon_+,{\bf k}_+)$ 
contain all vertex corrections from ``external'' interaction
(order parameter fluctuations, impurities, phonons etc.)
but {\em do not contain} vertex corrections from Hubbard interaction.

Thus one achieves a significant simplification of the problem.
To calculate optical conductivity within  DMFT+$\Sigma$ approach
we have only to solve  single-particle problem of obtaining the
local self-energy $\Sigma(\varepsilon_{\pm})$ with the help of DMFT+$\Sigma$
procedure described above,
while non-trivial contribution from non-local correlations
or ``external'' perturbations enters via ``blocks'' (\ref{fi0RA_func}),
which can be calculated in any suitable approximation
accounting  only for ``external'' interaction, with
``bare'' Green's functions (\ref{G0pg}), which contains the
local self-energy from  DMFT+$\Sigma$ procedure.
In fact Eq. (\ref{cond_final}) provides also an effective algorithm
to compute optical conductivity in framework of conventional DMFT
(neglecting all ``external'' non-local correlations). 
In this case (\ref{fi0RA_func}) could be easily found from the simple loop 
diagram, defined by two Green's functions and free {\em scalar} vertices.
To get optical conductivity there is actually no need to calculate
vertex corrections in the framework of DMFT itself, as it was first shown
for the loop with {\em vector} vertices in Refs. \cite{georges96,pruschke}. 

In the following, in Sections \ref{dmfts_use} and \ref{pg_real} 
we shall discuss in details some applications of the generalized  DMFT+$\Sigma$ 
approach to the solution of concrete physical problems.

\subsection{Some other generalizations of DMFT}
\label{clusterDMFT}

Up to now, a number of different theoretical approaches were suggested to 
account for non-local effects within generalizations of DMFT. Here we restrict 
ourselves to a brief review of some of these approaches.

First of all we shall refer to cluster methods already mentioned above.
Instead of an isolated Anderson impurity one can consider some
cluster containing several such impurities, treating single site correlations
within DMFT, while considering intersite correlations by some other method.
This is basic idea of the so called cluster DMFT methods
\cite{georges96,clusterDMFT,LichtensteinDCA}. 

A drawback of such methods is related to different treatment of
non-local correlation inside cluster and between clusters, though 
from the physical point of view
(for example because of translational invariance) they should be identical.
To overcome this difficulty it was proposed to average self-consistent
``cluster'' self-energy over pairs of sites connected by
translation vectors \cite{clusterDMFT}. Unfortunately such averaging
procedure does not work well within the sel-consistent cycle of DMFT, since it 
breaks down analytical properties of Green's function.

In some sense alternative approach, named dynamical cluster approximation
(DCA), was proposed in Refs. \cite{LichtensteinDCA} and \cite{DCA,DCASC}.
This approach conserves translational invariance and provides physical behavior 
of Green's function. Within DCA the Brillouin zone is divided into several cells
with centers defined by appropriate vectors ${\bf K}$.
The same time self-energy is assumed to be constant in each cell
$\Sigma_{\bf k}(\omega)=\Sigma_{\bf K}(\omega)$,
but these constant are different for different $K$.
Distinction between DCA and cluster DMFT is that in DCA DMFT-cluster satisfies
periodic boundary conditions, instead of open boundary conditions in cluster 
DMFT.

The choice of particular cluster method is dictated by physical problem under 
consideration. Naturally, the common difficulty of cluster approaches is 
essentially larger computer time consumption in contrast to the standard DMFT,
which is connected with the solution of the appropriate cluster problem.
Nevertheless, a number of successes were achieved on this way.
Cluster DMFT generalizations were applied to different models
as well as to studies of some realistic systems.
Detailed review of these works can be found in Ref. \cite{TMrmp}.

Recently a number of diagrammatic DMFT generalizations was proposed,
attempting for more or less consistent construction of perturbation theory
over the inverse powers of spatial dimensionality, with the
standard DMFT used as the zeroth order approximation.
We mention in this connection the Ref. \cite{Kus} and also the so called 
dynamical vertex approximation (D$\Gamma$A), developed in Ref. \cite{Kat}.
Most promising in this respect seems to the dual fermion approach
formulated in Refs. \cite{Rub1,Rub2}, which is claimed to be a consistent 
realization of such perturbation theory. 
Unfortunately up to now only some simple model problems were solved
by these methods, while realistic systems where not yet considered at all.

\section{Application of generalized DMFT+$\Sigma$ approach to model problems.}
\label{dmfts_use}

\subsection{Strongly correlated systems in the pseudogap state.}
\label{pg_model}

\subsubsection{Pseudogap fluctuations.}

Striking example of strongly correlated systems are high-T$_c$ cuprates.
Parent stoichiometric cuprate compounds are antiferromagnetic insulators
with well developed optical gap and antiferromagnetism due to
spin ordering on copper ions with Neel temperature of the order of hundreds of 
$K$. This insulating state is rapidly destroyed by introduction of rather few
doping impurities. Thus these systems can be classified as doped Mott insulators
with strong electronic correlations.

Among many anomalies of the normal phase of high temperature
superconductors special interest attracts observations of a pseudogap
in the electronic spectra of underdoped cuprates \cite{Tim,MS}.
Despite continuing discussions about pseudogap nature,
from our point of view, most preferable is the scenario of pseudogap formation
due strong scattering of charge carriers on antiferromagnetic (AFM, SDW)
short range order fluctuations \cite{MS,Pines04}. 
In momentum representation this scattering is characterized by
momentum transfer of the order of  ${\bf Q}=(\frac{\pi}{a},\frac{\pi}{a})$ 
($a$ is two-dimensional lattice constant).
This leads to formation of certain features in single-particle spectrum, which 
are precursors of changes in the spectra due to the appearance of AFM long range
order (period doubling). 
As a result we end up with non Fermi liquid behavior (dielectrization)
of spectral density in the vicinity of the so called ``hot-spots'' on the Fermi 
surface, appearing at intersections of Fermi surface with borders of 
AFM Brillouin zone \cite{MS}.
 
In the framework of this spin-fluctuation scenario in works a simplified model 
of the pseudogap state was studied in Refs. \cite{MS,Sch,KS99}. 
This model is based on the assumption that for high enough temperatures dynamics 
of spin fluctuations can be neglected and one can consider instead the 
scattering of charge carriers by static Gaussian random field (quenched 
disorder) of pseudogap fluctuations (short range order AFM fluctuations).
Scattering intensity on fluctuations is characterized by a narrow peak
near scattering vectors of the order of ${\bf Q}$ with a width
defined by inverse correlation length $\kappa=\xi^{-1}$ and
corresponding energy scale  $\Delta$ (of the order of pseudogap crossover 
temperature $T^*$). 

Thus for momentum dependent self-energy we shall concentrate on the case of
electron scattering on such (SDW-like) antiferromagnetic spin 
fluctuations (similar consideration works fine also for CDW-like charge 
fluctuations) with short range order. 
To calculate $\Sigma_{\bf k}(i\omega)$ for the case of electrons propagating
in quenched random field of Gaussian spin (or charge) fluctuations with
dominating scattering momentum close to characteristic vector ${\bf Q}$
(``hot-spot'' model \cite{MS}), we shall use the generalized version of 
recurrent procedure proposed in Refs. \cite{Sch,KS99,C79},
allowing to take into account {\em all} Feynman diagrams 
describing scattering of electrons by this random field.
This becomes possible because of remarkable property of simplified
``hot-spot'' model: {\em contribution of arbitrary diagram with crossing 
interaction lines is equal to contribution of some diagram of the same order 
without crossing of those lines} \cite{C79}. 
Thereby we can restrict ourselves to consideration of non-crossing diagrams 
only and take into account other diagrams by combinatorial prefactors 
attributed to interaction lines \cite{KS99,C79}.
Finally we obtain the following recurrent relation for the self-energy
(continuous fraction representation \cite{KS99,C79}):
\begin{equation}
\Sigma_{n}(i\omega{\bf k})=\Delta^2\frac{s(n)}
{i\omega+\mu-\Sigma(i\omega)
-\varepsilon_n({\bf k})+inv_n\kappa-\Sigma_{n+1}(i\omega,{\bf k})}\;\;. 
\label{rec9}
\end{equation}
Here, the term  $\Sigma_{n}(i\omega,{\bf k})$ of recurrent procedure
contains all diagrammatic contributions with number of interaction lines 
$\geq n$.
Recurrent procedure for $\Sigma_{n}(i\omega,{\bf k})$ converges rather fast,
we can put  $\Sigma_{n}(i\omega,{\bf k})$ for large enough $n$ equal to zero and
performing straightforward computations obtain the desired physical self-energy
for $n=1$ \cite{KS99}, which can be subsequently used in DMFT+$\Sigma$ 
computational scheme:
\begin{equation}
\Sigma_{\bf k}(i\omega)=\Sigma_{n=1}(i\omega,{\bf k})
\label{Sk}
\end{equation}

Parameter $\Delta$ characterizes the energy scale (width) of the pseudogap,
$\kappa=\xi^{-1}$ is the inverse correlation length of
SDW (CDW) fluctuations, $\varepsilon_n({\bf k})=\varepsilon({\bf k+Q})$ and  
$v_n=|v_{\bf k+Q}^{x}|+|v_{\bf k+Q}^{y}|$ 
for odd $n$, $\varepsilon_n({\bf k})=\varepsilon({\bf k})$ and $v_{n}=
|v_{\bf k}^x|+|v_{\bf k}^{y}|$ for even $n$, where velocities projections
$v_{\bf k}^{x}$ and $v_{\bf k}^{y}$ are defined by usual derivatives
with respect to corresponding momenta components of the bare electron
dispersion $\varepsilon ({\bf k})$. At last, $s(n)$ are combinatorial prefactors
defining the number of coinciding diagrams:
\begin{equation}
s(n)=n
\label{vcomm}
\end{equation}
for the case of commensurate charge (CDW-type) fluctuations
with ${\bf Q}=(\pi/a,\pi/a)$ \cite{C79}. 
For incommensurate CDW fluctuations \cite{C79} 
(when ${\bf Q}$ is not related to lattice period) we get:
\begin{equation} 
s(n)=\left\{\begin{array}{cc}
\frac{n+1}{2} & \mbox{for odd $n$} \\
\frac{n}{2} & \mbox{for even $n$}.
\end{array} \right.
\label{vinc}
\end{equation}
If we want to take into account spin (Heisenberg) structure of interaction
with spin fluctuations for nearly antiferromagnetic Fermi liquid
(spin--fermion model \cite{Sch}), the combinatorics becomes more complicated.
Scattering processes preserving spin projection are controlled by
commensurate combinatorics, while spin flip scattering 
is described by diagrams of incommensurate type
(``charged'' random field, according to Ref.~\cite{Sch}).
In this model the recurrent procedure  (\ref{rec9}) for single-particle
Green's function remains the same but with another combinatorial
prefactors $s(n)$ \cite{Sch}:
\begin{equation} 
s(n)=\left\{\begin{array}{cc}
\frac{n+2}{3} & \mbox{for odd $n$} \\
\frac{n}{3} & \mbox{for even $n$}.
\end{array} \right.
\label{vspin}
\end{equation} 

Obviously this procedure introduces an important length scale $\xi$,
missed in standard DMFT. Physically this length scale reflects the
influence of short range order fluctuations (SDW or CDW) on electronic bath
surrounding effective Anderson impurity in DMFT. 

After the self-consistent solution of DMFT+$\Sigma$ set of equations 
(\ref{Gloc}-\ref{G009}) is obtained, one can calculate the spectral density
$A(\omega,{\bf k})$:
\begin{equation}
A(\omega,{\bf k})=-\frac{1}{\pi}{\rm Im}\frac{1}{\omega+\mu
-\varepsilon({\bf k})-\Sigma(\omega)-\Sigma_{\bf k}(\omega)},
\label{specf10}
\end{equation}
where $\Sigma(\omega)$, $\Sigma_{\bf k}(\omega)$ and chemical potential $\mu$
are already computed in a self-consistent way. Density of states can be found
by integration of (\ref{specf10}) over the Brillouin zone.

Analogous approach can be developed also to determine two-particle vertices.
Basic idea employed here is the possibility to get arbitrary vertex diagram
by introducing ``external field'' line into corresponding self-energy diagram 
\cite{C1,S91,SS02}.
In the model under consideration we can again restrict ourselves to
non-crossing diagrams, while contribution of all other diagrams can be accounted
for by combinatorial prefactors $s(n)$ attributed to interaction lines 
\cite{Sch,KS99,C79}.
Thus, all vertex diagrams are obtained from the simple ladder diagrams with
additional prefactors  $s(n)$ on corresponding interaction lines 
\cite{S91,SS02} (see also \cite{Diagr}).
Then we obtain the following system of recurrent relations for the vertex
$\Gamma^{RA}(\varepsilon_-,{\bf k}_-;\varepsilon_+,{\bf k}_+)$ \cite{SS02},
where contribution of local DMFT self-energy (obtained within the DMFT+$\Sigma$ 
procedure) is already included:
\begin{eqnarray}
&& {\Gamma}_{n-1}^{RA}(\varepsilon_-,{\bf k}_-;\varepsilon_+,{\bf k}_+)=
1+\Delta^2s(n)G_{n}^A(\varepsilon_-,{\bf k_-})
G_{n}^R(\varepsilon_+,{\bf k_+})\times\nonumber\\
&& \times\left\{1+ 
\frac{2iv_n\kappa k}{\omega-\varepsilon_n({\bf k}_+)
+\varepsilon_n({\bf k}_-)-\Sigma^R(\varepsilon_+)+\Sigma^A(\varepsilon_-)
-\Sigma_{n+1}^R(\varepsilon_+,{\bf k_+})
+\Sigma^A_{n+1}(\varepsilon_-,{\bf k_-})}\right\}\times\nonumber\\
&& \times{\Gamma}_n^{RA}(\varepsilon_-,{\bf k}_-;\varepsilon_+,{\bf k}_+),
\label{JrecRA}
%\nonumber\\
\end{eqnarray}
where
\begin{equation}
G^{R,A}_{n}(\varepsilon_{\pm},{\bf k}_{\pm})=\frac{1}{\varepsilon_{\pm}-
\varepsilon_n({\bf k}_{\pm}) \pm inv_n\kappa-\Sigma^{R,A}(\varepsilon_{\pm})
-\Sigma^{R,A}_{n+1}(\varepsilon_{\pm},{\bf k}_{\pm})}.
\label{G11}
\end{equation}
``Physical'' vertex
$\Gamma^{RA}(\varepsilon_-,{\bf k}_-;\varepsilon_+,{\bf k}_+)$
is defined as $\Gamma^{RA}_{n=0}(\varepsilon_-,{\bf k}_-;\varepsilon_+,
{\bf k}_+)$. Recurrent procedure (\ref{JrecRA}) accounts for {\em all}
diagrams of perturbation theory for the vertex part.
In the limit of $\kappa\to 0\quad (\xi\to\infty)$ (\ref{JrecRA}) can be reduced
to a series investigated in Ref. \cite{C1} (see also \cite{Sch}),
which can be exactly summed in analytical form.
Standard ladder approximation is reproduced if all combinatorial factors
in (\ref{JrecRA}) are made equal to one for all $n$ \cite{S91}.
Recurrent procedure for 
$\Gamma^{RR}(\varepsilon_-,{\bf k}_+;\varepsilon_+,{\bf k}_+)$
differs from (\ref{JrecRA}) only by the evident change of $A\to R$, as well as
replacing the whole expression in figure brackets on the r.h.s. of
Eq. (\ref{JrecRA}) by 1.
Eqs. (\ref{Gk9}), (\ref{rec9}), (\ref{JrecRA}) together with
(\ref{fi0RA_func}) and (\ref{cond_final}) provide the complete self-consistent
procedure to calculate optical conductivity within our model in the
framework of DMFT+$\Sigma$ approach.

Important aspect of our theory is the possibility of microscopic calculation
of both effective parameters $\Delta$ and $\xi$.
For example, applying two-particle self-consistent theory of Ref.~\cite{VT},
together with approximations introduced in Refs. \cite{Sch,KS99} for
two-dimensional Hubbard model, we derived a microscopic expression for $\Delta$ 
\cite{cm05}, which can be calculated within the standard DMFT.
It can be shown that for wide range of hole doping the pseudogap amplitude 
$\Delta$ varies in the interval from $t$ to $2t$ ($t$ is the nearest neighbor 
hopping integral).

\subsubsection{Basic electronic properties in the pseudogap state.}
\label{pg_mod_res}

Let us discuss results for the standard single band Hubbard model
on a square lattice with electron dispersion
\begin{equation}
\varepsilon({\bf k})=-2t(\cos k_xa+\cos k_ya)-4t'\cos k_xa\cos k_ya\;\;,
\label{spectr9}
\end{equation}
with $t$ and $t'$ nearest and next nearest hopping integrals.

Energy scale in the following is defined by nearest neighbor hopping integral 
$t$, and length scale by the lattice constant $a$.
Impurity solver used was the numerical renormalization group
(NRG) \cite{NRG,BPH}. Detailed computational results on single particle
properties demonstrating pseudogap anomalies can be found in Refs. 
\cite{cm05,JTL05,FNT06}, and on optical conductivity in Ref. \cite{PRB07}.
Here we only discuss most typical results corresponding mostly to the case of
$t'/t=-0.4$ (characteristic for cuprates) and band filling $n=0.8$ (hole doping).

{\bf Density of states and spectral function.}

Lets start with results obtained within generalized
DMFT+$\Sigma$ approach for the densities of states (DOS) in case
of rather weak (compared to bandwidth) Coulomb interaction $U=4t$.
Characteristic feature of strongly correlated metallic state is coexistence of
lower and upper Hubbard subbands splitted by Coulomb interaction $U$ with 
quasiparticle peak at the Fermi level \cite{pruschke,georges96}.
Noninteracting DOS for the square lattice has Van-Hove singularity near the 
Fermi level, so that in general the peak on the Fermi level can not be treated 
simply as a quasiparticle one. Actually there are two contributions to this peak:
(i) from quasiparticle peak appearing in strongly correlated
metals because of manybody effects and (ii)
smoothed Van-Hove singularity of noninteracting DOS \footnote{With decrease of 
Coulomb repulsion Van-Hove singularity gradually transforms into 
quasiparticle peak at $U=(6\div 8)t$.}.

\begin{figure}
\includegraphics[clip=true,width=0.45\textwidth]{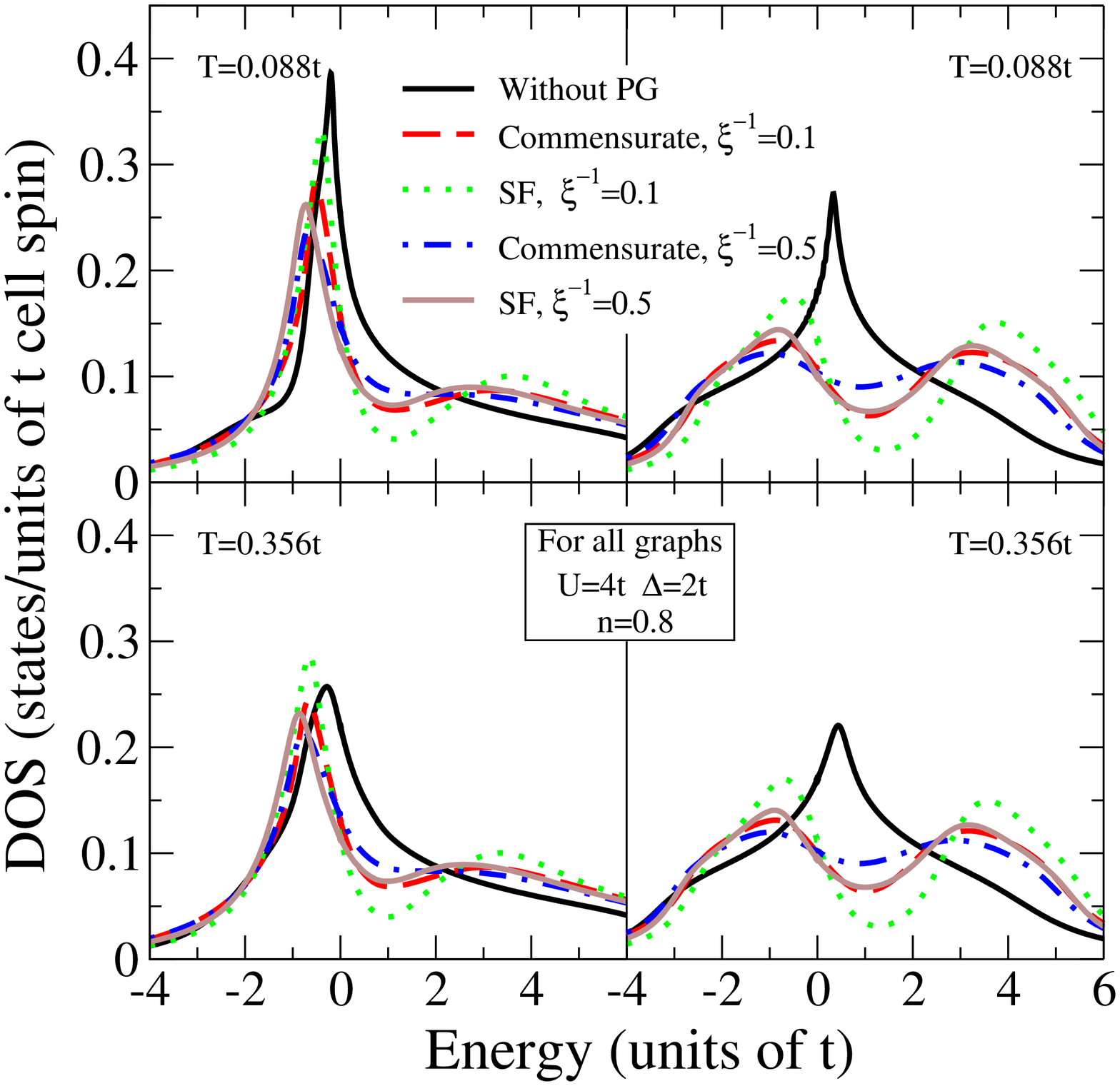}
\includegraphics[clip=true,width=0.55\textwidth]{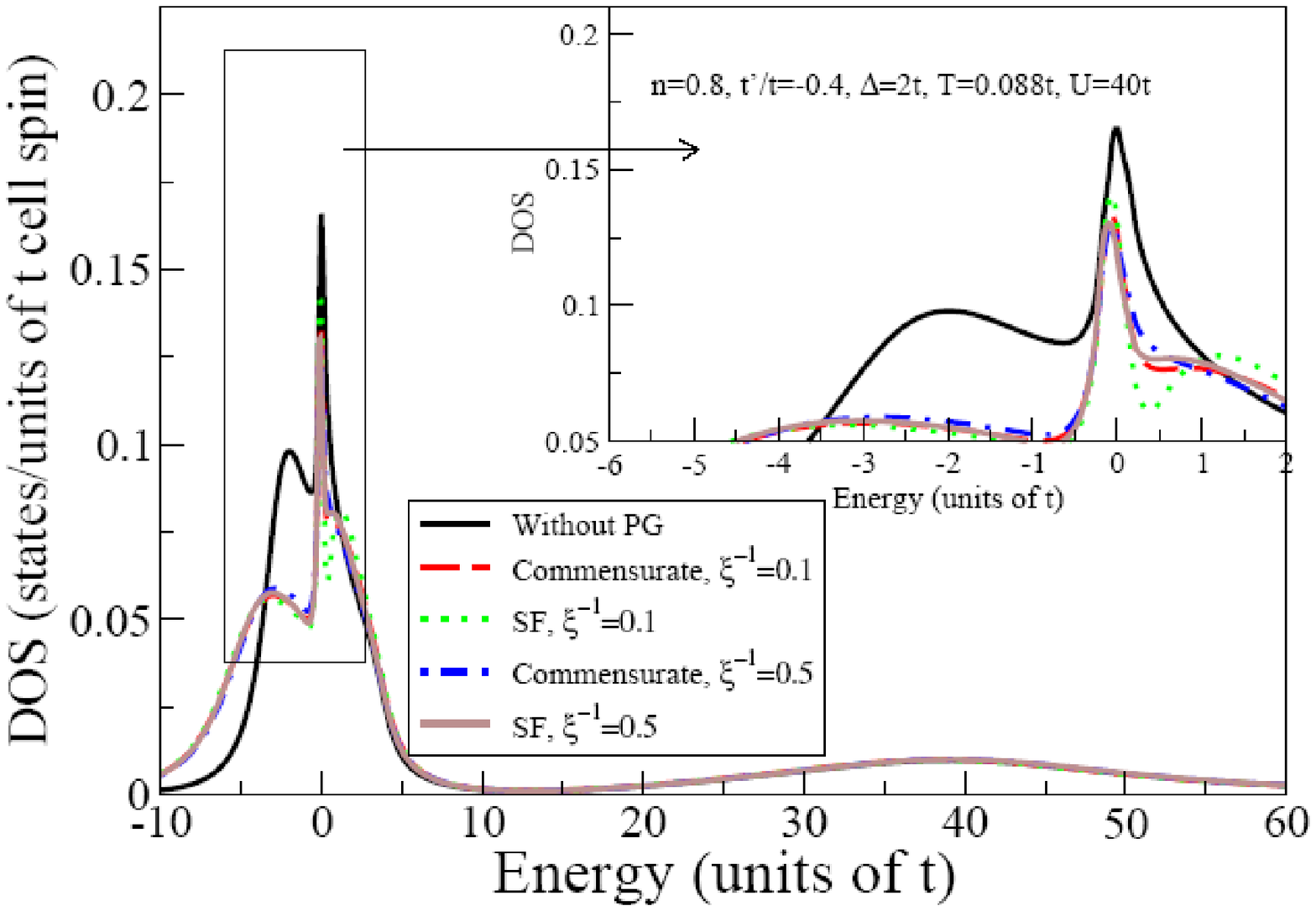}
\caption{\small Comparison of DMFT(NRG)+$\Sigma$ DOS calculated in Ref. \cite{cm05}
for different combinatorial prefactors 
(SF --- spin-fermion model, commensurate fluctuations),
inverse correlation lengths $\xi^{-1}$ (in units of lattice constant),  
pseudogap amplitude $\Delta=2t$ and band filling $n=0.8$.
On the left side -- $U=4t$,  $t'/t=-0.4$ (left column),  $t'=0$ (right column),  
temperature $T=0.088t$ (upper part) and $T=0.356t$ (lower part).
On the right side -- $U=40t$, $T=0.088t$.
Fermi level corresponds to zero energy.}
\label{DOS_4t_n08}
\end{figure}

On the left side of Fig.~\ref{DOS_4t_n08} we show 
DMFT+$\Sigma$ DOS with $n=0.8$ for the case of $t'/t=-0.4$ (left panel) and 
$t'=0$ (right panel) for two different temperatures $T=0.356t$ (lower panel) and 
$T=0.088t$ (upper panel).
Black curves are obtained in the absence of fluctuations.
Other curves on Fig.~\ref{DOS_4t_n08} present results for DOS
with nonlocal fluctuations with amplitude $\Delta=2t$.
For all parameter sets we see that introduction of non-local
fluctuations leads to pseudogap formation on a quasiparticle peak.
Width of the pseudogap (energy interval between corresponding peaks in the DOS)
is of the order of $\sim 2\Delta$.
Decrease of the value of $\Delta$ from $2t$ to $t$ gives twice smaller pseudogap 
width and makes it less deep. More pronounced pseudogap is found for
spin-fermion model combinatorial prefactors (see (\ref{vspin})), as compared
with the case of commensurate charge fluctuations 
(combinatorial prefactors (\ref{vinc})).
The influence of correlation length value corresponds to general expectations.
Decrease of correlation length or, for inverse correlation length, the change 
from $\xi^{-1}=0.1$ to $\xi^{-1}=0.5$ slightly smears the pseudogap. 
The rise of temperature from  $T=0.088t$ to $T=0.356t$ 
leads to general broadening of DOS structures.
One should note that DMFT+$\Sigma$ results for
$U=4t$ (which is less than the bandwidth $W$) are qualitatively similar
to results obtained earlier in the absence of Hubbard interaction \cite{Sch,KS99}.

Let us consider now the case of doped Mott insulator with Hubbard interaction 
value $U=40t$,  $t'/t=-0.4$ and band filling $n=0.8$  (right side of 
Fig.~\ref{DOS_4t_n08}).
Characteristic feature of DOS for such strongly correlated metals 
is strong splitting of lower and upper Hubbard bands with the Fermi
level within the lower Hubbard band (the case of hole doping)
In absence of nonlocal fluctuations again the quasiparticle peak is formed
on the Fermi level. However upper Hubbard band now is quite far away to the 
right and does not touch the quasiparticle peak (as it does for the case of weak 
Hubbard interaction).

For strong enough nonlocal fluctuations with $\Delta =2t$ pseudogap
appears in the middle of quasiparticle peak and the lower Hubbard band
is slightly broadened by fluctuations effects. Qualitatively
pseudogap anomalies behavior reminds that described above for the case of  
$U=4t$ -- decrease of $\xi$ smears the pseudogap and makes it less pronounced,
decrease of $\Delta$ from $\Delta =2t$ to $\Delta =t$ 
narrows the pseudogap and makes it more shallow (see.~\cite{cm05}).
Let us notice also that for the doped Mott insulator
pseudogap is more evident for spin SDW-like fluctuations
than for the charge CDW-like ones.

Nevertheless there are quite appreciable distinctions in contrast to
the $U=4t$ case. For example, the width of the pseudogap in DOS
is found to be essentially smaller than $2\Delta$
which is connected, in our opinion, with noticeable narrowing
of the quasiparticle peak itself caused by local correlations.

\begin{figure}
\includegraphics[clip=true,width=0.4\textwidth]{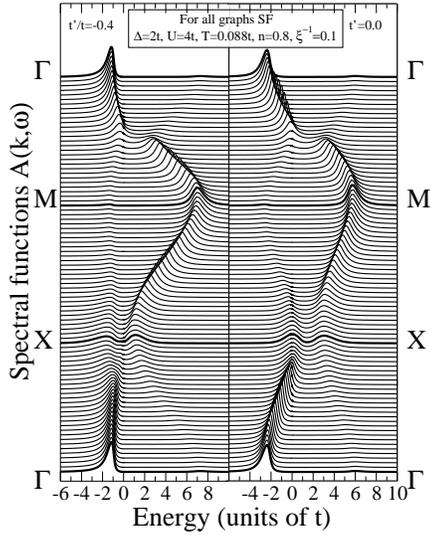}
\caption{\small DMFT(NRG)+$\Sigma$ spectral densities $A({\bf k},\omega)$
\cite{cm05} along high-symmetry directions of the first 
Brillouin zone $\Gamma(0,0)\!-\!\rm{X}(\pi,0)\!-\!\rm{M}(\pi,\pi)\!-\!\Gamma(0,0)$,
for spin-fermion combinatorics (SF).
The Fermi level is at zero energy.}
\label{n08_U4t_tri}
\end{figure}

In Fig.~ \ref{n08_U4t_tri} we show spectral densities $A(\omega,{\bf k})$, 
calculated within the DMFT+$\Sigma$ approach along high-symmetry directions of 
the first Brillouin zone: 
$\Gamma(0,0)\!-\!\rm{X}(\pi,0)\!-\!\rm{M}(\pi,\pi)\!-\!\Gamma(0,0)$.
In fact this figure shows the quasiparticle band of manybody system --- 
positions of maxima of spectral functions define
quasiparticle dispersion, while their width defines quasiparticle damping.
Also we clearly observe the partial reconstruction (``destruction'')
of this band by pseudogap fluctuations. 
One can see characteristic double-peak structure close to  $X$-point of the 
Brillouin zone.
In the middle of $M\!-\!\Gamma$ direction (the so called ``nodal''-point) one 
can observe the rise of the pseudogap, i.e. the ``memory'' of the AFM gap, 
which has maximum here in the case of AFM long range order. Generally speaking 
varying of filling leads to a shift of spectral functions with respect to 
the Fermi level.

{\bf  Fermi surface ``destruction''.}

Within conventional DMFT Fermi surface is not renormalized
by interaction i.e. it stays the same as for the bare quasiparticles 
\cite{vollha93}.
However in the case of nontrivial self-energy momentum dependence 
substantial renormalization of the Fermi surface appears due to pseudogap 
formation \cite{Sch}.
There are several ways to define Fermi surface for strongly correlated systems.
Below we shall exploit intensity map of the spectral function (spectral density)
(\ref{specf10}) for $\omega=0$, which is often called the Fermi surface map.
Such a map is directly measured by ARPES experiments and positions of its
intensity maxima specify the Fermi surface in a sense of the usual Fermi liquid 
theory, in case of quasiparticle damping becoming negligibly small.

\begin{figure}
\includegraphics[clip=true,width=0.5\textwidth]{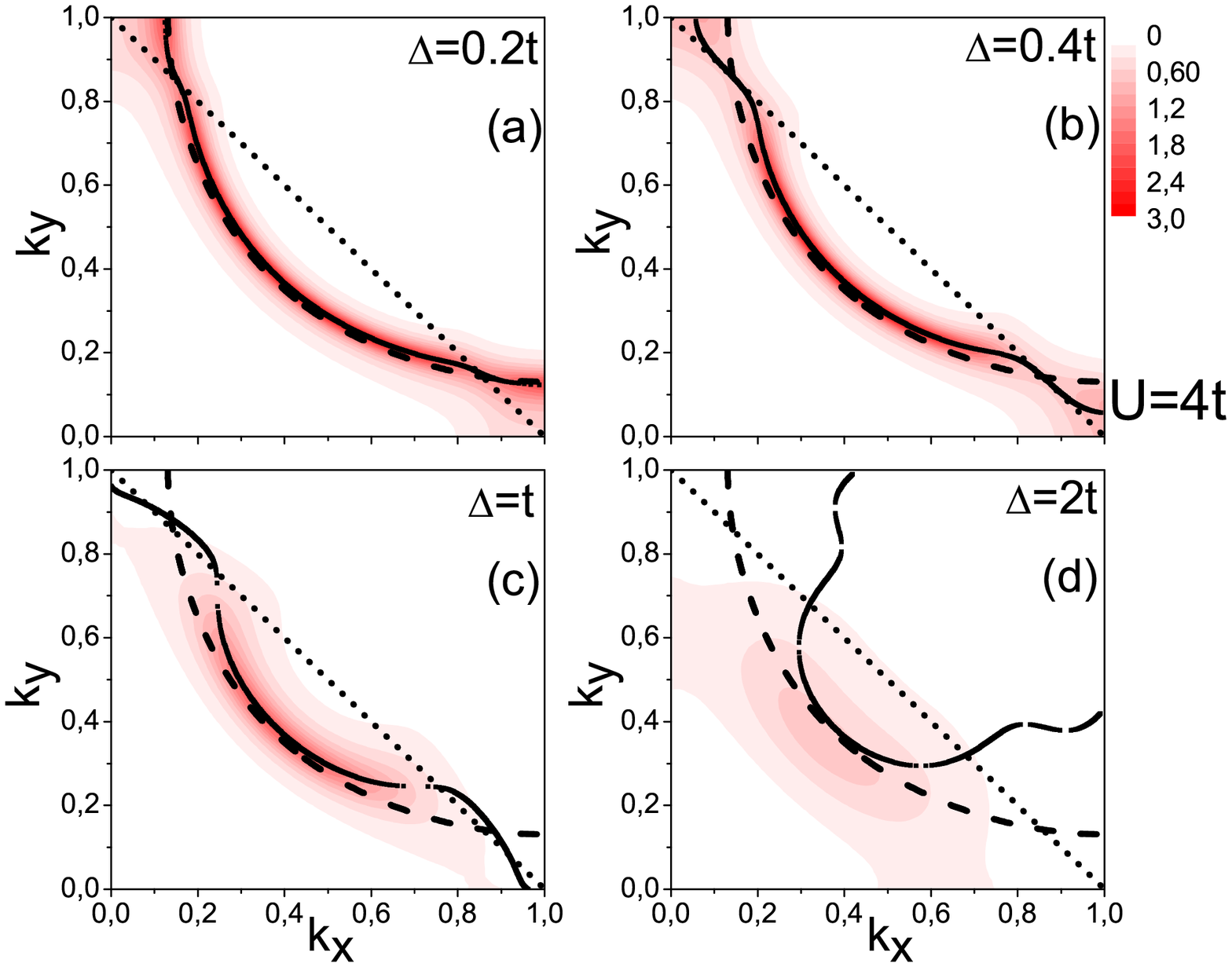}
\includegraphics[clip=true,width=0.5\textwidth]{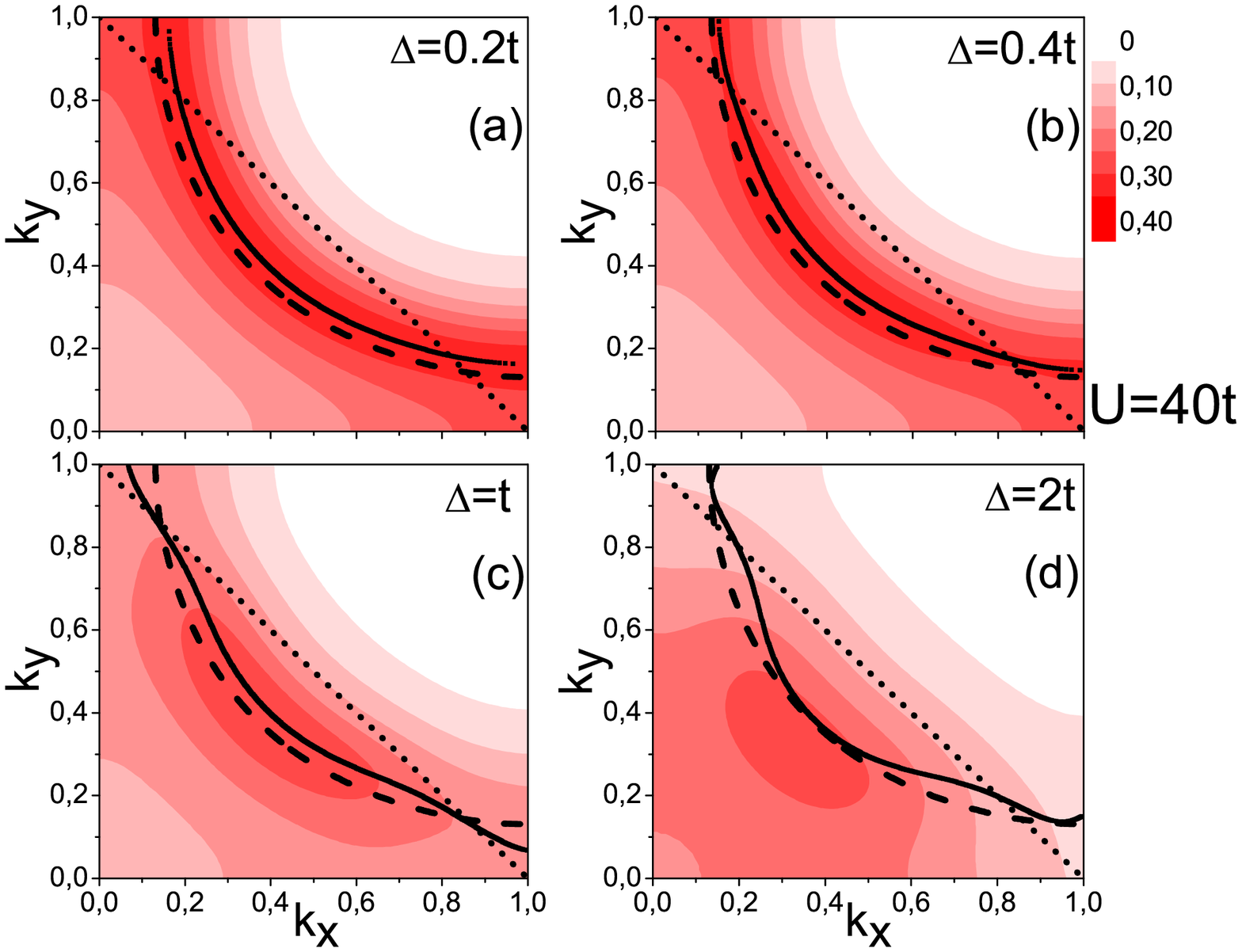}
\caption{\small DMFT(NRG)+$\Sigma$ picture of the Fermi surface ``destruction'' 
Ref. \cite{JTL05}
On the left -- for $U=4t$.
On the right -- for $U=40t$ 
(a) $\Delta=0.2t$;\ (b) $\Delta=0.4t$;\ (c) $\Delta=t$;\
(d) $\Delta=2t$.
Band filling is $n=0.8$. Dashed line --  ``bare'' Fermi surface.
Full lines -- solutions of Eq. (\ref{ReFS}).}
\label{FS_U4}
\end{figure}

On the right side of the Fig. \ref{FS_U4} there are displayed these maps
for case of strongly correlated metal with $U=4t$.
This contour plot of the spectral function clearly demonstrates
the ``destruction'' of the Fermi surface in the ``hot-spots'' together with
formation of the ``Fermi arcs'' upon  $\Delta$ growth
similar to that observed in pioneer works of Norman $et~al.$ \cite{Norm},
which were later confirmed in a large number of other works.
One should note that qualitatively analogous behavior is also found in the 
absence of local electronic correlations ($U=0$) \cite{JTL05,KS06}. 
The role of finite $U$ values add up to the decrease of spectral function 
intensity compared to the case $U=0$ and leads to additional ``smearing''
making ``hot-spots'' less visible.
``Destruction'' of the Fermi surface starts in the vicinity of ``hot-spots'' for 
small $\Delta$ values, but practically simultaneously it disappears in the whole
antinodal region (near points X($\pi$,0), Y(0,$\pi$)) of the Brillouin zone, 
while only ``Fermi arcs'' in the nodal region remain, with the shape close to 
bare Fermi surface.
Those results naturally explain why in ARPES the clear ``hot-spots'' behavior is
rather rarely observed \cite{Arm}.
In more details the question of possibility to observe ``hot-spots''
will be elucidated in the section below, devoted to LDA+DMFT+$\Sigma$ 
description of realistic cuprates.

In the case of doped Mott insulator with $U=40t$ shown in Fig.~\ref{FS_U4} we
see that ``Fermi surface'' is rather ill defined for all values of $\Delta$.
The profile of spectral function is significantly more ``smeared'' in contrast 
to smaller values of $U$ reflecting important role of local correlations. 
For comparison in Fig. \ref{FS_U4} we also show the renormalized
Fermi surfaces obtained within the model by formal solution of
Eq. (\ref{ReFS})
\footnote{This definition is used in the standard Fermi liquid theory.
In fact in our particular case influence of nonlocal pseudogap fluctuations
leads to qualitative changes of simple Fermi liquid picture.
Herewith we leave aside the question whether Fermi liquid is applicable
for {\em such} defined Fermi surface or not in the limit  $T\to 0$
since static approximation used is by construction the high temperature one --
short range order AFM fluctuations can be considered as quasistatic
only if $T\gg\omega_{sf}$, where $\omega_{sf}$ - characteristic frequency of
spin fluctuations \cite{Sch,KS99}.}:
\begin{equation}
\omega-\varepsilon({\bf k})+\mu-Re\Sigma(\omega)-Re\Sigma_{\bf k}(\omega)=0
\label{ReFS}
\end{equation}
for $\omega=0$ used for example in the work \cite{Sch}.
Obviously, this definition gives Fermi surface close to the one obtained from
intensity map for small $\Delta$, but does not account for significant damping
essentially important for large $\Delta$.
For large pseudogap amplitudes this definition of the Fermi surface is 
qualitatively adequate to true behavior, resulting from spectral function 
analysis, only in the nodal region.
Actually, the contour plot of spectral function (at $\omega=0$)
gives most complete and natural representation of the Fermi surface for the 
systems with strong correlations and nonlocal fluctuations of some order 
parameter, which are present in a wide region of the phase diagram of high-T$_c$ 
cuprates because of their low dimensionality. Results obtained in a such an
approach directly correspond to ARPES experiments, where exactly this 
definition of the Fermi surface is most conventional.

{\bf  Optical conductivity.}

Lets set about discussion of DMFT+$\Sigma$ results for optical conductivity.
\begin{figure}
\includegraphics[clip=true,width=0.5\textwidth]{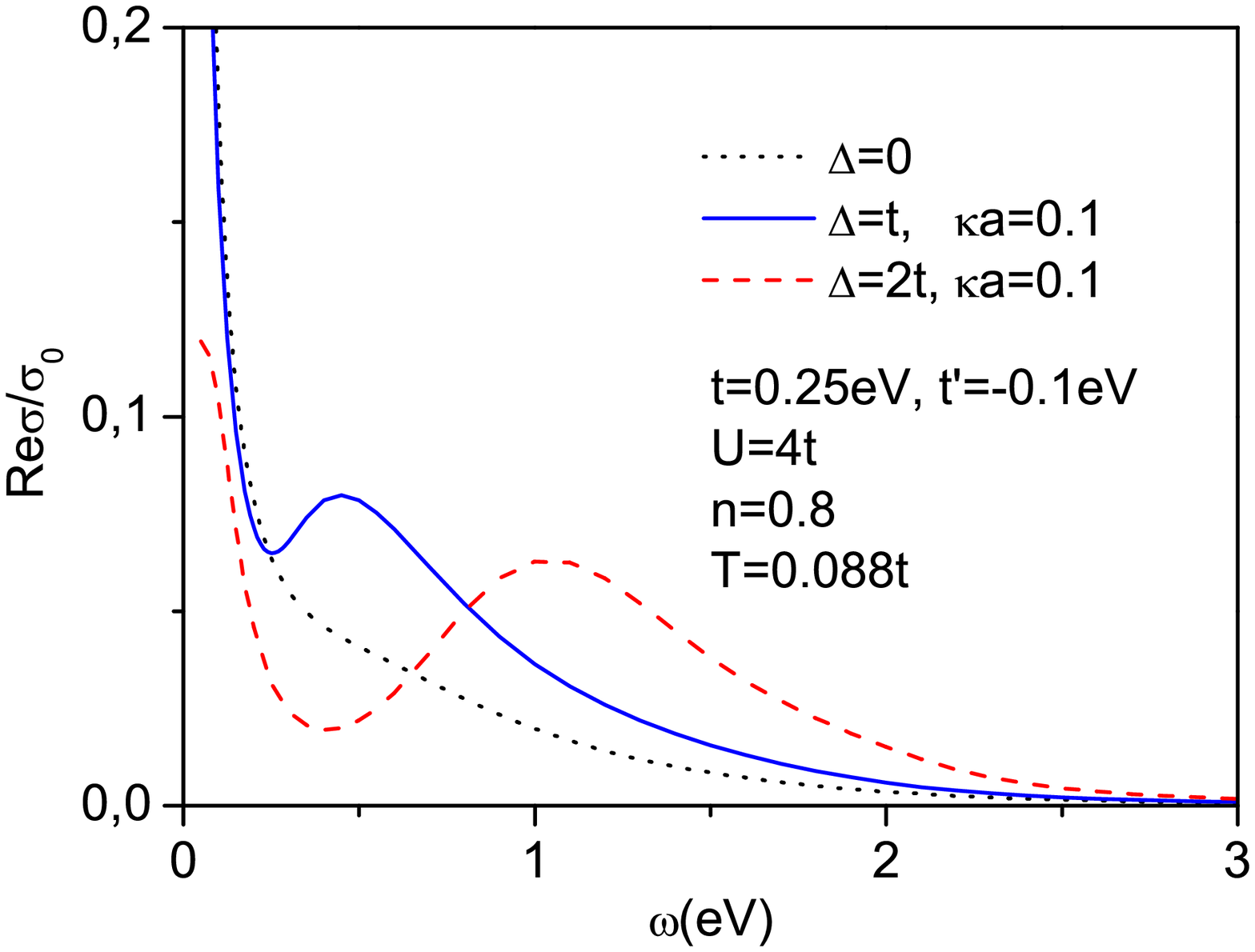}
\includegraphics[clip=true,width=0.5\textwidth]{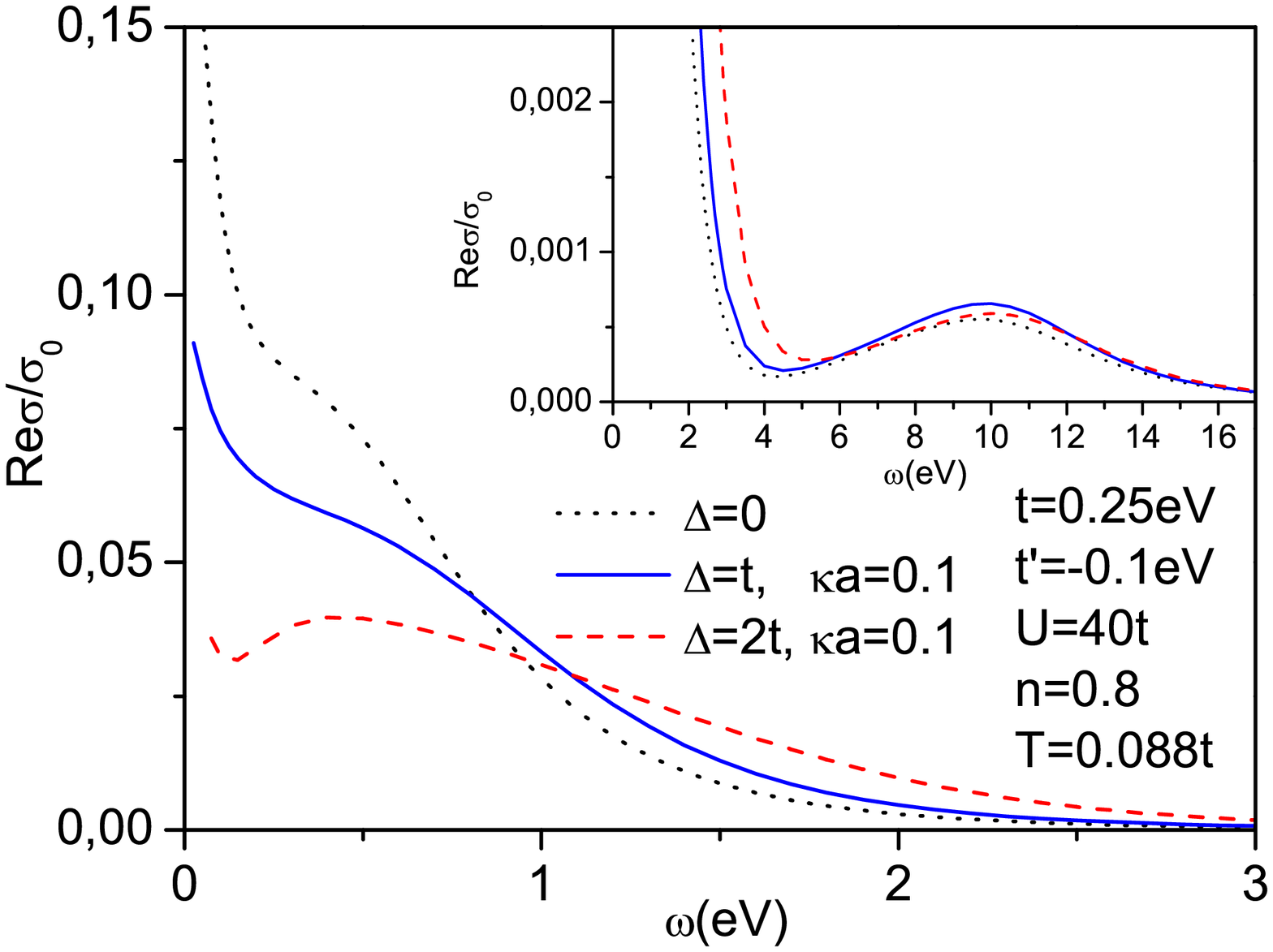}
\caption{\small Real part of DMFT+$\Sigma$ optical conductivity
($t'=-0.4t$, $t=0.25$ eV) obtained in Ref. \cite{PRB07}  
for different values of the pseudogap amplitude:
$\Delta=0$, $\Delta=t$, $\Delta=2t$ .
Temperature is $T=0.088t$, band filling -- $n=0.8$ and  
correlation length $\xi=10 a$. On the left side -- strongly correlated metal 
with $U=4t$. On the right side -- doped Mott insulator with $U=40t$. 
Inset: conductivity in a wide frequency range, which includes transitions to the
upper Hubbard band.} 
\label{DMFT_S_D} 
\end{figure}
On the left panel of  Fig. \ref{DMFT_S_D} we show DMFT+ $\Sigma$ results
for the real part of optical conductivity in the case of strongly correlated 
metal ($U=4t$) for different values of the pseudogap amplitude.
We clearly observe the formation of typical pseudogap anomaly on the ``shoulder''
of the Drude peak and it grows as $\Delta$ increases.
This behavior is rather similar to ``mid-infrared feature'' which is
observed in optical conductivity of cuprate superconductors \cite{Bas,Timu}. 
The rise of temperature and decrease of fluctuations correlation length 
wash off pseudogap, making this anomaly less pronounced \cite{PRB07}. 

The right panel of Fig. \ref{DMFT_S_D} demonstrates
DMFT+ $\Sigma$ optical conductivity of doped Mott insulator
($U=40t$) for several values of the pseudogap amplitude.
We see that frequency range where pseudogap anomaly is observed
gets narrower with the growth of local correlation strength and for
large $U$ values pseudogap anomalies are strongly suppressed.
Pseudogap fluctuations lead to noticeable changes of optical conductivity
only for relatively low frequencies, of the order of  $\Delta$.
For higher frequencies (e.g. of the order of $U$, where transitions to upper 
Hubbard band take place) pseudogap effects are not seen
(see also inset on the right panel of Fig. \ref{DMFT_S_D}).
For low frequencies we observe suppression of Drude peak with rather weak
anomaly at $\omega\sim\Delta$ which disappears for small $\Delta$ values
or for short correlation lengths.

\subsection{Mott-Anderson transition in disordered systems.}
\label{MA_trans_dis}

The importance of both electron interactions and disorder effects
in the condensed matter research is well known \cite{Lee85}.
Coulomb interaction and disorder are two driving forces
leading to metal-insulator transition, connected with localization and
delocalization of charge carriers. In particular Mott-Hubbard transition is 
induced by electron repulsion \cite{NFM,Mott90}, while Anderson metal-insulator transition
is related to scattering of noninteracting particles by
impurities \cite{Anderson58}. It is well known that a subtle competition between disorder 
effects and interaction has many manifestations \cite{Lee85,ma}. Most relevant this problem 
is in the case of strong disorder and strong electron correlations,
determining physical mechanisms of Mott-Anderson metal-insulator transition \cite{Lee85}.

One of the main models allowing for the account of both electronic correlations
(leading to Mott metal-insulator transition \cite{NFM,Mott90}) and strong
disorder effects (leading to Anderson metal-insulator transition)
is Anderson-Hubbard model
\cite{HubDis, Dobrosavljevic97, Dobrosavljevic03, BV, ShapiroPRB08, Shapiro08, PB09}.

In Refs. \cite{Dobrosavljevic97, Dobrosavljevic03, BV} three-dimensional 
Hubbard-Anderson model was investigated in the framework of dynamical mean-field 
theory  (DMFT) \cite{MetzVoll89,vollha93,georges96,pruschke}.
Influence of local disorder was taken into account through averaged density of states (DOS)
\cite{ulmke95,vlaming92} within the well known coherent potential approximation (CPA),
which does not describe Anderson localization. To overcome this difficulty
in Dobrosavljevic and Kotliar \cite{Dobrosavljevic97} has proposed a version of DMFT,
where the self-consistent solution of stochastic DMFT equations
for an ensemble of systems with given realizations of disorder, was used to calculate
the averaged logarithmic (geometric mean) density of states, which
gives information on critical disorder for Anderson transition.
Further this approach was developed in Refs. \cite{Dobrosavljevic03, BV}
where highly nontrivial phase diagram of three-dimensional paramagnetic
Anderson-Hubbard model \cite{BV} was obtained, containing correlated metal phase, 
Mott insulator phase and correlated Anderson insulator phase.
Main problem of the approach used in Refs. \cite{Dobrosavljevic97, Dobrosavljevic03, BV}
is the impossibility of direct computation of measurable physical properties
such as conductivity, which actually defines metal-insulator transition.

At the same time there exists the well developed self-consistent theory
of Anderson localization, based on the solution of equations for the
generalized diffusion coefficient.
Efficiency of this approach in the absence of interactions
is known for a long time \cite{VW,Diagr,MS,WV,MS86,VW92},
certain attempts to include interaction effects into this
approach with some promising results were undertaken in Refs. \cite{MS86,KSS95}.
However, up to now this approach was not extended to modern
theory of strongly correlated systems.
For the first time such investigation was performed in Ref. \cite{HubDis} for 
three-dimensional systems and later for two-dimensional case \cite{HubDis2}.

Let us consider disordered paramagnetic Anderson-Hubbard model (mostly) at half-filling
for arbitrary interaction and disorder strength.
Obviously this model contains both Mott-Hubbard and Anderson metal-insulator transitions.
Hamiltonian of the model is:
\begin{equation}
H=-t\sum_{\langle ij\rangle \sigma }a_{i\sigma }^{\dagger }a_{j\sigma
}+\sum_{i\sigma }\epsilon _{i}n_{i\sigma }+U\sum_{i}n_{i\uparrow
}n_{i\downarrow },  
\label{And_Hubb}
\end{equation}
where $t>0$ is nearest neighbor hopping amplitude, while $U$ is 
on-site Hubbard repulsion, $n_{i\sigma }=a_{i\sigma }^{\dagger }a_{i\sigma }^{{%
\phantom{\dagger}}}$ is particle number operator, $a_{i\sigma }$ 
($a_{i\sigma }^{\dagger}$) is annihilation (creation) operator of electron 
on site $i$ with spin $\sigma$.  Local energies $\epsilon _{i}$ are assumed to be
random and independent at different lattice sites.
To simplify diagram technique hereafter we assume the Gaussian distribution
for $\epsilon _i$:
\begin{equation}
\mathcal{P}(\epsilon _{i})=\frac{1}{\sqrt{2\pi}\Delta}\exp\left(
-\frac{\epsilon_{i}^2}{2\Delta^2}
\right).
\label{Gauss}
\end{equation}
Parameter $\Delta$ here is the measure of disorder.
Such Gaussian random field (``white noise'') of energy levels $\epsilon _{i}$
at different lattice sites is equivalent to impurity scattering
and can be described by standard diagram technique for the
averaged Green functions \cite{Diagr}.

Self-energy $\Sigma_{\bf p}(i\varepsilon)$ caused by scattering by disorder
calculated in simple one-loop approximation neglecting ``crossing'' diagrams
(i.e. in self-consistent Born approximation) \cite{Diagr},
in case the of Gaussian disorder (\ref{Gauss}) reduces to:
\begin{equation}
\Sigma_{\bf p}(i\varepsilon)=\Delta^2\sum_{\bf p}G(i\varepsilon,{\bf p})
\equiv \Sigma_{imp}(i\varepsilon),
\label{BornSigma}
\end{equation}
so that our ``external'' self-energy is independent of momentum ${\bf p}$ 
(local).

To analyze optical conductivity we shall apply the general DMFT+$\Sigma$ 
expression (\ref{cond_final}).
Most important block $\Phi^{0RA}_{\varepsilon}(\omega,{\bf q})$ can be
obtained using the ideology of self-consistent theory localization
\cite{VW,Diagr,WV,MS,MS86,VW92}, with some generalizations to account for
the role of Hubbard interaction via DMFT+$\Sigma$ approach \cite{HubDis,HubDis2}.
Main distinction from the standard derivation of the equations of self-consistent 
theory of localization is the use of Green's functions  (\ref{Gk9})
containing local contribution to self-energy from Hubbard interaction.

Following  standard derivation  \cite{VW,Diagr,MS,WV,MS86,VW92}
we obtain diffusion-like (for small $\omega$ and $q$) contribution to
$\Phi^{0RA}_{\varepsilon}(\omega,{\bf q})$ which takes the form:
\begin{equation}
\Phi^{0RA}_{\varepsilon}({\bf q},\tilde\omega)=\frac{2\pi iN(\varepsilon)}
{\tilde\omega+iD(\omega)q^2},
\label{FiRA} 
\end{equation}
where $D(\omega)$ is the generalized diffusion coefficient
and important difference from the single-particle case is
contained in:
\begin{equation}
\tilde\omega=\varepsilon_+-\varepsilon_-
-\Sigma^R(\varepsilon_+)+\Sigma^A(\varepsilon_-)=
\omega-\Sigma^R(\varepsilon_+)+\Sigma^A(\varepsilon_-)\equiv
\omega-\Delta\Sigma^{RA}(\omega),
\label{tomega}
\end{equation}
which substitutes for the usual $\omega$ term in the denominator of standard 
expression for $\Phi^{0RA}_{\varepsilon}(\omega,{\bf q})$. 
From general considerations it is clear that in metallic phase for $\omega\to 0$ 
we have
$\Delta\Sigma^{RA}(\omega=0)=2 i{\rm Im}\Sigma(\varepsilon)\sim Max\{T^2,\varepsilon^2\}$,
which reflects Fermi liquid behavior in DMFT 
(which is not violated by elastic impurity scattering). 
For finite $T$ it leads to usual phase decoherence caused by
(inelastic) electron--electron scattering \cite{Lee85,ma}.

Then Eq. (\ref{cond_final}) takes the form:
\begin{equation}
{\rm{Re}}\sigma(\omega)=\frac{e^2\omega}{2\pi}
\int_{-\infty}^{\infty}d\varepsilon\left[f(\varepsilon_-)
-f(\varepsilon_+)\right]{\rm Re}\left\{\frac{2\pi N(\varepsilon)D(\omega)}
{\omega^2} -
\phi^{0RR}_{\varepsilon}(\omega)\left[1-
\frac{\Delta\Sigma^{RR}(\omega)}{\omega}\right]^2\right\},
\label{con_fin}
\end{equation}
where the second term in figure brackets can be in fact neglected for small $\omega$,
while in case of necessity to describe a wide frequency range it can be calculated
using (\ref{fi0RA_func}) with $\Phi^{0RR}_{\varepsilon}(\omega,{\bf q})$ 
taken in the usual ladder approximation.

Now following the standard scheme of self-consistent theory of localization
\cite{VW,Diagr,MS,WV,MS86,VW92} we get the closed self-consistent equation
for generalized diffusion coefficient:
\begin{equation}
D(\omega)=i\frac{<v>^2}{d}\left\{\tilde\omega-
\Delta\Sigma_{imp}^{RA}(\omega)+
\Delta^4\sum_{\bf p}(\Delta G_{\bf p})^2\sum_{\bf q}
\frac{1}{\tilde\omega+iD(\omega)q^2}\right\}^{-1}
\label{Dsc}
\end{equation}
where $d$ is spatial dimensionality, $\Delta G_{\bf p}=G^R(\varepsilon_+,{\bf p})-G^A(\varepsilon_-,{\bf p})$, 
 $\Delta\Sigma^{RA}_{imp}(\omega)=\Sigma_{imp}^R(\varepsilon_+)
-\Sigma_{imp}^A(\varepsilon_-)$ and averaged velocity $<v>$, 
which can be well approximated just by the Fermi velocity, is
given by the following expression:
\begin{equation}
<v>=\frac{\sum_{\bf p}|{\bf v_p}|\Delta G_{\bf p}}
{\sum_{\bf p}\Delta G_{\bf p}},
\label{skor}
\end{equation} 
where ${\bf v_p}=\frac{\partial\epsilon({\bf p})}{\partial{\bf p}}$.
Eq. (\ref{Dsc}) should be solved together with self-consistent
DMFT+$\Sigma$ procedure (\ref{Gloc}--\ref{G009},  \ref{BornSigma}). 
In fact, this equation is a transcendental one and can be easily solved by
iterations for each $\tilde\omega$ value. 

In accordance with the usual applicability limit of diffusion approximation,
summation over $q$ in Eq. (\ref{Dsc}) should be restricted to \cite{MS86,Diagr}:
\begin{equation}
q<k_0=Min \{l^{-1},p_F\},
\label{cutoff}
\end{equation}
where $l=<v>/2\gamma$ is an elastic mean free path, 
$\gamma$ is Born scattering frequency by impurities, $p_F$  is the Fermi momentum.
It is well known that in two-dimensional case Anderson localization occurs
for any infinitely weak disorder.
However, in this case the localization radius is exponentially large and
sample size becomes essentially important.
Sample size L can be introduced into self-consistent localization theory
introducing the integration cutoff of diffusion pole at small $q$ \cite{VW, WV}, 
i.e. at:
\begin{equation}
q\sim k_L=1/L.
\label{cutoffd}
\end{equation}

For $\omega \rightarrow 0$ (on the Fermi surface ($\varepsilon=0$), and
obviously $\tilde\omega \rightarrow 0$)
in the Anderson insulator phase one gets localization behavior of
the generalized diffusion coefficient \cite{VW, WV,Diagr}:
\begin{equation}
D(\omega)=-i\tilde\omega {R_{loc}}^2.
\label{DR}
\end{equation}
After substitution of (\ref{DR}) into (\ref{Dsc}) 
one obtains equation defining localization radius
$R_{loc}$:
\begin{equation}
{R_{loc}}^2=-\frac{<v>^2}{d\Delta^4}
\left\{\sum_{\bf p}(\Delta G_{\bf p})^2\sum_{\bf q}
\frac{1}{1+{R_{loc}}^2 q^2}\right\}^{-1}.
\label{Rl2}
\end{equation}

\subsubsection{Three-dimensional systems.}
\label{3d_dis}

Below we present most interesting results for three-dimensional
Anderson-Hubbard model at half-filling on a cubic lattice
with semielliptic bare density of states with the bandwidth $W=2D$:
\begin{equation}
N_0(\varepsilon)=\frac{2}{\pi D^2}\sqrt{D^2-\varepsilon^2}.
\label{DOS13}
\end{equation}
Density of states is given in units of number of states in
energy interval for unit cell of the volume  $a^3$ 
($a$ is the lattice constant) and for one spin projection.
Conductivity values are given in natural units of
${e^2}/{\hbar a}$ ($a$ is the lattice constant).
For more detailed acquaintance with numerical results (also for the case of
deviations from half-filling) we refer the reader to Ref. \cite{HubDis}.

{\bf Evolution of the density of states}

Within the standard DMFT approach Hubbard model density of states at 
half-filling has a typical three-peak structure \cite{georges96,pruschke,Bull}
with a narrow quasiparticle peak (central peak) at the Fermi level
and wide upper and lower Hubbard bands situated at energies
$\varepsilon\sim\pm U/2$.
As $U$ grows quasiparticle band narrows within the metallic phase and disappears
at Mott-Hubbard metal-insulator transition at critical interaction 
value $U_{c2}\approx 1.5W$. With further increase of $U$ insulating gap
opens at the Fermi level.
\begin{figure}
\includegraphics[clip=true,width=0.5\textwidth]{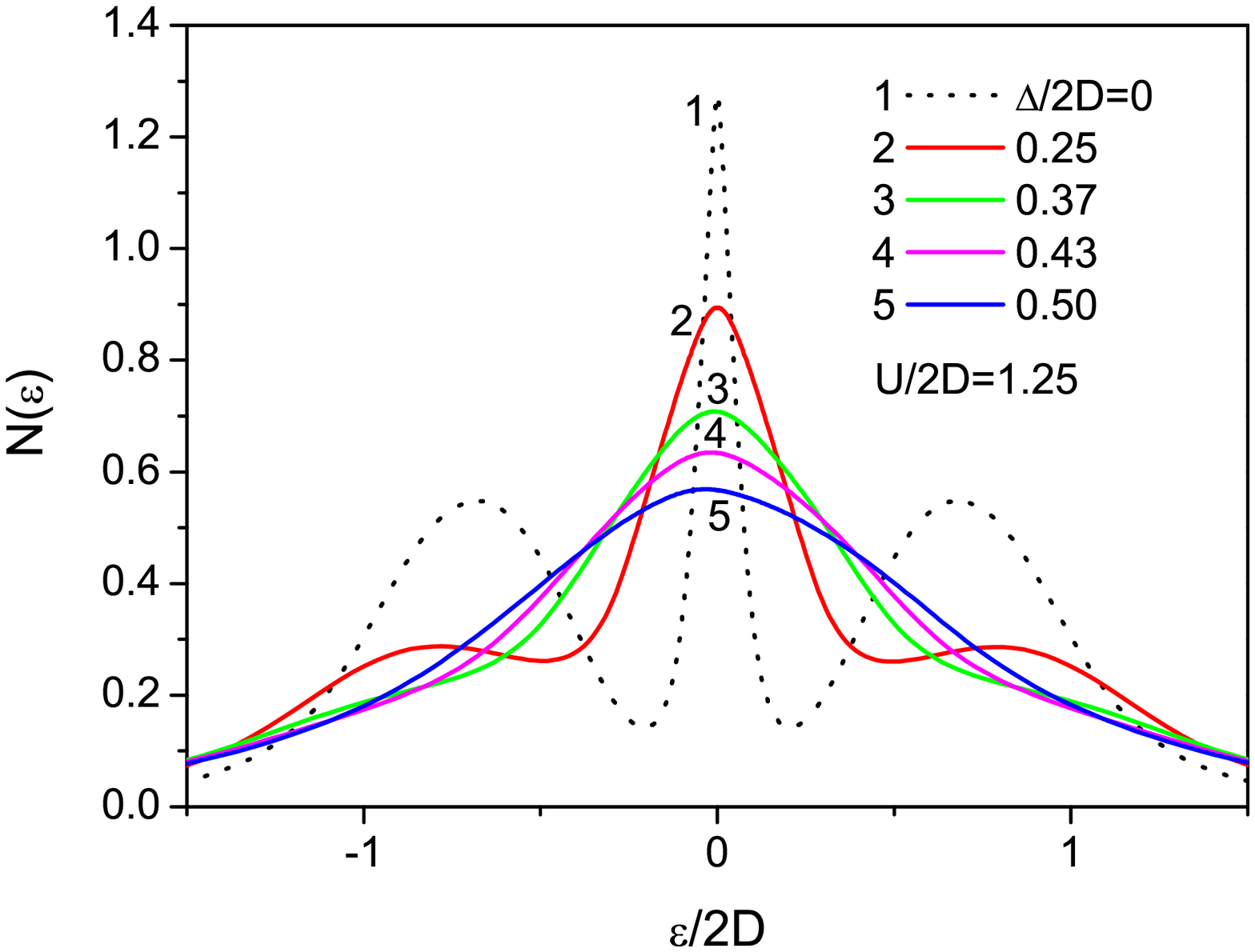}
\includegraphics[clip=true,width=0.5\textwidth]{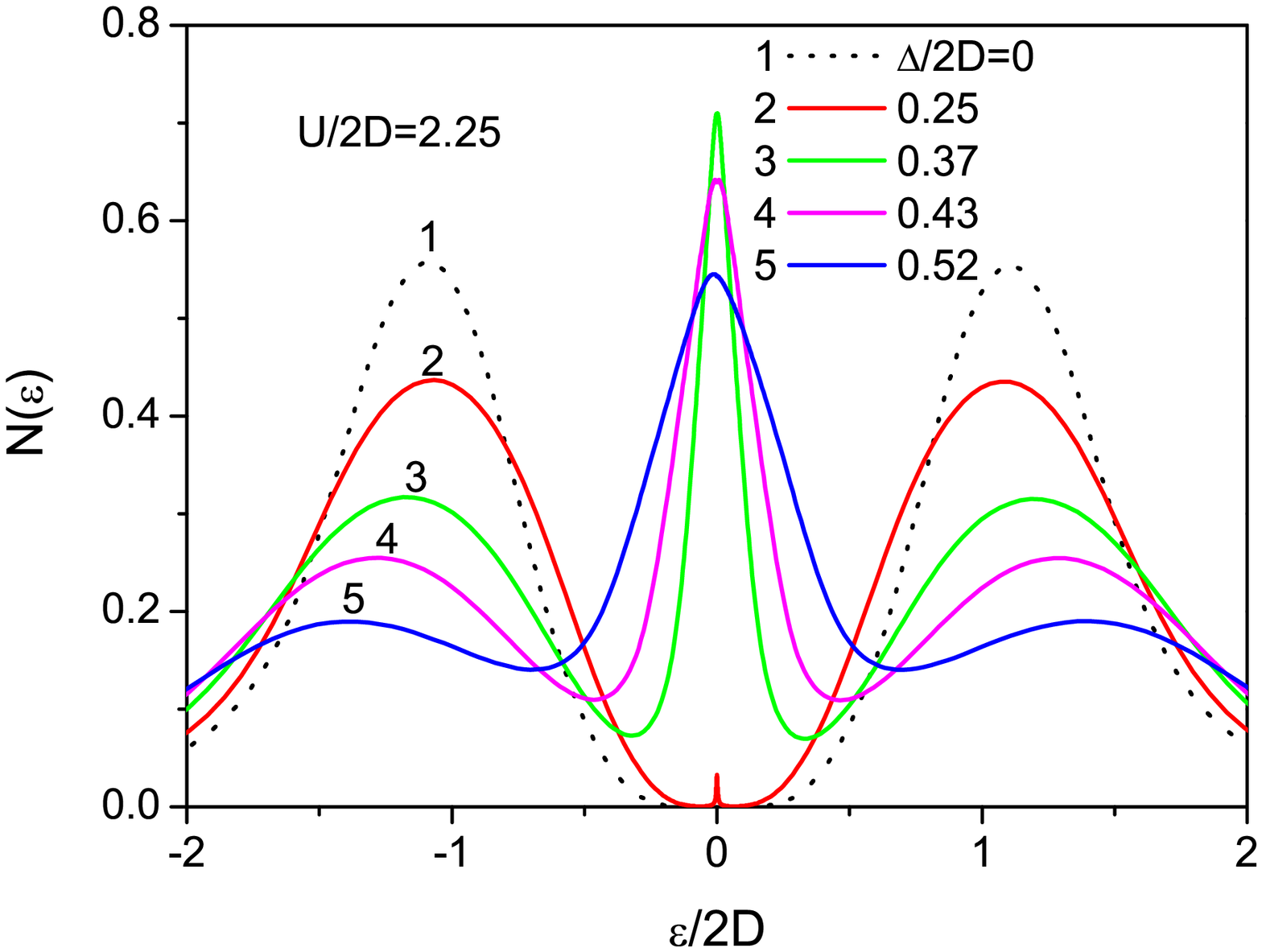}
\caption{\small Hubbard-Anderson model density of states at half-filling
for different disorder levels $\Delta$ \cite{HubDis}.
On the left side --- correlated metal with $U=2.5D$, 
On the right side --- Mott insulator with $U=4.5D$.} 
\label{metDOS} 
\end{figure} 

In Fig. \ref{metDOS} we present our results for DMFT+$\Sigma$ densities of 
states for typical strongly correlated metal with $U=2.5D=1.25W$, 
in the absence of disorder and for different values of disorder $\Delta$,
including strong enough disorder, transforming correlated metal to
correlated Anderson insulator (see also next section on conductivity).
As one can expect disorder leads to typical broadening and suppression
of the density of states.

More unexpected is the result obtained for $U=4.5D=2.25W$, typical for
Mott insulator and shown on right panel of Fig. \ref{metDOS}.
Here we observe restoration of the central peak (quasiparticle band)
in DOS with the increase of disorder, transforming Mott insulator
to correlated metal or to correlated Anderson insulator.
Similar DOS behavior was reported also in Ref. \cite{BV}. 

Physical origin of such quite unexpected central peak restoration
is pretty clear. Controlling parameter of metal-insulator transition
in DMFT is the ratio of Hubbard interaction $U$ to bare bandwidth $W=2D$.
With disordering (in the absence of Hubbard interaction) new effective 
bandwidth $W_{eff}$ appears which grows with disorder. Semielliptic form of DOS 
with well defined band edges within the self-consistent Born
approximation (\ref{BornSigma}) is preserved.
This leads to diminishing values of the ratio  $U/W_{eff}$, which in its turn
causes restoration of the quasiparticle band.
In more details this is discussed below, when we discuss the phase diagram of
Hubbard-Anderson model.

In absence of disorder characteristic feature of Mott-Hubbard metal-insulator 
transition is hysteresis DOS behavior appearing with the decrease of $U$
starting from insulating phase  \cite{georges96,Bull}.
Mott insulator phase is conserved (metastable) down to rather
small $U$ values deep within the correlated metal phase.
Metallic phase is restored only at about $U_{c1}\approx 1.0W$.
Corresponding interval $U_{c1}<U<U_{c2}$ typically is considered
as a coexistence region of metallic and Mott insulating phases,
where, from a thermodynamic point of view, metallic phase is more stable
\cite{georges96,Bull,Blum}. Such hysteresis DOS behavior (see \cite{HubDis})
is observed also in the presence of disorder and will be described below during
the discussion of the phase diagram of Anderson-Hubbard model.

{\bf Optical conductivity: Mott-Hubbard and Anderson transitions}

Without disorder our calculations reproduce conventional DMFT results
\cite{pruschke,georges96}, where optical conductivity is characterized
by the usual Drude peak at low frequencies and wide maximum at about
$\omega\sim U$, which corresponds to optical transitions to upper Hubbard band.
As $U$ grows Drude peak diminishes and disappears at Mott transition.
Introduction of disorder leads to qualitative change of the frequency dependence
of optical conductivity.

\begin{figure}
\includegraphics[clip=true,width=0.54\textwidth]{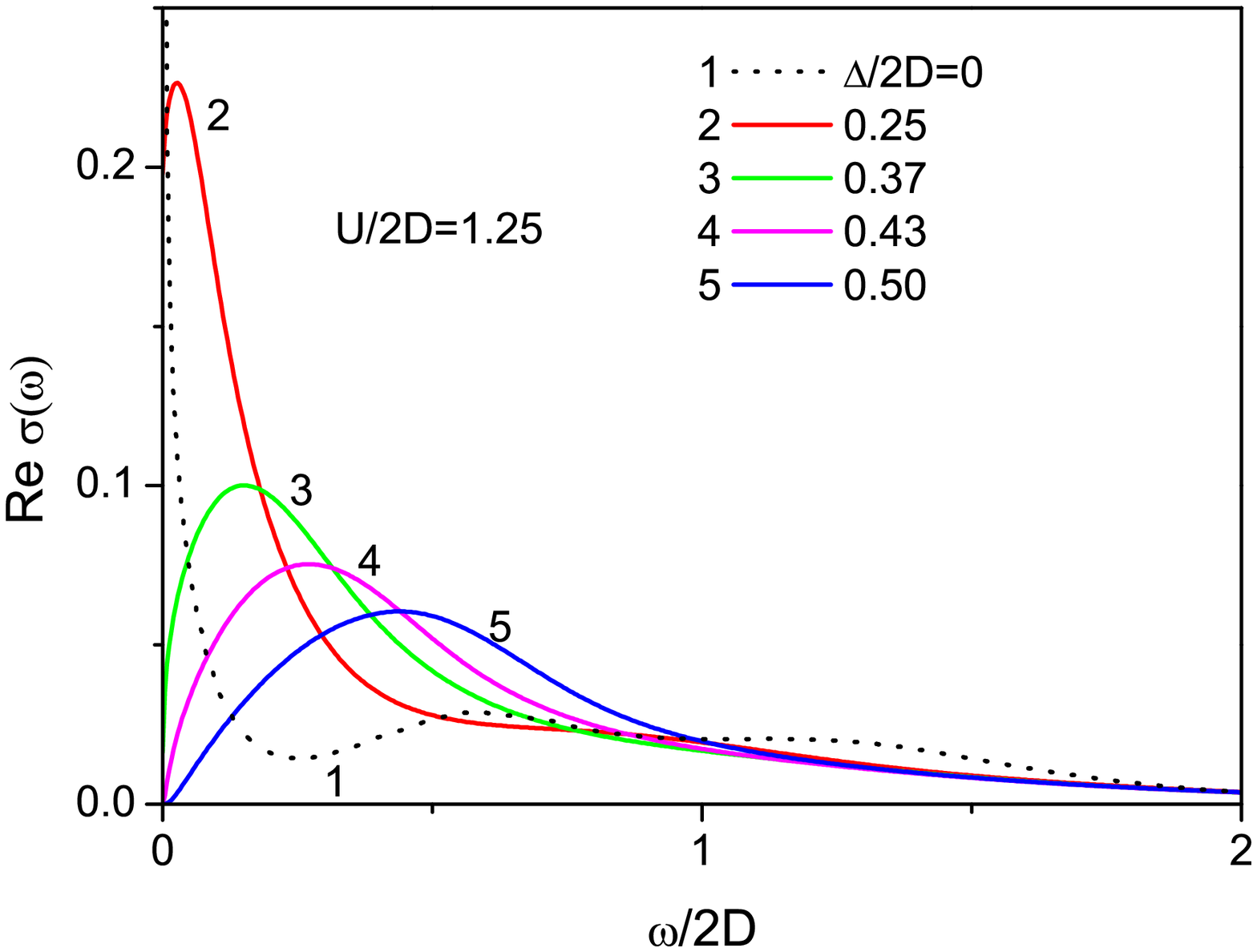}
\includegraphics[clip=true,width=0.46\textwidth]{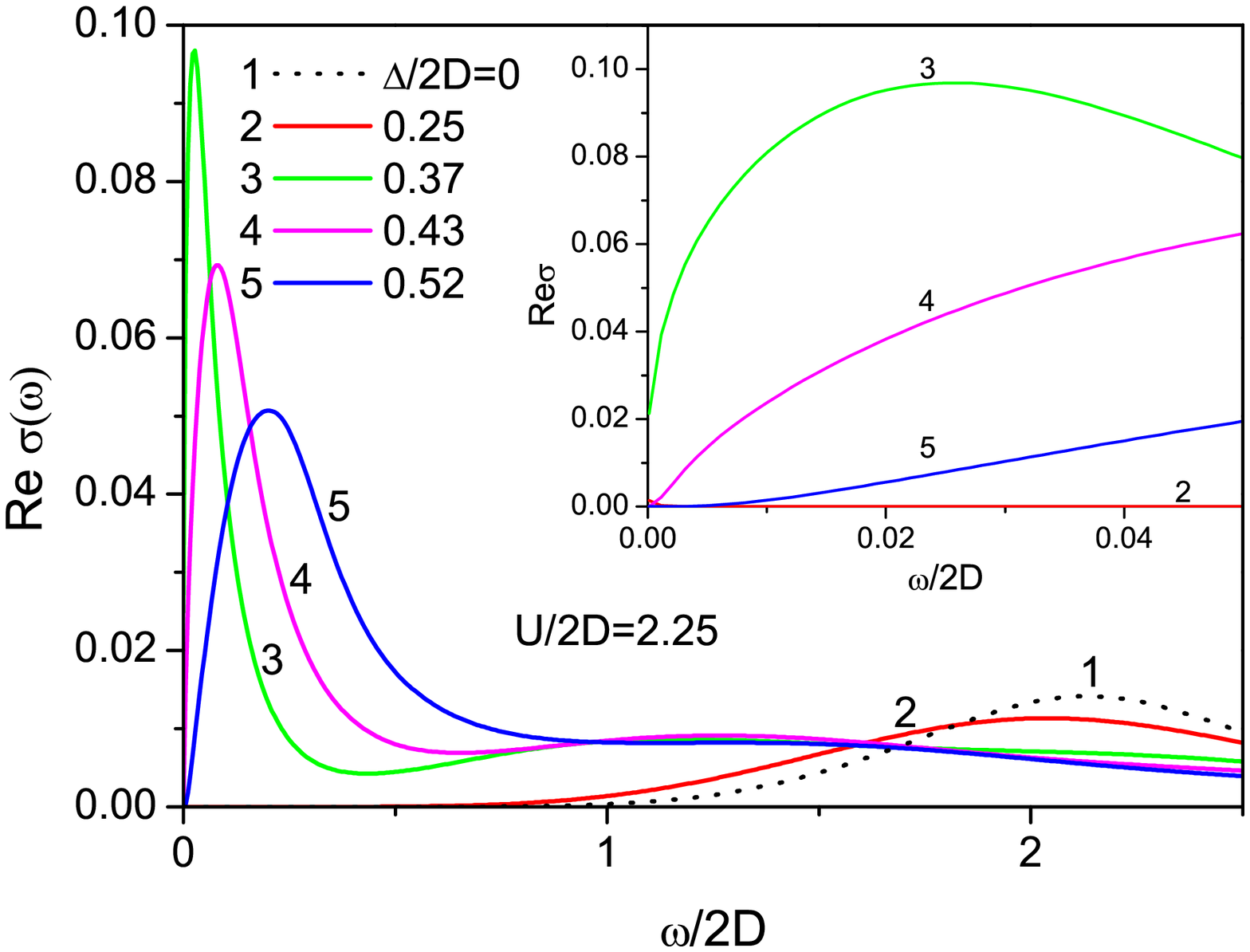}
\caption{\small Real part of optical conductivity of 
Hubbard-Anderson model at half-filling for different
disorder levels $\Delta$ \cite{HubDis}.
On the left side -- typical correlated metal with $U=2.5D$.
Curves 1,2, -- metallic phase, curve 3 corresponds to the mobility
edge (Anderson transition), curves 4,5 -- correlated Anderson insulator.
On the right side typical Mott insulator with $U=4.5D$.
Curves 1,2 correspond to Mott insulator, curve 3 -- mobility edge
(Anderson transition), curves 4,5 -- correlated Anderson insulator.
Inset -- magnified low frequency region.} 
\label{met_cond} 
\end{figure}

On the left panel of Fig.~\ref{met_cond} we show the real part of optical 
conductivity of Hubbard-Anderson model at half-filling for different disorder 
levels $\Delta$ and $U=2.5D$ typical for correlated metal. 
Transitions to the upper Hubbard bands at energies $\omega\sim U$
are almost unobservable. However it is clearly visible that
metallic Drude peak typically centered at zero frequency is broadened and 
suppressed by disorder, gradually transforming into a peak at finite frequency 
because of Anderson localization effects.
Anderson transition takes place at $\Delta_c\approx 0.74D=0.37W$
(corresponding to the curve 3 on all figures here and also for DOS).
Notice that this value depends on the cutoff (\ref{cutoff}), 
which is defined up to the coefficient of the order of unity \cite{MS86,Diagr}.
Naive expectations can bring us to a conclusion that narrow quasiparticle
band at the Fermi level (formed in a strongly correlated metal) may be 
localized much easily than normal conduction band. However we see that these 
expectations are wrong and the band localizes only at rather large disorder
$\Delta_c\sim D$, similar to that for conduction band of the width $\sim W$.
It agrees with the known analysis of localization in a two-band model \cite{ErkS}.

In the DMFT+$\Sigma$ approach critical disorder value $\Delta_c$
does not depend on $U$ as interaction effects enter Eq. (\ref{Dsc})
only through 
$\Delta\Sigma^{RA}(\omega)\to 0$  for $\omega\to 0$ (for $T=0$, $\varepsilon=0$), 
and the influence of interaction at $\omega=0$ disappears.
In fact this is the main shortcoming of DMFT+$\Sigma$ approach originating 
from the neglect of interference effect between interaction
and impurity scattering.
Significant role of these interference effects is known for a long time \cite{Lee85,ma}.
On the other hand, the neglect of these effects allows to perform the
reasonable physical interpolation between two main limits -- that of 
Anderson transition because of disorder and Mott-Hubbard transition because of
strong correlations. One can consider this approximation as a reasonable
first step to a future complete theory of metal-insulator transition
in strongly correlated disordered systems.

On the right panel of Fig. \ref{met_cond} we show the real part of optical 
conductivity of Mott-Hubbard insulator with $U=4.5D$ at different disorder 
levels $\Delta$. In the inset we show low frequency data, demonstrating 
different types of conductivity behavior, especially close to Anderson 
transition and within the Mott insulator phase. On the main part of the figure 
contribution to conductivity from transitions to upper Hubbard band at about 
$\omega\sim U$ is distinctly seen. 
Disorder growth results in the rise of finite conductivity for the frequencies
inside Mott-Hubbard gap, correlating with the restoration of quasiparticle
band in DOS within the gap as shown in the right panel of Fig. \ref{metDOS}.
This conductivity for $\Delta<\Delta_c$ is metallic (finite in the static 
limit $\omega=0$), and for $\Delta>\Delta_c$ at low frequencies we get 
${\rm Re}\sigma(\omega)\sim\omega^2$, which is typical for Anderson insulator 
\cite{VW,Diagr,MS,WV,MS86,VW92}. 

A bit unusual is the appearance in ${\rm Re}\sigma(\omega)$ of a peak
at finite frequencies even in the metallic phase.
This happens because of importance of localization effects.
In the ``ladder'' approximation for $\Phi^{0RA}_{\varepsilon}(\omega,{\bf q})$
which neglects all localization effects we obtain the usual Drude peak
at $\omega=0$  \cite{HubDis}, while account of localization effects
shifts the peak in ${\rm Re}\sigma(\omega)$ to a low (finite)
frequencies. As is well known \cite{NFM}, metallic state is defined
by finite static ($\omega=0$) conductivity at zero temperature.

Above we presented the data for conductivity data obtained for the case
of increase of $U$ from metallic to Mott insulator phase.
As $U$ decreases from Mott insulator phase we observe hysteresis of conductivity
in coexistence region defined (in the absence of disorder) by inequality
$U_{c1}<U<U_{c2}$. Hysteresis of conductivity is also observed in the 
coexistence region in the presence of disorder. Details of this behavior of
optical conductivity can be found in Ref. \cite{HubDis}. 

{\bf Phase diagram of Anderson-Hubbard model at half-filling}

Phase diagram of Anderson-Hubbard model at half-filling was studied
in Ref. \cite{BV}, using direct DMFT calculations for the lattice with finite 
number of sites with random realizations of energies $\epsilon_i$ in 
(\ref{And_Hubb}) and averaging over these realizations to get averaged DOS and
geometric mean local DOS which allows one to define critical disorder for
transition into Anderson insulator phase.
Below we present our results on Anderson-Hubbard model phase diagram
obtained from DOS and optical conductivity calculations within the 
DMFT+$\Sigma$ approach. One should emphasize that conductivity analysis is most 
direct way to distinguish metallic and insulating phases \cite{NFM}.

\begin{figure}
\includegraphics[clip=true,width=0.5\textwidth]{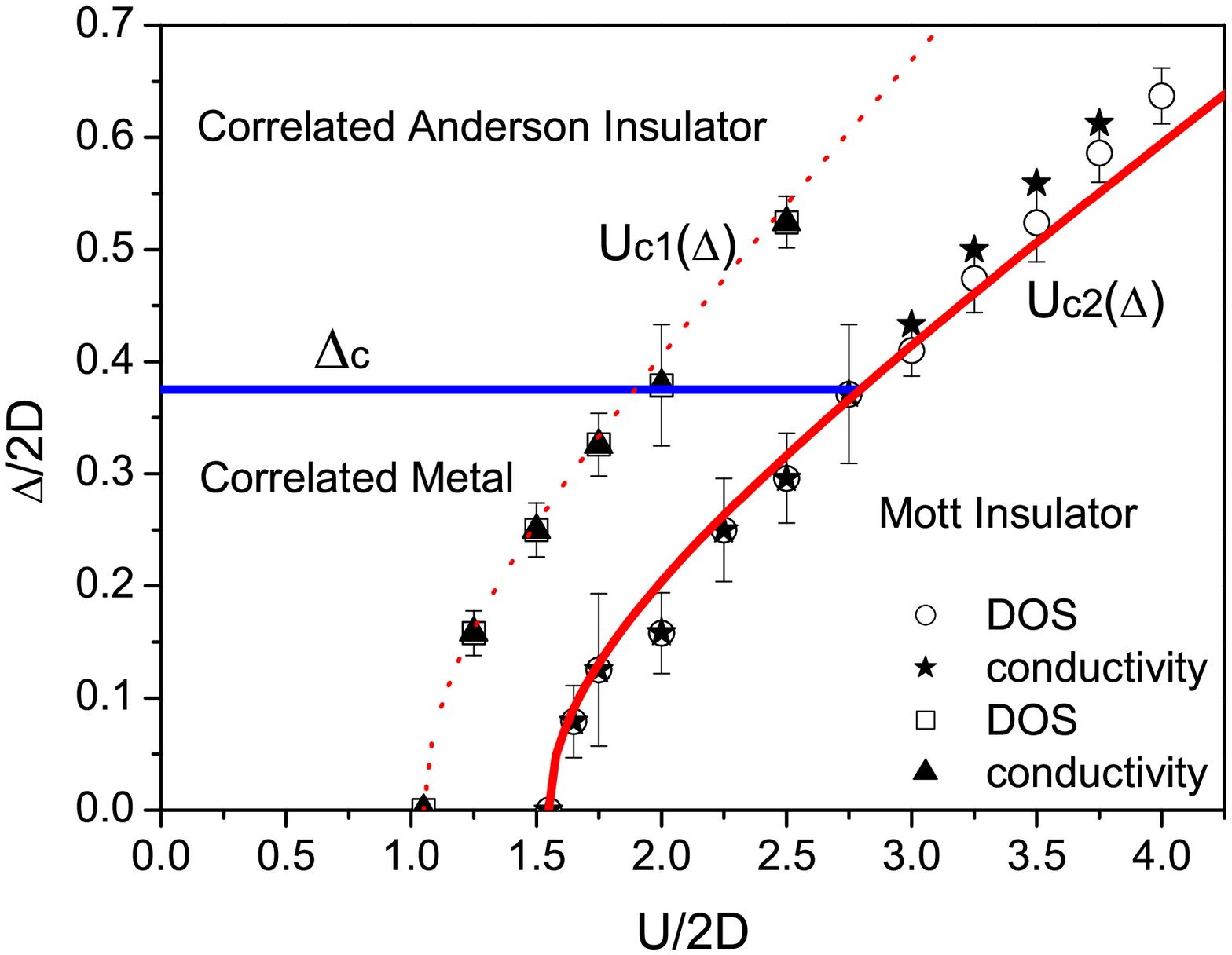}
\includegraphics[clip=true,width=0.5\textwidth]{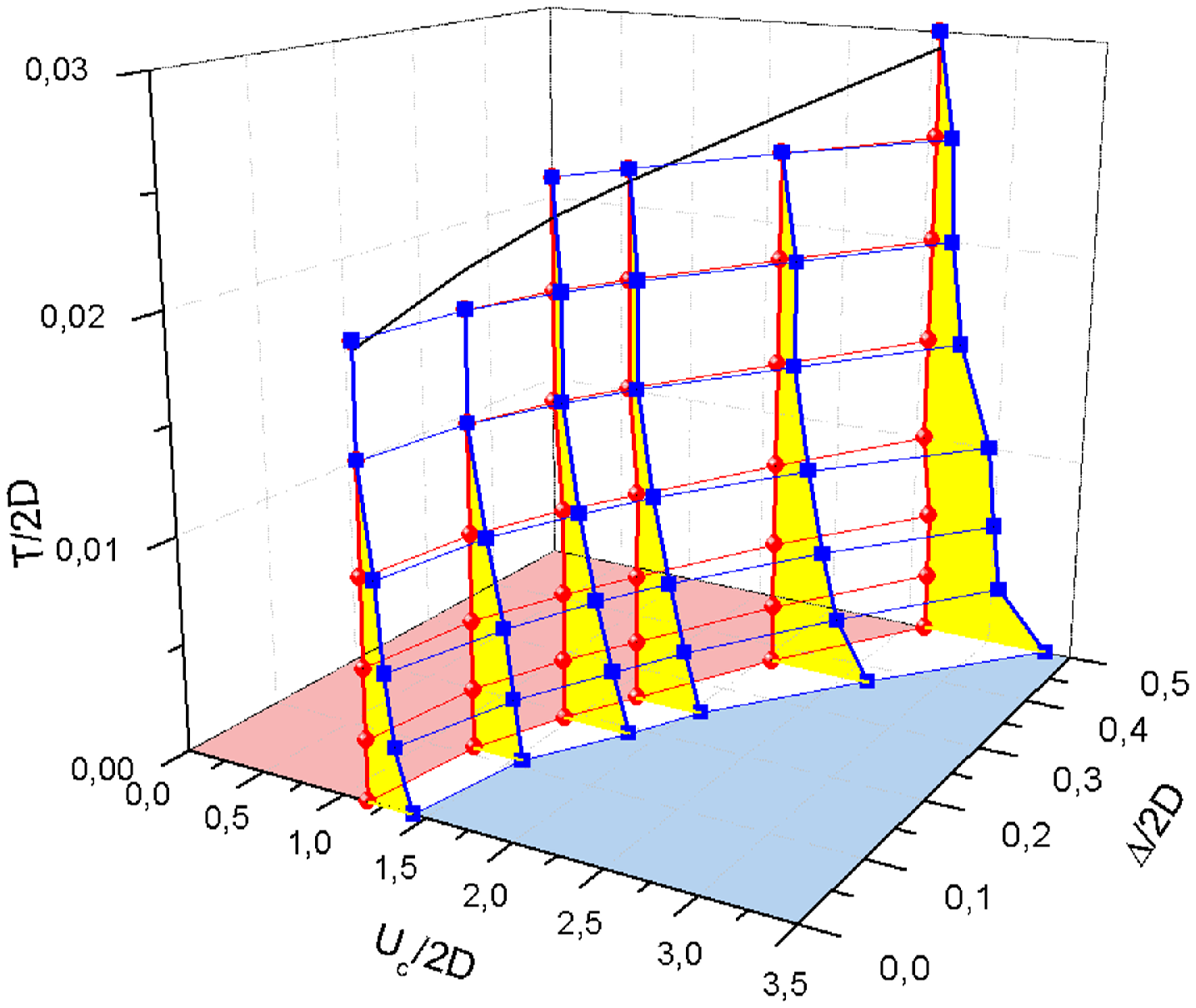}
\caption{\small Phase diagram of paramagnetic Anderson-Hubbard model.
On the left side --  zero temperature case \cite{HubDis}.
Continuous curves are Mott insulator phase boundaries $U_{c1,c2}(\Delta)$
obtained from analytical estimate of Eq. (\ref{Uc}), different symbols
represent results for these boundaries obtained from calculations from DOS and 
optical conductivity. 
Line of Anderson transition is given by $\Delta_c=0.37$.
On the right side -- finite temperature case.
Points are obtained from DOS calculations. Solid black curve
is linear fit ($T_c/2D=0.02(1+\Delta /2D$)) to the $T_c$ points
where coexistence region disappears.}
\label{ph_diag} 
\end{figure} 

Calculated disorder-correlation $(\Delta,U)$ phase diagram at zero temperature 
is shown on the left panel of Fig. \ref{ph_diag}.
Anderson transition line $\Delta_c\approx 0.37W=0.74D$ is defined
as a disorder strength for which static conductivity becomes zero at  $T=0$.
Mott-Hubbard transition can be detected from central (quasiparticle) peak
disappearance in DOS or from optical conductivity by observation of
gap closing in the insulating phase or from static conductivity disappearance
in the metallic phase.

We have already noticed that DMFT+$\Sigma$ approximation gives
universal ($U$ independent) value of critical disorder $\Delta_c$
because of neglect of interference between disorder scattering
and Hubbard interaction. This leads to differences
between phase diagram of Fig.  \ref{ph_diag} and the one obtained in 
Ref. \cite{BV}. At the same time influence of disorder scattering
on Mott-Hubbard transition is highly nontrivial and
qualitatively coincide with results of Ref. \cite{BV}.
Main difference is conservation of Hubbard bands in our results even in the 
limit of high enough disorder, while in the Ref. \cite{BV} they just disappear.
Moreover coexistence region in Fig. \ref{ph_diag} 
slowly widens with disorder growth instead of vanishing
at some ``critical'' point as on phase diagram of Ref. \cite{BV}.
Coexistence boundary regions, which are defined by Mott insulator
phase boundaries, obtained with decrease on increase of $U$,
represented by curves $U_{c1}(\Delta)$ and $U_{c2}(\Delta)$
on Fig. \ref{ph_diag},  can be obtained from the simple equation:
\begin{equation}
\frac{U_{c1,c2}(\Delta)}{W_{eff}}=\frac{U_{c1,c2}}{W},
\label{UcW13}
\end{equation}
where effective bandwidth in the presence of disorder was calculated
for $U=0$ within self-consistent Born approximation (\ref{BornSigma}):
\begin{equation}
W_{eff}=W\sqrt{1+16\frac{\Delta^2}{W^2}}.
\label{Weff}
\end{equation}
Thus the boundaries of coexistence region which define also
Mott insulator phase boundaries are given by:
\begin{equation}
U_{c1,c2}(\Delta)=U_{c1,c2}\sqrt{1+16\frac{\Delta^2}{W^2}}
\label{Uc}
\end{equation}
which are shown in Fig. \ref{ph_diag} by dotted and solid lines.
Phase transition points detected from disappearance of quasiparticle peak
as well as points following from qualitative changes of conductivity
behavior are shown in Fig. \ref{ph_diag} by different symbols.
These symbols demonstrate very good agreement with analytical results
supporting the choice of ratio (\ref{UcW13})
as a control parameter of Mott transition in presence of disorder.

On the right panel of Fig. \ref{ph_diag} we show temperature dependence of
Mott insulator phase boundaries $U_{c2}(\Delta)$ (squares) and coexistence 
region $U_{c1}(\Delta)$ (circles)
\footnote{These results for different temperatures were calculated
by N.A. Kuleeva.}.
It is well known that in ``pure'' DMFT without disorder coexistence region
narrows with temperature growth and vanishes at some critical temperature $T_c$.
In the DMFT+$\Sigma$ approach disorder increase at zero temperature
widens the coexistence region and this behavior remains for finite
temperatures. It is also seen that disorder growth leads to a practically 
linear growth (solid black line) of this critical temperature.
Notice also very weak temperature dependence of
coexistence boundary $U_{c1}(\Delta)$.

\subsubsection{Two-dimensional systems.}
\label{2d_dis}

According to scaling theory of localization \cite{AALR79}
metallic state in two-dimensional (2D) systems does not exists,
electrons are localized already at any infinitely small
disorder. Despite this prediction for 2D systems was made for noninteracting
particles, later it was found that in the simplest case weak interaction 
between electrons also favors localization \cite{AAL80}. 
In early 80th experiments done on different 2D systems \cite{2Dexp}
mostly confirmed these predictions.
However, some theoretical works \cite{ma} pointed that this point of view in 
general is incorrect, since in the limits of weak disorder and large enough 
interaction 2D systems can have finite conductivity at zero temperature.
Experimental discovery of metal-insulator transition in 2D weakly disordered
systems at low carrier concentration, absent in the single particle theory,
stimulated new direction of theoretical studies
(see introduction to this field in review papers \cite{Krav04,AKS01}).

In DMFT+$\Sigma$ approximation, as we shall see, for infinite
2D system ($L\to\infty$) localization radius defined by Eq. (\ref{Rl2})
remains finite (but exponentially large) for any infinitely weak disorder,
signalling the absence of Anderson transition in such system, similarly to the
case of the usual single particle theory. 
However, as we shall discover below, localization radius
in finite size systems diverges at some critical disorder, which is defined
by the system size $L$. Qualitatively, this critical disorder is determined
by the condition that localization radius of infinitely large system
becomes comparable to characteristic sample size $R_{loc}^{L\to\infty}\sim L$.
Thus for finite two-dimensional systems Anderson transition in fact exists,
as well as metallic phase for disorder below some critical value.
In the following, under the term ``correlated metal'' phase we shall imply 
precisely such phase for finite 2D systems.

Below we discuss most relevant results of DMFT+$\Sigma$ approach 
for 2D Anderson-Hubbard model at half-filling on a square lattice with
rectangular bare density of states with the bandwidth $W=2D$:
\begin{equation}
N_0(\varepsilon)=
\left\{
\begin{array}{ll}
\frac{1}{2D} & \quad |\varepsilon|\leq D\\ 
0 & \quad |\varepsilon| > D
\end{array}.
\right.
\label{DOS14}
\end{equation}
which corresponds right to a 2D case.

{\bf Density of states and optical conductivity.}

Calculations show that qualitative behavior of the density of states in 2D is
completely analogous to that discussed above in three-dimensional case.
Some quantitative distinctions are due to the different model of the ``bare''
density of states (\ref{DOS14}), leading in particular 
to  larger (than in three-dimensional case) critical Hubbard interaction
$U_{c2}\approx1.83W$, corresponding to Mott metal-insulator transition in 
absence of disorder and $U_{c1}\approx1.42W$, which limiting below the region 
coexistence of metallic and insulating phases. Similar to three-dimensional 
model for $U>U_{c2}$ (i.e. for Mott insulator without disorder)
increase of disorder leads to restoration of quasiparticle peak in the density 
of states. However, in this case such behavior does not signal in general the
transition to a correlated metal state, as for infinitely large systems we 
are dealing here with correlated Anderson insulator.

Optical conductivity behavior in a wide frequency range is also
qualitatively is quite similar to that in three-dimensional model.
But for infinite 2D model zero frequency conductivity always
disappears (in the zero temperature limit) and, in contrast
to  $d=3$ case \cite{HubDis}, even at very weak disorder
the peak in optical conductivity lies at finite frequency.
In the ladder approximation, which does not contain localization
corrections, the usual Drude peak is observed at zero frequency
and conductivity at $\omega=0$ is finite.

In more detail results for DOS and optical conductivity in 2D model
can be found in Ref. \cite{HubDis2}.

{\bf Localization radius and phase diagram of
2D Anderson-Hubbard model at half-filling.}

\begin{figure}
\includegraphics[clip=true,width=0.5\textwidth]{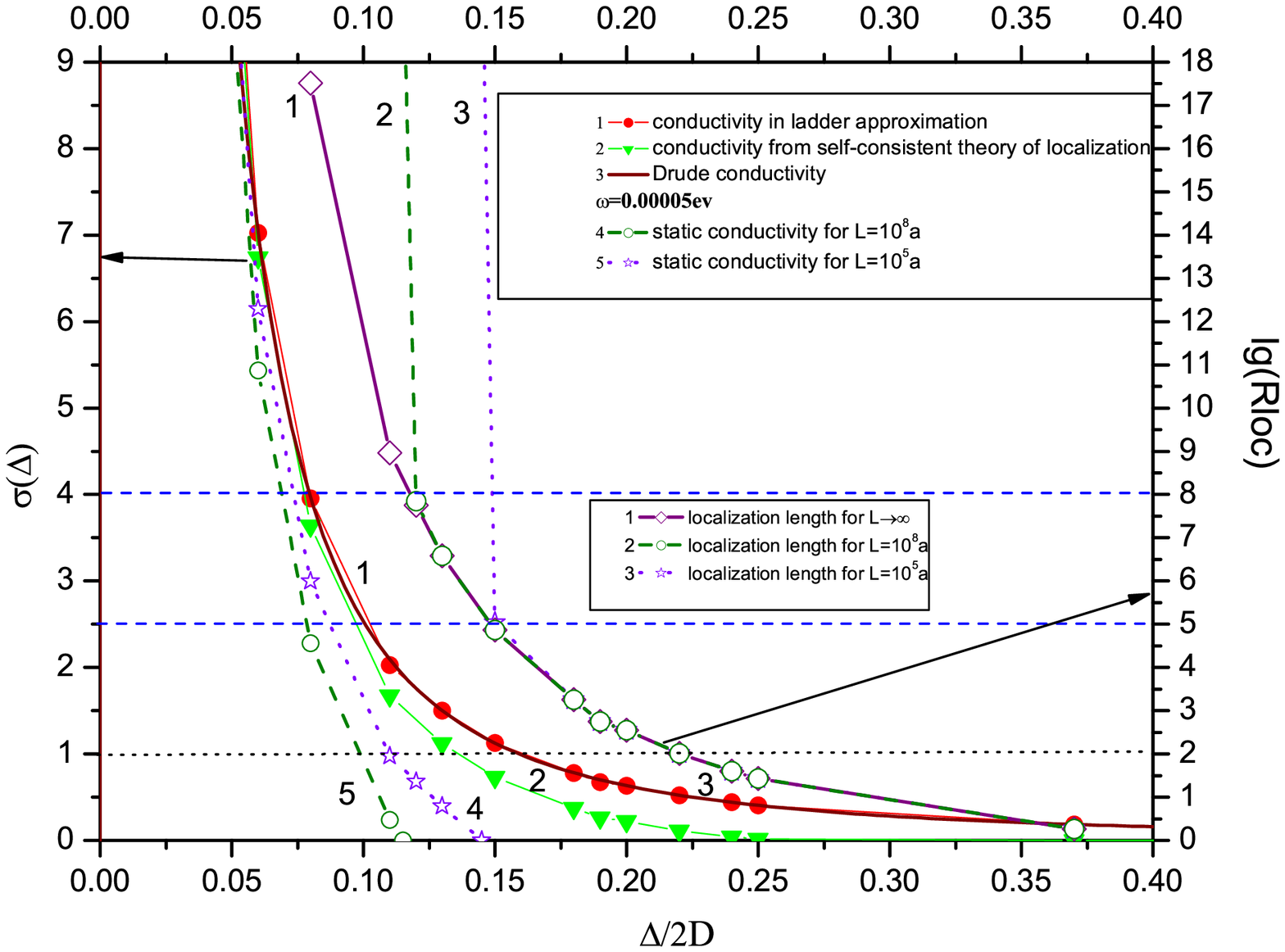}
\includegraphics[clip=true,width=0.5\textwidth]{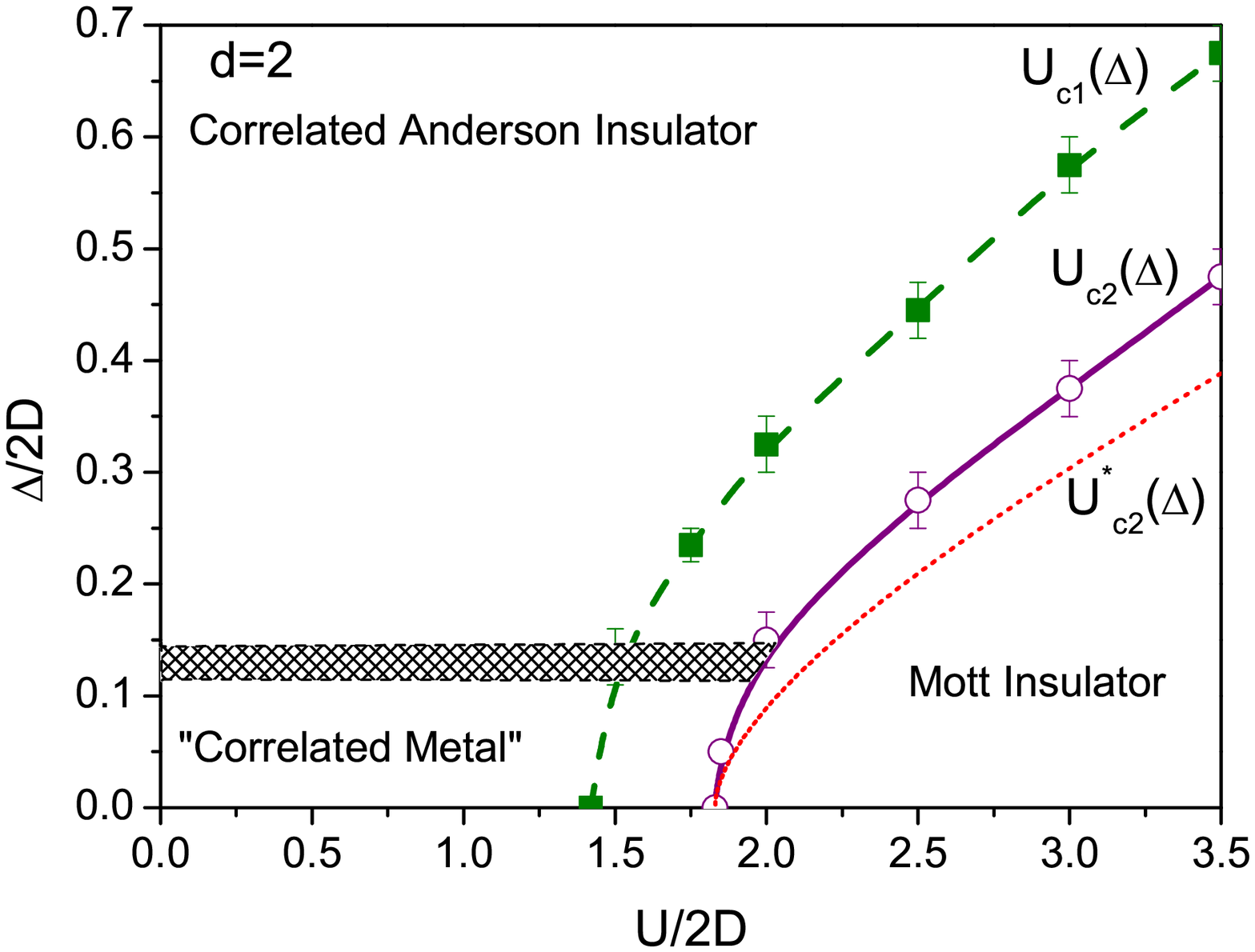}
\caption{\small ON the left side -- low finite frequency ($\omega=0.00005D$) conductivity 
dependence on disorder strength $\Delta$ at $U/2D=1$.
Circles (curve 1) show results of ladder approximation, triangles (curve 2) -- 
self-consistent theory of localization. Curve 3 (practically coinciding with 
ladder approximation) is obtained from Drude formula (\ref{Dr}).
Curves 4 and 5 -- static conductivity of finite samples with sizes
$L=10^8a$ and $L=10^5a$ correspondingly. On the right scale localization radius 
logarithm versus disorder strength $\Delta$: infinite size sample -- curve 1, 
finite samples with sizes $L=10^8a$ and $L=10^5a$ -- curves 2 and 3 \cite{HubDis2}.} 

{\small On the right -- phase diagram of 2D paramagnetic Anderson-Hubbard model
at zero temperature  \cite{HubDis2}.
Mott insulator region boundary  $U_{c2}(\Delta)$ and coexistence region boundary
$U_{c1}(\Delta)$ were obtained from density of states behavior.
Hatched part -- region of an effective Anderson metal--insulator transition
in finite size systems.} 

\label{ph_diag2}
 
\end{figure}

On the left side of Fig. \ref{ph_diag2} we show dependence of conductivity on 
disorder $\Delta$ for a finite but quite low frequency  $\omega=0.00005D$.
Circles show results of ladder approximation, triangles -- self-consistent 
theory of localization. Curve 3 qualitatively coinciding with ladder 
approximation was obtained from classical Drude expression:
\begin{equation} 
\sigma(\omega)=\sigma(0) \frac{\gamma^2}{\gamma^2+\omega^2},
\label{Dr}
\end{equation}
where static conductivity  
$\sigma(0)=e^2N(0)D_0\approx \frac{e^2}{\hbar}\frac{\varepsilon_F}{2\pi\gamma}$, 
$N(0)$ -- density of states at the Fermi level, $D_0$ is Drude diffusion 
coefficient. Impurity scattering rate was taken as 
$\gamma=\pi N(0)\Delta^2\approx \frac{\pi}{2D}\Delta^2$.
Significant contribution from localization corrections to conductivity
at finite frequency (noticeable distinction of curve 2 from 1 and 3)
appears only when conductivity reaches values of the order of minimal
metallic conductivity $\sigma_0=\frac{e^2}{\hbar}$
(which is taken as unit of conductivity on figures).
One should note that exactly in this range of disorder, as we shall see below,
Anderson metal-insulator transition takes place (localization radius diverges)
for 2D systems of reasonable finite sizes.

Also in Fig. \ref{ph_diag2} we show dependences of localization radius logarithm 
following from (\ref{Rl2}) (right scale) as function of disorder: 
curve 1 -- infinite sample,
curves 2 and 3 are for the finite size samples with $L=10^{8}a$ and 
$L=10^{5}a$ correspondingly.
It is seen that localization radius grows exponentially as disorder
diminishes and remains finite in the infinite 2D system, where Anderson
transition is absent. For the finite systems localization radius
{\em diverges} at critical disorder determined by  system size, demonstrating
the existence of an effective Anderson transition.
As can be seen from Fig. \ref{ph_diag2}, qualitatively the critical disorder
is defined by condition that localization radius of infinite system
becomes comparable with characteristic sample size $R_{loc}^{L\to\infty}\sim L$.
It should be noted that within our approach localization radius practically
does not depend on $U$ (in contrast to e.g.  \cite{Shapiro08}),
which leads to independence of critical disorder of correlation strength $U$
in 2D finite size systems. Analogous situation is realized also for
three-dimensional systems \cite{HubDis}. In general, it is of course a drawback 
of our approximations.

On the same left part of Fig. \ref{ph_diag2} we plot the dependence of static 
conductivity on disorder strength in finite samples with sizes
$L=10^{8}a$ and $L=10^{5}a$ (curves 4 and 5 correspondingly).
In the finite size systems with weak disorder static conductivity
is not zero (metal) and gradually goes down with increase of disorder.
It becomes zero at critical disorder, where localization
radius diverges in the sample of corresponding size.
Static conductivity of finite size samples within our approximation
does not practically depend on correlation strength $U$.
Significant difference of static conductivity from low finite frequency
conductivity observed in Fig. \ref{ph_diag2} is related to
exponential smallness of the frequency range with localization behavior of 
conductivity, mentioned above.

Let us consider now the phase diagram of 2D paramagnetic Anderson-Hubbard model
at half-filling, obtained from DMFT+$\Sigma$ calculated densities
of states and from the analysis of localization radius behavior in finite size 
2D systems. Such phase diagram in coordinates disorder of $\Delta$ and
correlation strength $U$ is shown on the right part of Fig. \ref{ph_diag2}.

Hatched stripe corresponds to the region of effective ``metal''- Anderson 
insulator transition. Boundaries of this region are defined by divergence of 
localization radius in finite samples with characteristic sizes  $L=10^{5}a$ 
(upper boundary) and $L=10^{8}a$ (lower boundary) (see left side of Fig. 
\ref{ph_diag2}).
One should emphasize that further increase of the system size,
e.g. ten times, up to $L=10^{9}a$, leads only to quite insignificant drop of 
critical disorder. In other words, it slightly shifts down low boundary of 
hatched stripe (on the right part of Fig. \ref{ph_diag2}) --
characteristic region of an effective Anderson transition in finite size 
systems.

Curve $U_{c2}(\Delta)$ computed from density of states behavior
defines the boundary of Mott transition.
Transition criteria is the disappearance of central quasiparticle peak
in the density of states $N(\varepsilon )$ together with gap opening
on the Fermi level. Similarly to three-dimensional model, decrease of $U$
starting from insulating phase leads to Mott transition $U=U_{c1}(\Delta)<U_{c2}(\Delta)$
and coexistence (hysteresis) region is observed on the phase diagram between
curves $U_{c1}(\Delta)$ and $U_{c2}(\Delta)$ (Fig. \ref{ph_diag2}).
In analogy with three-dimensional case we can guess that the ratio of Hubbard 
interaction and effective bandwidth $\frac{U_{c1,c2}(\Delta)}{W_{eff}(\Delta)}$ 
controls Mott metal-insulator transition and is a universal constant, which does 
not depend on disorder and obtain qualitative dependencies 
$U^{\ast}_{c2}(\Delta)$ for 2D model, which is plotted by dotted curve on 
Fig. \ref{ph_diag2}. It is seen that in contrast to $d=3$ case \cite{HubDis} 
the dependence of $U_{c2}(\Delta)$ obtained from straightforward calculations 
of densities of states significantly differs from the qualitative 
$U^{\ast}_{c2}(\Delta)$ dependence. Apparently it is related to important 
change of density of states lineshape (at $U=0$) as disorder $\Delta$ grows, 
which is absent for semielliptic band in $d=3$ case. 

\subsection{Singularities of electron dispersion in
strongly correlated systems in DMFT and DMFT+$\Sigma$ approaches.}
\label{dmft_disp}

\subsubsection{Cusps (``kinks'') in electron spectra.}

Electronic properties of crystalline solids are determined by single-particle
and collective excitations of electron subsystem and their interactions
with each other. These excitations are characterized by energy $E$ and momentum
$\bf{k}$ which are related to each other via \emph{dispersion} (spectrum) $E_{\bf{k}}$.
Interaction between single-particle and collective modes can result in noticeable
bends (cusps) of dispersion $E_{\bf{k}}$ -- the so called ``kinks''.
The lineshape and energy position of these features characterize interactions
in the manybody system. For example, kinks in electron dispersions experimentally observed by ARPES
in copper oxides, with energies 40-70 meV below the Fermi level, are evidence of electron-phonon
\cite{Valla1999,Hengsberger1999,Lanzara2002,Shen2002,Rotenberg2000,Higashiguchi2005,Sun2005} 
or electron-magnon (spin-fluctuation) interactions \cite{He2001,Hwang2004}. 

At the same time ARPES experiments detect kinks in electron dispersion
for a number of different systems at essentially higher energies (up to 800 meV)
\cite{Ronning2003,Yoshida2005,Graf2006}. Physical nature of these kinks remains unknown.
Among other attempts to find the explanation of these electron spectra
anomalies a novel purely electronic mechanism of kinks formation was proposed in Ref. \cite{Nature}.
This mechanism is applicable to strongly correlated metals, where
spectral function contains well developed Hubbard subbands, together with
central quasiparticle peak as, for example, in transition metal oxides.
Energy location of these kinks apparently determines the  range of applicability
of Landau Fermi liquid theory.

As is well known \cite{AGD} in general case interaction results in finite life time 
of excitations in the system, so that $E_{\bf{k}}$ becomes a complex function.
For electron systems with Coulomb interaction Landau Fermi liquid theory
proves the existence of weakly damped fermionic quasiparticles for
low enough temperature and in the narrow energy interval around the Fermi surface \cite{AGD}.
Beyond the Fermi liquid regime the concept of quasiparticles with well
defined dispersion, strictly speaking, is not applicable since
quasiparticle lifetime is too small.  However, in recent years ARPES experiments convincingly 
show the existence of essentially $\bf{k}$-dependent (though with rather broad maxima) single-particle 
spectral function behavior, pretty far from the Fermi level, despite the fact that st these energies
one can not speak about well defined quasiparticles.
In this case we understand as particle dispersion precisely this $\bf k$-dependence
of spectral function maxima which replaces the usual notion of quasiparticle spectrum.

\subsubsection{Kinks of purely electronic nature}
\label{ekink}

Let us consider in more details the new mechanism of kinks formation in
electronic dispersion of strongly correlated systems,
which does not assume an interaction of electrons with phonons
or other excitations \cite{Nature}.
In order to understand the nature of this mechanism at the beginning we
shall examine weakly correlated system described by standard Fermi liquid theory.
Because of large enough quasiparticles lifetime close to the Fermi level
and weakness of correlations, in the first approximation interaction
leads to a simple renormalization of initial dispersion of (noninteracting)
quasiparticles $\epsilon_{\bf{k}}$, which is replaced by
$E_{\bf k}=Z_{\text{FL}}\epsilon_{\bf{k}}$, where
$Z_{\text{FL}}$ is electron Fermi liquid mass renormalization coefficient.
It changes the slope of dispersion in the vicinity of the Fermi level.
compared to a bare one. However, if we consider an electron with energy
far away from the Fermi level, then in case of weak interaction one can
expect that its dispersion practically does not change: 
$E_{\bf{k}}\approx\epsilon_{\bf{k}}$, though the damping here can be pretty large.
In this sense one can say that electron-electron interaction by it self can result 
in formation of bends (kinks) in the generalized dispersion (of the spectral 
function), and position of such kink on the
energy scale is defined right as an energy, where simple Fermi liquid picture
becomes inapplicable. However, in weakly correlated metals $Z_{\text{FL}}\lesssim 1$,
so that the slope of $E_{\bf{k}}$ will be changed insignificantly,
making such kinks hard to observe.

This picture can be essentially different in strongly correlated systems,
where the value of $Z_{\text{FL}}$ can be much less than one, thus
making kinks more pronounced. Strong correlation interactions
give rise to strong spectral weight redistribution within the
single-particle spectral function due to formation of Hubbard subbands.
Moreover, as we know in strongly correlated metals there is also
quasiparticle peak which appears close to the Fermi level, between
Hubbard subbands. Below it will be shown that ``usual'' Fermi liquid
quasiparticles exist in such systems only in a very narrow vicinity
of the Fermi level. while beyond the Fermi liquid regime (but still
``inside'' of quasiparticle peak) exists some intermediate
regime with dispersion $E_{\bf k}\approx Z_{\text{CP}}\epsilon_{\bf{k}}$, where
renormalization factor $Z_{\text{CP}}$ is defined by central peak
spectral weight and its value essentially differs from $Z_{\text{FL}}$.
At these intermediate energies, which are much smaller than interaction
energy, electrons or holes are strongly interacting and their
dispersion differs from both Fermi liquid one and noninteracting one.
In this energy range one can speak about ``intermediately'' correlated
situation, when $Z_{\text{FL}}<Z_{\text{CP}}<1$.
Consequently, at some energies $\pm\omega_\star$ inside the quasiparticle
peak transition from renormalization $Z_{\text{FL}}$ to $Z_{\text{CP}}$ takes place.
That leads to an emergence of kink in electronic dispersion and the energy
position of this kink is directly related to limits of applicability of the usual
Fermi liquid theory.
Lets emphasize that this mechanism gives rise to kink formation
without any additional interaction with phonons or other collective modes.
The only necessary condition for such kinks formation is the presence
of strong electron-electron correlations in the system.

\begin{figure*}
\begin{center}\includegraphics[clip,height=90mm]{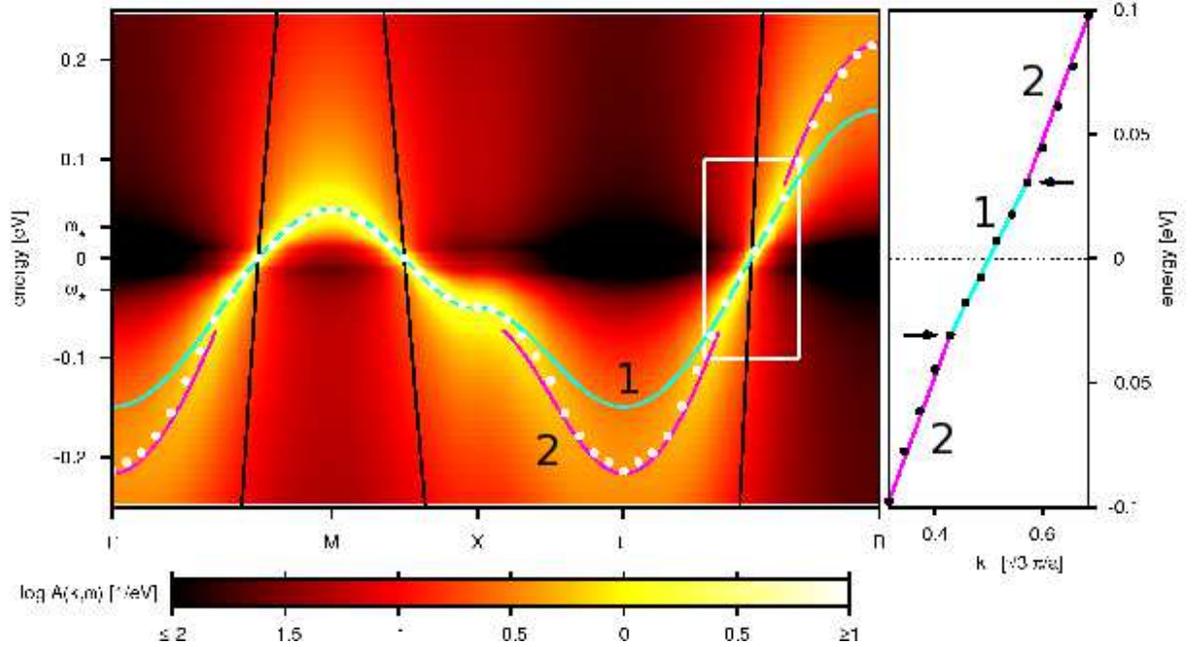}
\vspace*{-5mm}\end{center}
\caption{\textbf{Cusps (kinks) in electronic dispersion $E_{\bf{k}}$ for
the case of strongly correlated systems.}
On the contour plot intensity map of DMFT spectral function $A(\bf{k},\omega)$
for the Hubbard model on a cubic lattice is presented.
Hubbard interaction value is $U$=3.5 eV, bandwidth $W\approx$ 3.64 eV, band filling $n$=1,
calculated value of Fermi liquid renormalization is $Z_{\text{FL}}$=0.086,  temperature is $T=5$ K.
Close to Fermi level maxima of $A(\bf{k},\omega)$ (white dots) correspond to
renormalized dispersion $E_{\bf{k}}=Z_{\text{FL}}\epsilon_{\bf{k}}$
(line 1). At energies $|\omega|>\omega_\star$ spectral function 
$A(\bf{k},\omega)$ keeps its shape but with different renormalization
$E_{\bf{k}}=Z_{\text{CP}}\epsilon_{\bf{k}}-c\;\text{sgn}(E_{\bf{k}})$
(line 2). Values of $\omega_\star$=0.03 eV, $Z_{\text{CP}}=0.135$,
and $c=0.018$ eV are calculated in \cite{Nature} from the values of
$Z_{\text{FL}}$ and $\epsilon_{\bf{k}}$ corresponding to black line.
Inset on the right shows in details part
of dispersion in the  $\Gamma$-R direction marked out with white rectangle,
kinks (cusps in dispersion) at energies $\pm\omega_\star$ are pointed by arrows.
Black lines -- bare (noninteracting) quasiparticle spectra.
\vspace*{2mm}% 
}
\label{fig:strong}
\end{figure*}

For microscopic description of electronic kinks lets consider
Hubbard model, which will be analyzed in the framework of standard DMFT,
using numerical renormalization group (NRG) to solve effective
Anderson impurity problem.
For simplicity we consider single-band Hubbard model at half-filling.
Strongly correlated regime in Hubbard model occurs when
interaction value becomes of the order of bare bandwidth $U$ $\approx W$.
Consider as an example computational results shown
in Fig.~\ref{fig:strong}. 
It is clearly visible that dispersion undergoes from Fermi liquid regime
(line 1 on Fig.~\ref{fig:strong}) into described above ``intermediate''
regime (line 2 in Fig.~\ref{fig:strong}) with formation of well defined
cusps in dispersion (kinks) at energies $\omega_\star=0.03$ eV.
In some of the high symmetry directions (e.g. around X-point, Fig.~\ref{fig:strong})
dispersion has quite small slope close to the Fermi level making kinks 
less pronounced. 

Let us have a look on this situation from the point of view of single-particle
Green's function behavior. From general considerations it is clear that
any cusps of dispersion caused by interaction appear because of corresponding
behavior of the self-energy or, more precisely, that of $\text{Re}\Sigma_{\bf{k}}(\omega)$.
In the majority of real physical systems $\bf{k}$-dependence of self-energy, apparently,
is less important in comparison with $\omega$-dependence and thus the
neglect of self-energy $\bf{k}$-dependence $\Sigma_{\bf{k}}(\omega)=\Sigma(\omega)$ 
is, more or less, good approximation. In the framework of DMFT this statement is exact.
Self-consistent expression for self-energy of Hubbard model within DMFT
can be written as:
\begin{equation}
\Sigma(\omega)=\omega+\mu-1/G(\omega)-\Delta(G(\omega)),
\label{SigmaDMFT}
\end{equation}
where $G(\omega)=\frac{1}{N}\sum_{\bf k} G({\bf{k}},\omega)$ is the
local (averaged over $\bf{k}$) Green's function,
$\Delta(G)$ is frequency dependent hybridization function
expressed via $G(\omega)$ \footnote{In DMFT $\Delta(G)$ is defined by
$G(\omega)=G_0(\omega+\mu-\Sigma(\omega))$, i.e.
$G_0(\Delta(G)+1/G)=G$, where $G_0(\omega)$ is noninteracting local
Green's function.}. Hybridization function describes quantum-mechanical
coupling between an electron on a given site and other sites of the system.

\begin{figure}
\includegraphics[clip,width=86mm]{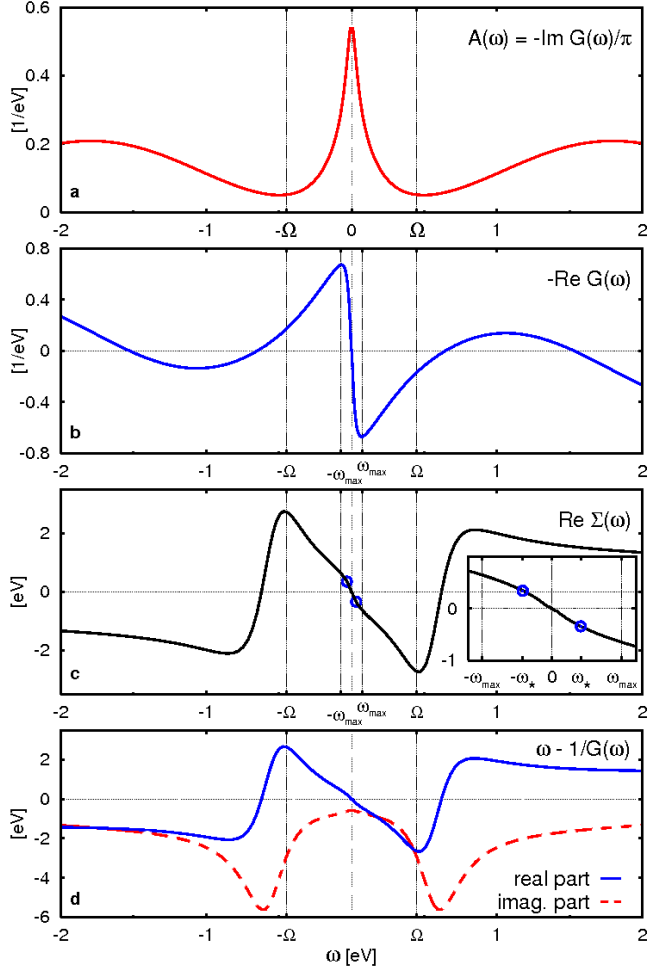}\vspace*{-5mm}
\caption{
Green's function characteristic for strongly correlated system with parameters
given in the caption of Fig.~\ref{fig:strong}.
\textbf{a} -- three-peak structure of spectral function
$A(\omega)=-\text{Im}G(\omega)/\pi$ with minima at energies $\pm\Omega=0.45$ eV.
\textbf{b} -- real part of Green's function with relative minima and maxima
at energies $\pm\omega_{\text{max}}$ located ``inside'' quasiparticle peak.
\textbf{c} -- real part of self-energy with cusps at energies
$\pm\omega_\star$ (circles) situated in the points of maximal curvature
of $\text{Re}G(\omega)$ ($\omega_\star=0.4\omega_{\text{max}}=0.03$ eV).
\textbf{d} -- real part of $\omega-1/G(\omega)$ (solid line) contributing to self-energy
which is linear, in general, in the energy interval $|\omega|<\Omega$.
Change of energy dependence of $-\text{Re}[G(\omega)]$ at energies  $\pm\omega_\star$
defines the location of kinks (cusps) in electronic dispersion.
\vspace*{-5mm}%
}
\label{fig:kinks}
\end{figure}

In Fig.~\ref{fig:kinks} we show frequency dependence of spectral function
(DOS) $A(\omega)=-\text{Im}G(\omega)/\pi$ calculated for the same model
parameters as in Fig.~\ref{fig:strong} and demonstrating a typical
three-peak structure. Corresponding real parts of Green's function
$G(\omega)$ and self-energy $\Sigma(\omega)$ are shown in
Fig.~\ref{fig:kinks}b and Fig.~\ref{fig:kinks}c.

Kinks in $\text{Re}\Sigma(\omega)$ are directly related to three-peak structure of
integrated spectral function (DOS) $A(\omega)$ (its minima are at energies $\pm\Omega$)
and define some new quite small energy scale.
To this end $\text{Re}[G(\omega)]$ should have maxima and minima in the
energy interval $\pm\omega_{\text{max}}$, i.e. ``inside'' the central peak 
(Fig.~\ref{fig:kinks}b). In its turn it directly leads to
kinks formation in  $\text{Re}\Sigma(\omega)$. Self-energy $\Sigma(\omega)$ 
consists of two contributions:
$\omega+\mu-1/G(\omega)$ and $-\Delta(G(\omega))$.
Expression $\text{Re}[\omega+\mu-1/G(\omega)]$ is linear 
within quite large energy range $|\omega|<\Omega$ (Fig.~\ref{fig:kinks}d),
while the value of $-\text{Re}[\Delta(G(\omega))]$ is proportional
to $-\text{Re}[G(\omega)]$, at least in the first order of corresponding
momenta expansion, only in the small energy interval $|\omega|<\omega_{\text{max}}$.
The sum of these two contributions gives rise to cusps in real part of self-energy
at energies $\pm\omega_\star$, where $\omega_\star=(\sqrt{2}-1)\omega_{\text{max}}$.
At this energy $\text{Re}[G(\omega)]$ has maximal curvature
(these points are marked out by circles in Fig.~\ref{fig:kinks}c).
Thus the Fermi liquid regime, when the slope of the real part of self-energy
is described by $\partial \text{Re}\Sigma(\omega)/\partial\omega =  1-1/Z_{\text{FL}}$,
exists only within a narrow part of the central peak, namely in the energy interval
$|\omega |<\omega_\star$.
At higher (intermediate) energies the slope of $\text{Re}\Sigma(\omega)$
will be given by $\partial\text{Re}\Sigma(\omega)/\partial\omega =   1-1/Z_{\text{CP}}$.
As a result, effective dispersion $E_{\bf{k}}$ will manifest
kinks at energies $\omega_\star$.

This analysis also helps to understand why outside Fermi liquid region
$E_{\bf{k}}$ demonstrates another value of renormalization of noninteracting 
electron dispersion, given by $Z_{\text{CP}}$ with small offset $c$. 
This behavior is determined
by the value of main contribution to self-energy $\omega+\mu-1/G(\omega)$
at energies $\omega_\star<|\omega|<\Omega$ i.e. within the
central peak of DOS.
Values of $\omega_\star$, $Z_{\text{CP}}$, and $c$
can be expressed via $Z_{\text{FL}}$ together with characteristics
of noninteracting electron density of states.
One can show that $\omega_\star=Z_{\text{FL}}(\sqrt{2}-1)D$, where $D$ is the
halfwidth of the bare band (details see in work \cite{Nature}).
If correlations are weak and $Z_{\text{FL}}\lesssim1$
kinks positions in $E_{\bf{k}}$ practically coincide
with the edges of bare electron band, which makes them almost
unobservable. On the other hand, in the strongly correlated regime
($Z_{\text{FL}}\ll 1$) kinks energy $\omega_\star/D\propto Z_{\text{FL}}$ 
approaches the Fermi level inside the central peak, which width diminishes with
the increase of correlations as $\Omega/D\propto\sqrt{Z_{\text{FL}}}$ \cite{Bulla99}.

For the first time, these purely electronic kinks were observed in
LDA+DMFT calculations for SrVO$_3$ system \cite{Nekrasov06}.
Definition of energy scale $\omega_\star$ contains only parameters
of initial band structure, which can be obtained (for realistic systems)
via band structure calculations, together with Fermi liquid mass
renormalization $Z_{\text{FL}}=1/(1-\partial
\text{Re}\Sigma(0)/\partial\omega)\equiv m/m^*$, which can be experimentally
determined from specific heat or spin susceptibility measurements 
In particular, in Ref. \cite{Gamkink} it was shown that kinks of electronic 
nature can cause corresponding cusps in the linear (in temperature) term of
specific heat of strongly correlated metals, which was,
apparently, observed in LiV$_2$O$_4$ heavy fermion system.

\subsubsection{Role of electron-phonon interaction}
\label{dmft_eph}

The previous section material inevitably brings us to the question
about the relation and mutual influence of kinks of electronic nature and
the ``usual'' kinks in the electron dispersion induced by electron-phonon
interaction. This is related to a more general problem of joint influence
between strong electronic correlations and electron-phonon interaction.
In fact, the history of such investigations is relatively long and
one of the most popular models of electron-phonon interaction in
strongly correlated systems is Hubbard-Holstein model.
Hubbard model describes local Coulomb interaction on a lattice.
On the other hand, Holstein model describes linear interaction
of conduction electrons with {\em local} (Einstein) phonon modes 
\cite{Holstein59}. Studies of Hubbard-Holstein model
were performed in the framework of conventional DMFT  \cite{georges96},
in particular with the use of numerical renormalization group (NRG) \cite{BPH}
as ``impurity solver''. Reducing of Hubbard-Holstein model to
Anderson-Holstein impurity problem was first performed by Hewson and Mayer  \cite{HewMay02}.
They showed that using NRG one can calculate the total electron-phonon contribution
to self-energy, thus achieving nonperturbative solution of Hubbard-Holstein 
model not only with respect to Hubbard interaction, but also with
respect to electron-phonon interaction. Let us note that the general structure
of DMFT equations in this approach is preserved.

However, until recently there were no studies of strongly correlated electrons
interacting with Debye phonons. It is quite surprising
in view of rather wide discussion of the physics of kinks in electron dispersion
observed in ARPES experiments in high-T$_c$ superconducting oxides \cite{Lanzara2002}.
The origin of these kinks is typically attributed to electron-phonon interaction 
\cite{Shen2002}.
Problem of kinks formation in electron dispersion caused by electron-phonon
interaction in strongly correlated systems was shortly discussed in the 
framework of Hubbard-Holstein model Ref. \cite{Koller05,Hague}.
In this section we overview DMFT+$\Sigma$ results for Hubbard model
with added interaction with Debye phonons, under the assumption
of the validity of Migdal theorem (adiabatic approximation).
This approximation is reasonable for electron-phonon coupling constant
$\lambda <E_F/\omega _{D}\sim 10$, where $E_F$ is Fermi energy, 
$\omega _{D}$ is Debye frequency. 

To consider electron-phonon interaction in the strongly correlated system
we introduce self-energy $\Sigma_{\bf k}(\varepsilon)=\Sigma_{ph}(\varepsilon,{\bf k})$,
appearing in the usual Froehlich model. According to Migdal theorem,
in adiabatic approximation we can restrict ourselves to the simplest first order
contribution to $\Sigma_{ph}(\varepsilon,{\bf k})$.
Main advantage of adiabatic approximation is the possibility to neglect any
vertex corrections from electron-phonon interaction, which are small over
adiabatic parameter $\frac{\omega_D}{E_F}\ll 1$ \cite{Migdal}.

In fact, $\Sigma_{ph}(\varepsilon,\textbf{k})$ in this approximation
has only weak momentum dependence, which can be neglected and we have to account 
only for essential frequency dependence. Direct calculations (see e.g. \cite{Diagr})
in the case of Debye phonon spectra
$\omega_0(\textbf{k})=u|\textbf{k}|$ for $|\textbf{k}| < \frac{\omega_D}{u}$, 
where $u$ is sound velocity, give: 
\begin{equation}
\Sigma_{ph}(\varepsilon)=\frac{-ig^2}{4\omega_c^2} \int_{-\infty}^{+\infty}\frac{d\omega}{2\pi}
\bigl\{\omega_D^2+\omega^2ln\bigl|\frac{\omega_D^2-\omega^2}{\omega^2}\bigr|+
i\pi\omega^2\theta(\omega_D^2-\omega^2)\bigr\}I(\varepsilon+\omega),
\label{phdb}
\end{equation}
where $g$ is the usual electron-phonon interaction constant 
$I(\epsilon)=\int_{-D}^{+D}d\xi\frac{N_0(\xi)}{E_{\varepsilon}-\xi}$, 
$E_{\varepsilon}=\varepsilon-\Sigma(\varepsilon)-\Sigma_{ph}(\varepsilon)$
and $\omega_c=p_Fu$ is characteristic frequency of the order of Debye one.
In case of semielliptic bare DOS $N_0(\varepsilon)$ with
halfwidth $D$ one obtains
$I(\epsilon)=\frac{2}{D^2}(E_{\varepsilon}-\sqrt{E_{\varepsilon}^2-D^2})$.
It is convenient to introduce dimensionless electron-phonon coupling
constant, which for this model can be written as \cite{Diagr}: 
\begin{equation}
\lambda=g^2N_0(\varepsilon_F)\frac{\omega_D^2}{4\omega_c^2}.
\label{lambda}
\end{equation}
To simplify calculations we neglect renormalization of phonons
because of electron-phonon interaction \cite{Diagr}, assuming in the following
that phonon spectrum is fixed by experiment.

\subsubsection{Electronic and phonon kinks within the DMFT+$\Sigma$ approach}
\label{kink_res}

Let us focus on most interesting DMFT+$\Sigma$ results in this model,
referring the reader for details to Refs. \cite{e_ph_DMFT,e_ph_DMFTS}.
Here we present results obtained for the case of interaction of electrons
with Debye phonons (results for Einstein phonons is analogous \cite{e_ph_DMFTS}).
 
\begin{figure}
\includegraphics[clip=true,angle=270,width=0.42\columnwidth]{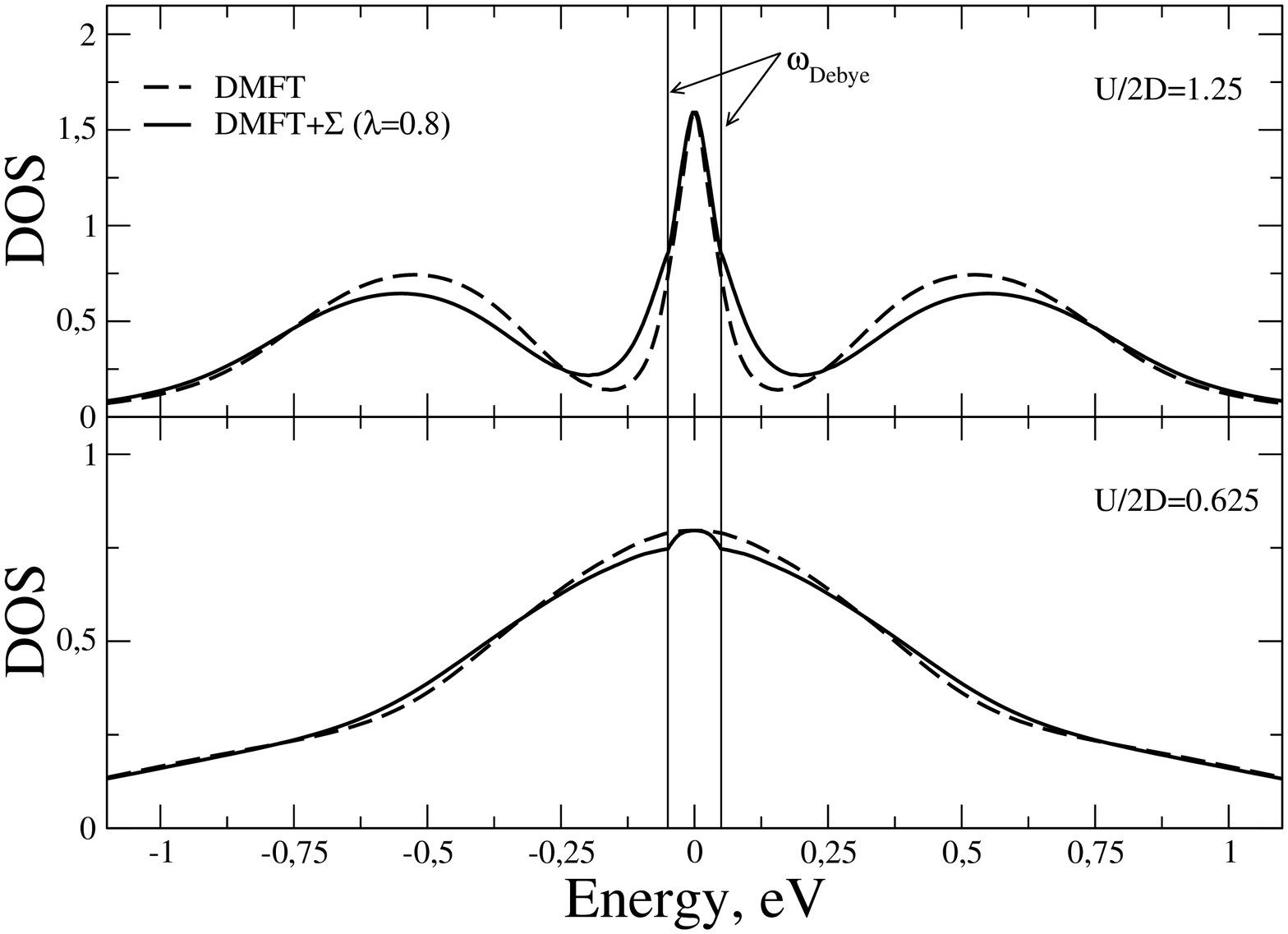}
\includegraphics[clip=true,angle=270,width=0.42\columnwidth]{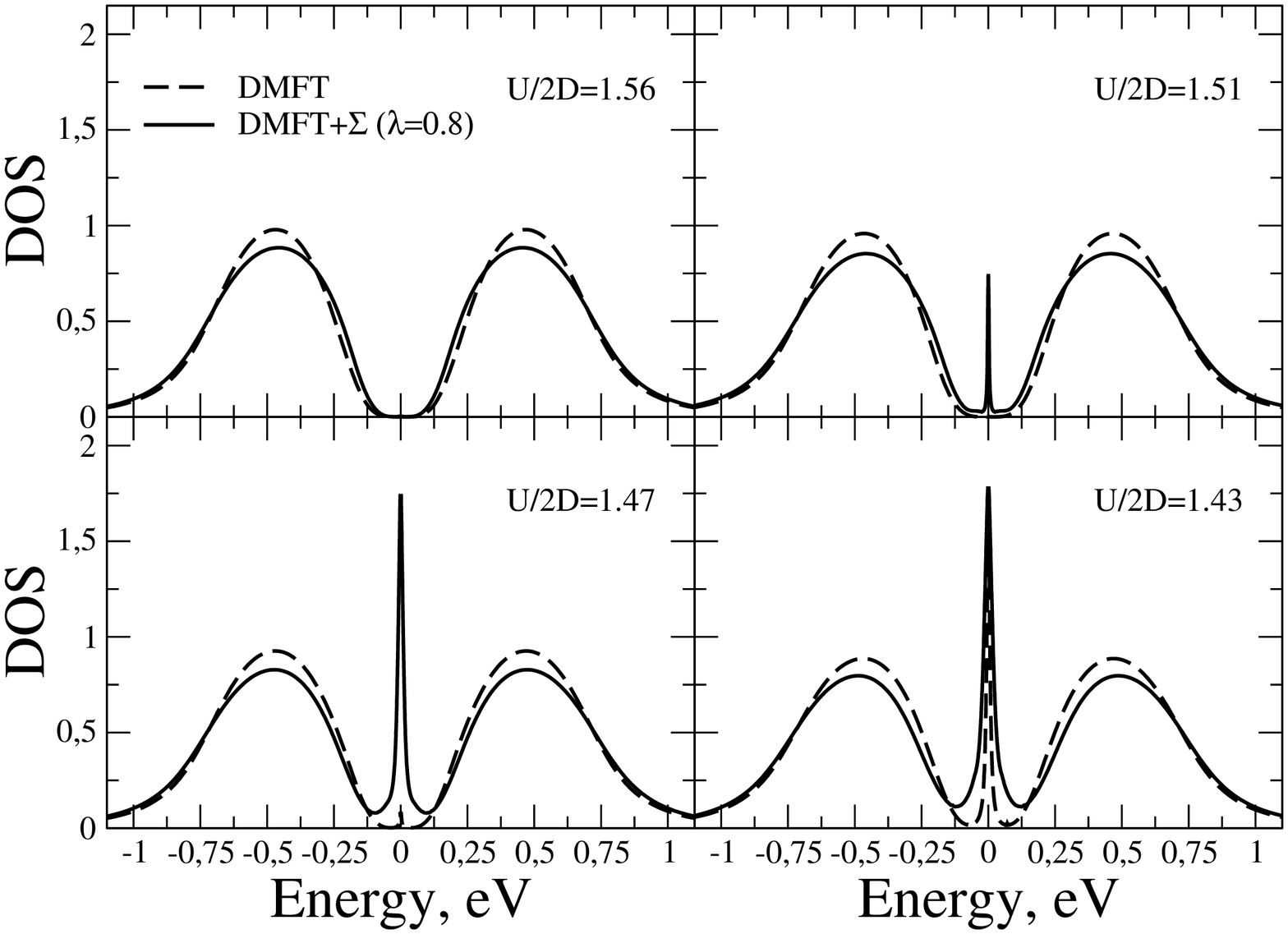}
\caption{\small  Comparison of DMFT (dashed line) and DMFT+$\Sigma_{ph}$ (solid line)
densities of states. 
On the left side -- for strong (upper panel, $U/2D=$1.25) 
and weak (lower panel, $U/2D=$0.625) Hubbard interactions. 
On the right side -- densities of states evolution close to metal-insulator
transition. Dimensionless electron-phonon coupling constant is $\lambda$=0.8 
\cite{e_ph_DMFT,e_ph_DMFTS}.}
\label{DOSes}
\end{figure}

Comparison of DMFT and DMFT+$\Sigma_{ph}$ densities of states with
electron-phonon interaction for strong (U/2D=1.25) and 
weak (U/2D=0.625) Hubbard interaction is presented on upper and lower
panels on the left side of Fig.~\ref{DOSes}.
Dimensionless constant (\ref{lambda}) exploited in these calculations
was chosen to be $\lambda$=0.8 and Debye frequency  $\omega_D$=0.125D.
In both cases we observe some spectral weigh transfer caused by
electron-phonon interaction. 
For U/2D=1.25 (upper panel of Fig.~\ref{DOSes})
we see well developed three-peak structure typical for strongly
correlated systems. In the energy interval $\pm \omega_D$ around the Fermi level
(which is zero energy in all figures) practically there is no difference
in the shape of quasiparticle peak in the DOS obtained within  
DMFT and DMFT+$\Sigma_{ph}$. 
But beyond this interval DMFT+$\Sigma_{ph}$ quasiparticle peak becomes
significantly wider because of partial transfer of spectral weight from 
Hubbard bands. This broadening of quasiparticle peak in DMFT+$\Sigma_{ph}$ 
leads to a delay of metal-insulator transition as we shall see later.

In case of U/2D=0.625 no clearly distinguishable Hubbard bands are formed and
we observe only some side wings in the DOS.
Redistribution of spectral weight on the lower panel of Fig.~\ref{DOSes}
is not very dramatic despite qualitative difference with the U/2D=1.25 case.
Main distinction between DMFT and DMFT+$\Sigma_{ph}$ results
occurs here in the interval $\pm \omega_D$, where formation of a ``cap'' in DOS
is observed, due to electron-phonon interaction.
Corresponding spectral weight goes to the energies around
$\pm$U where Hubbard bands start to form.

On the right panel of Fig..~\ref{DOSes} we compare behavior
of densities of states in DMFT without phonons and in DMFT+$\Sigma_{ph}$,
for different values of U/2D  parameter nearby Mott-Hubbard transition.
At U/2D=1.56 both standard DMFT and DMFT+$\Sigma_{ph}$ give
insulating solution. However, there are some distinctions between these solutions.
In DMFT+$\Sigma_{ph}$ Hubbard bands are lower and wider than in DMFT since
additional (electron-phonon) interaction is included.
With decrease of U at U/2D=1.51 and 1.47 we observe that DMFT+$\Sigma_{ph}$
results correspond to metallic state
(with narrow quasiparticle peak on the Fermi level),
while DMFT without phonons still shows insulating solution.
Only about U/2D=1.43 both DMFT and DMFT+$\Sigma_{ph}$ results for DOS 
correspond to the metallic state.
Thus, under increase of U the finite value of electron-phonon interaction 
slightly delays Mott-Hubbard transition from metallic to insulating phase.
This result is analogous to one obtained within the Hubbard-Holstein model
for weak electron-phonon coupling \cite{KolMayHew04,Jeon,Mayer04}.

Therefore, moderate electron-phonon interaction brings to rather
insignificant changes of electron density of states both in
correlated metal and in Mott insulating state, only slightly
delaying transition from metal to insulator with the growth of U.

Now we turn to the source of sharp slope changes
of electron dispersion (kinks). It is well known that in general case kinks 
are formed because of interaction of electrons with bosonic modes.
In case of electron-phonon interaction, typical energy of the kink
is about Debye (or Einstein) frequency. Above we have shown that in strongly
correlated metal kinks of purely electronic nature can arise~\cite{Nature}.
Energy of such kink for semielliptical bare DOS is $\omega^*=Z_{FL}(\sqrt{2}-1)D$,
where D is the halfwidth of the ``bare''  band and
$Z_{FL}=(1-\frac{\partial Re\Sigma)}{\partial \varepsilon}\bigr|_{\varepsilon=
E_F})^{-1}$ is Fermi liquid renormalization factor.
Roughly speaking $\omega^*$ is defined by halfwidth of the quasiparticle
peak in DOS.

Kink of electronic nature is quite smooth and its observation is rather 
difficult. DMFT+$\Sigma_{ph}$ calculations show that electronic kinks are 
hardly detectable on the background of phonon kinks and fine ``tuning'' of 
model parameters is necessary pick them out.
First of all, it is necessary to guarantee that $\omega_D \ll \omega^*$
(in other cases smooth electronic kinks will be practically
indiscernible against kinks from electron-phonon interaction).
For U/2D=1 and U=3.5~eV we have $\omega^*\sim$ 0.1D, while Debye
frequency can be taken quite small e.g. $\omega_D\sim$ 0.01D.
In order to make phonon kink pronounced enough at such relatively
low Debye frequency one needs to increase electron-phonon coupling constant
up to $\lambda$=2.0. 

\begin{figure}
\includegraphics[clip=true,width=0.42\columnwidth]{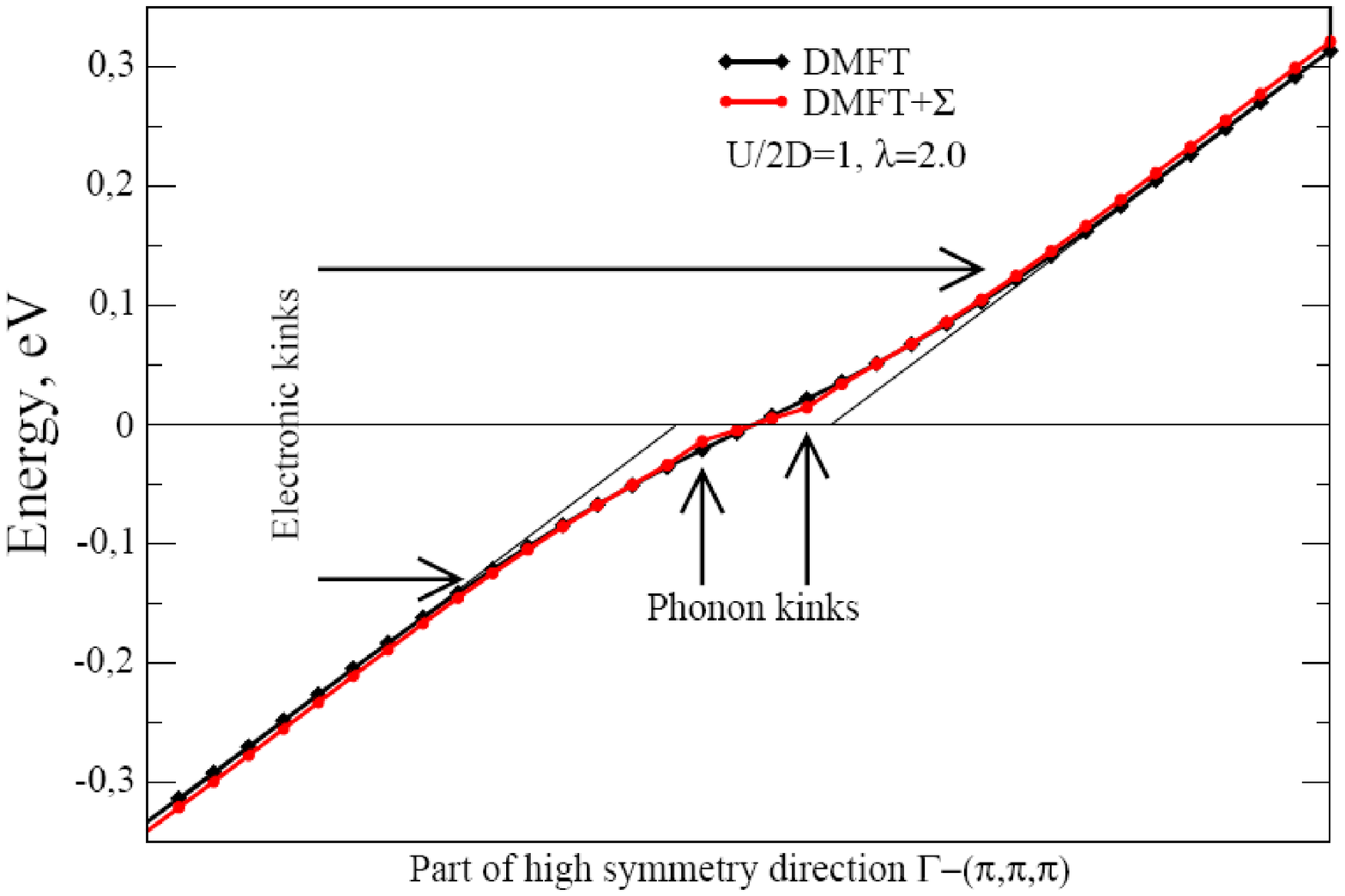}
\includegraphics[clip=true,width=0.42\columnwidth]{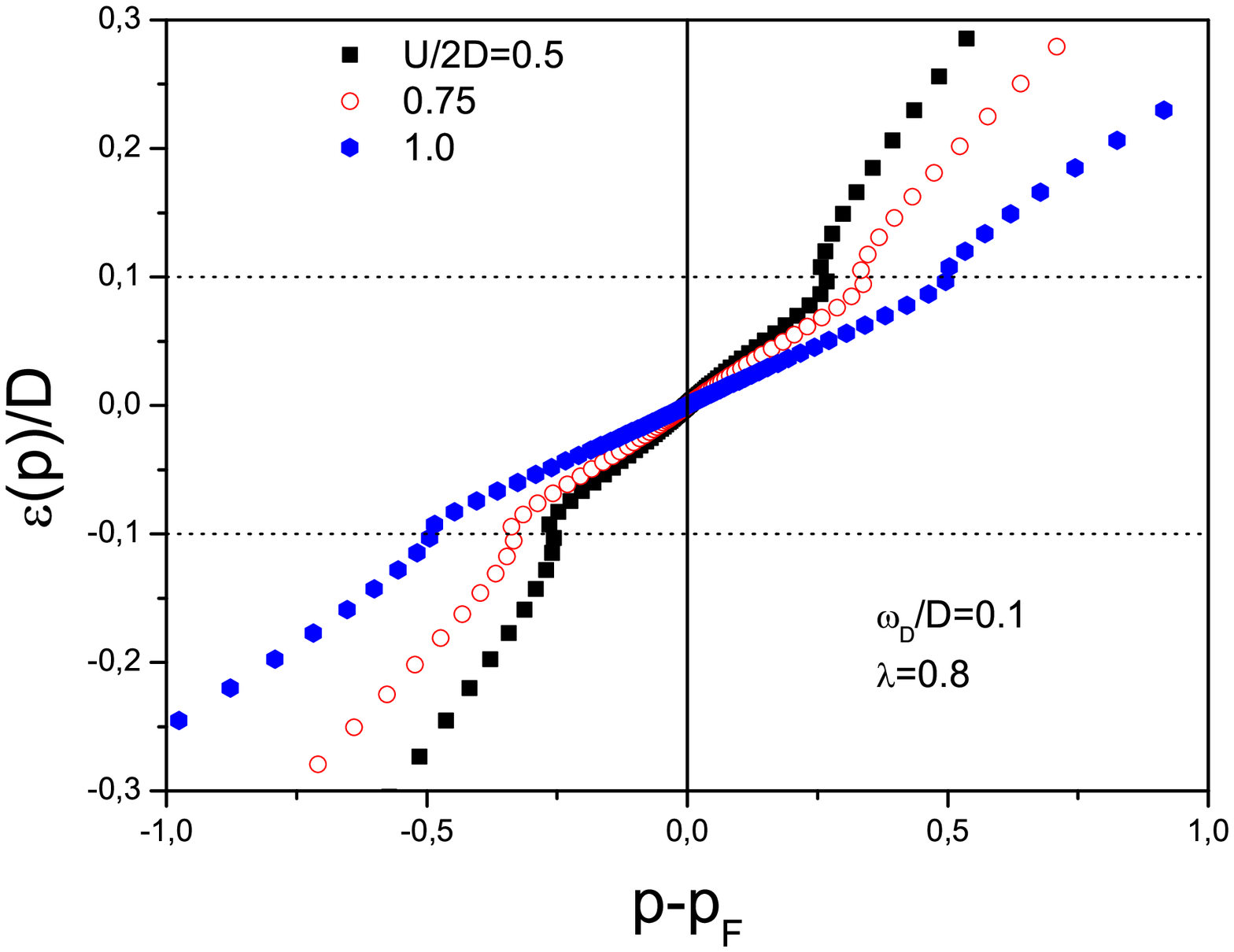}
\caption{\small Quasiparticle dispersion with kinks around the Fermi level.
On the left side -- dispersion along  $\Gamma-(\pi,\pi,\pi)$
high symmetry direction in the Brillouin zone for the case of
simple cubic lattice and bare dispersion with nearest neighbors hopping only:
DMFT (curve with diamonds) and DMFT+$\Sigma_{ph}$ (curve with circles) 
($\lambda=2$, $\omega_D=0.01D$). On the right side -- 
DMFT+$\Sigma_{ph}$ phonon kinks evolution for different
values of Hubbard interaction $U/2D=$\ 0.5,\ 0.75,\ 1.0;
\ $\lambda=0.8$, $\omega_D=0.1D$ \cite{e_ph_DMFT}.}
\label{ephkinks}
\end{figure}

To demonstrate the possibility of coexistence of both types of kinks
in the spectra, let us consider the energy dispersion for simple cubic
lattice with nearest neighbors hopping only. Most convenient is to discuss the
high symmetry direction
$\Gamma-(\pi,\pi,\pi)$ of the Brillouin zone \cite{Nature}.
On the left panel of Fig.~\ref{ephkinks} we show electronic dispersion
along this direction close to the Fermi level.
Line with diamonds is the electron spectrum of standard DMFT without phonons.
Lines with circles presents  DFMT+$\Sigma_{ph}$ results.
Electronic and phonon kinks are marked by arrows.
In general case, kinks from electron-phonon interaction dominate
for most typical model parameters making electronic kink observation
predicted in \cite{Nature} quite difficult.

In conclusion we give the picture of phonon kinks evolution 
depending on the value of Hubbard interaction $U$. With decrease of $U/2D$ 
ratio Fermi velocity goes down and kink position on the momentum axis
shifts farther away from $p_F$, while the kink energy remains about $\omega_D$.
This behavior follows from direct  DMFT+$\Sigma_{ph}$ calculations
\cite{e_ph_DMFT,e_ph_DMFTS} and is shown on the right panel of Fig.~\ref{ephkinks}.
In the case of interaction with Einstein phonons results are quite
analogous \cite{e_ph_DMFTS}.  

\section{Electronic structure of real strongly correlated systems: 
LDA+DMFT and LDA+DMFT+$\Sigma$}
\label{pg_real}

At present the most advanced {\sl ab initio} (i.e. ideally
without any fitting parameters) computational scheme of electron
spectra of {\it realistic} strongly correlated systems is 
LDA+DMFT method \cite{IzAn,poter97}.
LDA band structure in this method is used to obtain ``noninteracting'' starting 
Hamiltonian, while strong correlations are accounted afterwards within DMFT. 
In fact, LDA+DMFT computational scheme combines
two scientific areas: ``realistic'' band structure calculations
and traditional model approaches, which were essentially separated from
each other before.
Without DMFT just in LDA there is no way to describe strongly correlated
systems, while without LDA calculations manybody methods can not
be material specific. Below we briefly discuss the basics of LDA+DMFT and
generalized LDA+DMFT+$\Sigma$ methods.

\subsection{Density functional theory
(DFT). Local density approximation (LDA).}
\label{dft_lda}

In Born-Oppenheimer adiabatic approximation~\cite{Born27a},
neglecting relativistic effects, electronic properties in
solid state physics are described by the Hamiltonian: 
\begin{eqnarray}
\hat{H} &=&\sum_{\sigma }\int \!d^{3}r\;\hat{\Psi}^{+}({\bf
r},\sigma )\left[
{-\frac{\hbar ^{2}}{2m_{e}}\Delta +V_{{\rm ion}}({\bf r})}\right] \hat{\Psi}(%
{\bf r},\sigma )  \nonumber \\
&&+\frac{1}{2}\sum_{\sigma \sigma ^{\prime }}\int \!d^{3}r \, d^{3}r^{\prime }\;
\hat{\Psi}^{+}({\bf r},\sigma )\hat{\Psi}^{+}({\bf r^{\prime
}},\sigma
^{\prime })\;{V_{{\rm ee}}({\bf r}\!-\!{\bf r^{\prime }})}\;\hat{\Psi}({\bf %
r^{\prime }},\sigma ^{\prime })\hat{\Psi}({\bf r},\sigma ).
\label{abinitioHam}
\end{eqnarray}%
Here $ \hat{\Psi}^{+}({\bf r},\sigma )$ and $\hat{\Psi}({\bf r},\sigma )$ 
are creation and annihilation operators of electron with coordinate 
${\bf r}$ and spin $\sigma $, $\Delta $ is the Laplace operator, $m_{e}$ 
is electron mass, $e$ is electron charge, 
\begin{eqnarray}
V_{{\rm ion}} ({\bf r})=-e^2\sum_i \frac{Z_i}{|{\bf r}-{\bf R}_i|}
&\;\; {\rm ,}\;\; & V_{\rm ee}({\bf r}\!-\!{\bf r'})=\frac{e^2}{2}
\sum_{{\bf r} \neq {\bf r'}} \frac{1}{|{\bf r}-{\bf r'}|}
\end{eqnarray}
denote single particle potential created by all ions  $i$ with charge
$eZ_{i}$ located at given positions ${\bf R}_i$ and electron-electron 
interaction. 

Although the ``first principle'' Hamiltonian (\ref{abinitioHam}) is easy to 
write down, it is impossible to solve corresponding quantum mechanical problem 
exactly. This is the reason to make substantial physical approximations.
In particular, density functional theory (DFT) is based on
Hohenberg-Kohn theorem ~\cite{Hohenberg64a} 
(see also the review~\cite{JonesGunn}), which asserts that ground state energy
is the unique functional of electron charge density, which is minimal for
the equilibrium electron density in the ground state:
\begin{equation}
E[\rho ]=E_{{\rm kin}}[\rho ]+E_{{\rm ion}}[\rho ]+E_{{\rm
Hartree}}[\rho ]+E_{{\rm xc}}[\rho ],  \label{Erho}
\end{equation}%
where the Hartree energy
$E_{{\rm Hartree}}[\rho ]=\frac{1}{2}\int d^{3}r^{\prime }\,d^{3}r\;
V_{{\rm ee}}({\bf r}\!-\!{\bf r^{\prime }})\;\rho ({\bf r^{\prime })\rho (r)}$ 
and potential energy of ions 
$E_{{\rm ion}}[\rho ]=\int d^{3}r\;V_{{\rm ion}}({\bf r})\;\rho({\bf r)}$, 
can be directly expressed via electron charge density.
The term $E_{{\rm kin}}[\rho ]$ denotes the kinetic energy of electrons and
$E_{{\rm xc}}[\rho ]$ is unknown, in general, exchange-correlation term,
containing electron-electron interaction energy beyond the Hartree term.
In fact, all peculiarities of the manybody problem are transferred into
the computation of $E_{{\rm xc}}[\rho ]$. 

In practice, instead of minimization of $E[\rho ]$ over $\rho $,
minimization is performed usually over some set of orthonormal functions
$\varphi _{i}$, related to $\rho $ via expression: 
\begin{equation}
\rho ({\bf r})=\sum_{i=1}^{N}|\varphi _{i}({\bf r})|^{2}.
\label{rhophi}
\end{equation}%
Introducing arbitrary Lagrange parameters $\varepsilon _{i}$ and requiring
\begin{equation}
\frac{\delta}{\delta \varphi_{i}(\bf r)}\left\{E[\rho ]+\varepsilon _{i}[1-\int d^{3}r
|\varphi _{i}({\bf r})|^{2}]\right\}=0, 
\label{var_pr}
\end{equation}
one gets Kohn-Shem equations \cite{DFT_LDA}:
\begin{equation}
\left[ -{\frac{\hbar ^{2}}{2m_{e}}\Delta +V_{{\rm ion}}({\bf r})}+
\int d^{3}{r^{\prime }}\,
{V_{{\rm ee}}({\bf r}\!-\!{\bf r^{\prime }})} {\rho ({\bf r^{\prime }})}
+{{\frac{\delta {E_{{\rm xc}}[\rho ]}}{\delta \rho ({\bf r)}}}}%
\right] \varphi _{i}({\bf r})=\varepsilon _{i}\;\varphi _{i}({\bf r}). \label{KohnSham}
\end{equation}
Formally this equation coincides with single particle stationary Schroedinger 
equation. Electron kinetic energy corresponding to charge density of the ground 
state is given now by the expression
\begin{equation}
E_{{\rm kin}}[\rho _{{\rm min }}]=-\sum_{i=1}^{N}
\langle \varphi _{i}|{\hbar ^{2}\Delta }/{(2m_{e})}|\varphi _{i}\rangle,
\label{kinen}
\end{equation} 
where $\varphi _{i}$ are self-consistent (spin degenerate) solutions of
equations (\ref{KohnSham}) and (\ref{rhophi}), corresponding 
to the smallest single particle energy $\epsilon_{i}$~\cite{Janak}.

Most common way to calculate $E_{\rm xc}[\rho ]$ is to use the {\em local 
density approximation} (LDA). It approximates the functional 
$E_{\rm xc}[\rho ]$ by the local charge density functional:
\begin{equation}
E_{\rm xc}[\rho ]\;{\rightarrow }\;\int d^{3}r\;\rho ({\bf r})\epsilon_{\rm xc}^{{\rm LDA}}(\rho ({\bf r}))
\end{equation}
Explicit expression for $\epsilon_{\rm xc}^{{\rm LDA}}(\rho ({\bf r}))$
can be found in the framework of different models e.g.
from numerical analysis of ``jellium'' model
(electronic gas on a positive ionic background) \cite{jellium2}.

In fact, LDA approximation corresponds to the replacement of
the Hamiltonian (\ref{abinitioHam}) by
\begin{eqnarray}
\hat{H}_{{\rm LDA}} &=&\sum_{\sigma }\int \!d^{3}r\;\hat{\Psi}^{+}({\bf r}%
,\sigma )\left[ -\frac{\hbar ^{2}}{2m_{e}}\Delta +V_{{\rm ion}}({\bf r}%
)+\int d^3{r^{\prime }}\,{\rho ({\bf r^{\prime }})}{V_{{\rm ee}}({\bf r}\!-\!%
{\bf r^{\prime }})}\right.   \nonumber \\ &&\left. \phantom
{\sum_{\sigma \sigma'} \int\! d^3r\; \hat{\Psi}^+({\bf r},\sigma)
\big[ \;}+{{\frac{\delta {E_{\rm xc}^{{\rm LDA}}}[\rho ]}{\delta \rho
({\bf r)}}}}\right] \hat{\Psi}({\bf r},\sigma ).  \label{HLDA0}
\end{eqnarray}%
In practical calculations electron field operators are expanded over
some set of atomic-like functions (basis set) $\Phi _{ilm}$ 
($i$ denotes lattice site, $l$ is orbital quantum number, 
$m$ is magnetic quantum number). 
In this representation
\begin{eqnarray}
\hat{\Psi}^+({\bf r},\sigma) &=& \sum_{i l m} \hat{c}_{i l
m}^{\sigma \dagger} \Phi_{i l m}({\bf r})^{\phantom{+}}
\end{eqnarray}
and the Hamiltonian (\ref{HLDA0}) is rewritten as
\begin{eqnarray}
 \hat{H}_{\rm LDA}  &=& \sum_{ilm,{\rm }jl^{\prime }m^{\prime },\sigma }(\delta
_{ilm,jl^{\prime }m^{\prime }} \;
{\varepsilon_{ilm}}^{\phantom{\sigma}}{\hat{n}}_{ilm}^{\sigma
}+ {t_{ilm,jl^{\prime }m^{\prime }}}\;{\hat{c}}_{ilm}^{\sigma \dagger }{\hat{c}}%
_{jl^{\prime }m^{\prime }}^{\sigma }). \label{HLDA}
\end{eqnarray}
Here ${\hat{n}}_{ilm}^{\sigma }=\hat{c}_{ilm}^{\sigma \dagger }{\hat{c}}%
_{ilm}^{\sigma }$ is electron density operator on a given orbital, while matrix 
elements:
\begin{equation}
t_{ilm,jl^{\prime }m^{\prime }}=\Big\langle\Phi
_{ilm}\Big|-\frac{\hbar
^{2}\Delta }{2m_{e}}+V_{{\rm ion}}({\bf r})+\int d^3{r^{\prime }}{\rho ({\bf %
r^{\prime }})}{V_{{\rm ee}}({\bf r}\!-\!{\bf r^{\prime }})}+{{\frac{\delta {%
E_{\rm xc}^{{\rm LDA}}}[\rho ]}{\delta \rho ({\bf r)}}}}\Big|\Phi
_{jl^{\prime }m^{\prime }}\Big\rangle
\label{matel}
\end{equation}%
in case of $ilm\neq jl^{\prime }m^{\prime }$  define effective hopping integrals
and single particle energies $\varepsilon _{ilm}$ are given by corresponding
diagonal expressions in case of identical indices.
On this step purely analytic work ends and numerical calculations follow
within the chosen basis set, e.g. linearized muffin-tin orbitals (LMTO).
Specific expressions for matrix elements (\ref{matel}) within the LMTO basis
are presented in Ref.~\cite{LMTO}.

\subsection{LDA+DMFT computational scheme}
\label{lda_dmft_sch}

The primary importance for strongly correlated materials is the
onsite Coulomb interaction between $d$- or $f$-electrons,
since this contribution to interaction energy is the largest one.
The largest nonlocal contribution is density-density type
interaction between nearest neighbors, where the main contribution
comes from Hartree term (see~\cite{MuHa89} and~\cite{Wahle98}),
which is already taken into account in the LDA.
Moreover in LDA some part of exchange-correlation
interactions is taken into account via  effective potential
${{\frac{\delta {E_{\rm xc}^{{\rm LDA}}}[\rho ]}{\delta \rho ({\bf r)}}}}$.

In order to take into account strong local Coulomb interaction one should
supplement Hamiltonian  (\ref{HLDA}) with approximate Coulomb matrix
with most important parameters only \cite{poter97}: $U$ -- onsite intraband
Coulomb repulsion, $J$ -- exchange interaction and interorbital Coulomb
repulsion $U^\prime$, acting on different electronic orbitals of the same site 
$i_d$, where sits an atom with partially filled $d$-shell 
($l$ -- orbital quantum number,
$m$ --  magnetic quantum number, $\sigma$ -- spin index): 
\begin{eqnarray} \hat{H} =\hat{H}_{\rm LDA}
&+& U\sum_{m}\sum_{i=i_d, l= l_d} \hat{n}_{ilm\uparrow}\hat{n}_{ilm\downarrow}\label{H} \nonumber \\ 
&+&\;\sum_{i=i_d, l=l_d}\sum_{m\neq m'}\sum_{\sigma \sigma'}\;(U'-\delta_{\sigma \sigma'}J)\; \hat{n}_{ilm\sigma}\hat{n}_{ilm'\sigma'} \\
&-&{\sum_{i=i_d, l= l_d}}\sum_{m\sigma} \Delta\epsilon_d \,\hat{n}_{ilm\sigma},
\nonumber
\label{Hint}
\end{eqnarray}
where for simplicity only density-density type interactions are left,
and so called Kanamori parametrization is applied, when for the same
orbitals ($m=m'$) direct Coulomb interaction is taken as $U$, while for
different orbitals ($m\neq m'$) this interaction is equal to $U^\prime$.
Because of rotational invariance of the single atom problem $U^\prime=U-2J$
and exchange interaction parameter does not depend on orbital index
and is equal to $J$.

Moreover, in (\ref{Hint}) the last $\Delta\epsilon_d$ - term is added (the so 
called double counting correction), which should correct for double counting, 
as some part of local Coulomb interaction is already
included into $\hat{H}_{{\rm LDA}}$.
General expression for $\Delta\epsilon_d$ via $U$ and $\rho$ is unknown.
However, there are several qualitative recipes to determine the value of
$\Delta\epsilon_d$, and which are employed in different modern LDA+DMFT
calculations (detailed discussion can be found in Refs. \cite{Held07,Korolak}). 
The simplest physical assumption is that within the DFT Coulomb interaction
energy can be written as:
\begin{equation}
E_{DFT}=\frac{1}{2}{\bar U}n_d(n_d-1),
\label{CoulDFT}
\end{equation}
where $n_d$ is the total number of electrons on $d$-shell and $\bar U$ is the 
average Coulomb interaction (here we assume averaging over all orbital pairs
$m\sigma$, $m'\sigma'$ on a given site). Thus $\Delta\epsilon_d$ is taken as:
\begin{equation}
\Delta\epsilon_d=\frac{\partial E_{DFT}}{\partial n_d}=\bar U \left(n_d-
\frac{1}{2}\right).
\label{EDC}
\end{equation} 

The values of interaction parameters $U$, $J$ and $U^\prime$ can be obtained 
from the averaged Coulomb interaction $\bar U$ and Hund exchange parameter $J$. 
The averaged interaction $\bar U$ is related to $U$ and 
$U'$ parameters via the following relation:
\begin{equation} 
\bar U = \frac{U+(N_{\rm orb}-1)U'+(N_{\rm orb}-1)(U'-J)}{2N_{\rm orb}- 1},
\label{ubar}
\end{equation}
where N$_{\rm orb}$ is the number of interacting orbitals. 
Since $U$ and $U^\prime$ are not independent parameters $\bar{U}$ and $J$ are 
sufficient to determine $U$~\cite{LDADMFT,Zoelfl00}.

For microscopic calculations of averaged Coulomb interaction
different methods were developed, such as ``constrained LDA'' 
\cite{Gunnarsson88} or ``constrained RPA'' \cite{CRPA,RPAC}.
Generalizations to calculate Hund exchange parameter also exist.
Unfortunately, there are rather large discrepancies between the values of
parameters obtained with these methods.
The value of $\bar U$ appears to be strongly dependent on the basis set
used (e.g. in the problem of screening of long-range part
of Coulomb interaction). It is clear that introduction of all these
essentially model parameters takes us quite far away from 
``first principle'' ideal, though it is the best one can do
at the moment to calculate band structure of solids with
transition metal atoms, where electron-electron interactions
play the crucial role. In that sense it is probably more correct to
speak about ``modelling'' of electron structure of such systems.
 
Matrix elements of ``noninteracting'' Hamiltonian in the reciprocal space
$H^{0}_{{\rm LDA}}({\bf k})$ can be calculated numerically at every point of
the Brillouin zone, then the integrals over the Brillouin zone usually are 
calculated with tetrahedron method \cite{Lambin}.
In case of relatively simple band dispersions, when analytical
expression for $H^{0}_{{\rm LDA}}({\bf k})$ dependence on ${\bf k}$ can be 
written explicitly, the values of hopping integrals can be found from LDA by 
projecting on corresponding Wannier functions \cite{NMTO,Anis}.
Matrix elements of this Hamiltonian, i.e. single particle LDA energies
without local Coulomb interaction can be written in a following way:
\begin{eqnarray}
\!(H^{0}_{{\rm LDA}}({\bf k}))_{qlm,q^{\prime }l^{\prime }m^{\prime
}}\!&\!\! =\!\!\! &\!(H_{{\rm LDA}}({\bf k}))_{qlm,q^{\prime}l^{\prime}m^{\prime }}
 -\delta _{qlm,q^{\prime }l^{\prime}m^{\prime }}\delta _{ql,q_{d}l_{d}}
\Delta \epsilon_d n_d.
\label{LDAHam}
\end{eqnarray}
where $q$ is index of an atom in the primitive unit cell.

The essence of the next step is to use in DMFT or DMFT+$\Sigma$ loop 
(see section \ref{secDMFT}) the local lattice Green's function 
(\ref{Gloc}), determined by momentum integrated Dyson's equation of the form:
\begin{eqnarray}
G_{qlm,q^{\prime }l^{\prime }m^{\prime }}(\omega )=\!\frac{1}{V_{B}}\int
{{d}{\bf k}} \!&\!\left[ \;\omega \;\delta _{qlm,q^{\prime }l^{\prime
}m^{\prime }}-(H_{{\rm LDA}}^{0}({\bf k}))_{qlm,q^{\prime
}l^{\prime }m^{\prime }}\right.   &\nonumber \\ &\;\;+\;\delta
_{ql,q_{d}l_{d}}\;\Sigma _{qlm,q^{\prime }l^{\prime }m^{\prime
}}(\omega )]^{-1},&  \label{Dyson}
\end{eqnarray}
where $[...]^{-1}$ denotes the inverse matrix with indices
$n$(=$qlm$), $n^{\prime }$(=$q^{\prime }l^{\prime }m^{\prime }$),
while the integration is performed over the Brillouin zone of the volume 
$V_{B}$.

Significant simplification of computations is achieved for the case of cubic 
lattice symmetry, when the crystal field strongly splits $d$-orbitals into the 
three-fold degenerate  $t_{2g}$-states and two-fold degenerate $e_g$-states.
In this special case both Green's function and self-energy become diagonal
over orbital and spin indices. Then the calculation of local Green's function
of the lattice problem can be performed as energy integration with the use of
unperturbed density of states, which allows to avoid tedious
integration over the Brillouin zone  in (\ref{Dyson}) and write:
\begin{equation}
G(\omega)=G^0(\omega-\Sigma(\omega))=\int d\epsilon\frac{N^0(\epsilon)}
{\omega-\Sigma(\omega)-\epsilon}.
\label{Hilb}
\end{equation}
In this case double counting correction  $\Delta\epsilon_d$ reduces to
immaterial shift of the chemical potential and its particular
mathematical form is irrelevant.

\subsection{Examples of LDA+DMFT calculations.}
\label{LDADMFT_res}

\subsubsection{Cubic perovskites CaVO$_3$ and SrVO$_3$}
\label{srcavo}

In this section we consider examples of some LDA+DMFT calculations of
electronic band structure of realistic compounds with strong enough
electronic correlations. Transition metal oxides are ideal testing area
to study electronic correlations in solids. Among these materials cubic 
perovskites have simplest crystal structure and thus can be viewed as a 
starting point to understand electronic properties of more complex systems.
Usually 3$d$-states in such materials form comparatively narrow bands of
the width $W\!\!\sim \!2\!-\!3\,$~eV, leading to strong 
electron-electron correlations.

Modern stage of experimental investigations of spectral and transport
properties of strongly correlated 3$d^1$ transition metal oxides
started from the work of Fujimori {\it et al.}~\cite{Fujimori92a}.
The authors, apparently for the first time, discovered strongly pronounced
lower Hubbard band in photoemission spectra, which could not be explained
by standard methods of band structure calculations.
In many of earlier works ~\cite{old_experiments1,old_experiments2,old_experiments4,Morikawa95},
devoted to the properties of the series of compounds Sr$_{1-x}$Ca$_x$VO$_3$ with different 
values of $x$ rather controversial results were  reported.
While thermodynamic characteristics (Sommerfeld coefficient, electric resistivity
and magnetic susceptibility) appeared to be more or less $x$ independent,
spectroscopic measurements data rather strongly changed as system transformed 
from $x\!=\!0$ (SrVO$_3$) to $x\!=\!1$ (CaVO$_3$).
These data indicated a transition from strongly correlated metal (SrVO$_3$) to
practically ideal insulator (CaVO$_3$), with  concentration range 
$x\!\rightarrow \!1$ in Sr$_{1-x}$Ca$_x$VO$_3$ being the boundary of
Mott-Hubbard transition. Analysis of this problem was performed using the 
high penetration depth photoemission experiments by Maiti {\it et al.}~\cite{Maiti01},
and similar experiments with high resolution photoemission by Sekiyama {\it et al.}~\cite{Sekiyama02}.
In particular, in the last work it was shown that
(1) surface preparation technique is very important (preferable is cleavage method)
and (2) energy of X-ray beam should be big enough to provide penetration depth
of several elementary cells. At the same time high instrumental resolution 
should be guaranteed (about 100~meV in the work~\cite{Sekiyama02}).
Such improvement of photoemission methods lead to observation
of almost identical spectra for Sr(Ca)VO$_3$~\cite{Maiti01,Sekiyama02},
demonstrating agreement of spectroscopic and thermodynamic measurements.
Results of these experiments agree also with earlier
1$s$ X-ray absorption spectra (XAS) obtained by Inoue {\it et al.}~\cite{Inoue94}, 
which differ only for energies slightly below the Fermi level in contrast to
BIS data~\cite{Morikawa95}.
In the framework of single band Hubbard model with neglect of orbital structure 
of 3$d$-shell of V, Rozenberg {\it et al.}~\cite{Rozenberg96} modelled
Sr$_{1-x}$Ca$_x$VO$_{3}$ spectra obtained by high penetration depth
photoemission \cite{Maiti01} using adjustable parameters.
Later in Ref.~\cite{Sekiyama02} it was demonstrated that data 
Ref.~\cite{Maiti01} contained quite significant surface contribution.

Below we present results of LDA+DMFT(QMC) calculations, performed without 
any adjustable parameters, both for spectral function and density of states of 
cubic SrVO$_3$ and orthorhombic CaVO$_3$ perovskites. According to these 
both systems are strongly correlated metals, which are quite far away from 
metal-insulator transition boundary. Despite significantly smaller V--O--V bond 
angle in CaVO$_3$, photoemission spectra of both systems are very similar and 
their quasiparticle peaks are almost identical. The results obtained agreed very 
well with modern high resolution bulk sensitive photoemission data, mentioned
above. In the spectral function of SrVO$_3$, obtained from LDA+DMFT(QMC) 
calculations, kinks of purely electronic nature at about 200 meV were observed,
and later these kinks were observed experimentally.

\paragraph{Results of LDA+DMFT calculations}

First of all from LDA calculated band structure we extract single electron 
Hamiltonian $\hat{H}_{\mathrm{LDA}}^{0}$ with subtracted average Coulomb 
interaction (to avoid double counting) \cite{poter97}. Supplementing 
$\hat{H}_{\mathrm{LDA}}^{0}$ with local Coulomb interaction
between electrons we obtain the Hamiltonian (\ref{H}) for the material of 
interest. Since symmetry of CaVO$_{3}$ is close to cubic one, it is possible to 
simplify the calculations and use integration with LDA density of states 
$N^{0}(\epsilon )$, instead of integration over the Brillouin zone.
In the Hamiltonian (\ref{H}) local intraorbital and interorbital
repulsions and exchange interactions are taken into account explicitly as
$U$, $U^{\prime }$ and $J$.
The values of these interactions for SrVO$_{3}$ were calculated
by constrained LDA method~\cite{Gunnarsson88} with  
$e_{g}$-states included into screening~\cite{Solovyev96}.
Obtained value of averaged Coulomb interaction is
$\bar{U}=3.55$~eV ($\bar{U}=U^{\prime }$ for $t_{2g}$
orbitals~\cite{LDADMFT1,Zoelfl00}) and $J=1.0$~eV.
Intraorbital Coulomb repulsion $U$ is fixed by rotational invariance
$U=U^{\prime }+2J = 5.55$~eV. 
For CaVO$_{3}$ $\bar U$ was not calculated,
since standard procedure of calculation of Coulomb interaction
parameters between two $t_{2g}$ electrons screened by
$e_{g}$ states is not applicable for distorted crystal structure,
where $e_{g}$ and $t_{2g}$ orbital are not separated by symmetry.
On the other hand it is known that changes of {\em local} Coulomb
interaction are usually much smaller than changes in density of states,
which as shown above are weakly dependent on bond angle V-O-V.
It means that $\bar{U}$ for CaVO$_{3}$ should be practically the same
as for SrVO$_{3}$. Correspondingly, the values  $\bar{U}=3.55$~eV and $J=1.0$ eV
were used for both SrVO$_{3}$ and CaVO$_{3}$.
These values agree with other band structure calculations for
vanadium compounds ~\cite{Solovyev96} and experimental data~\cite{Makino98}.

\begin{figure}
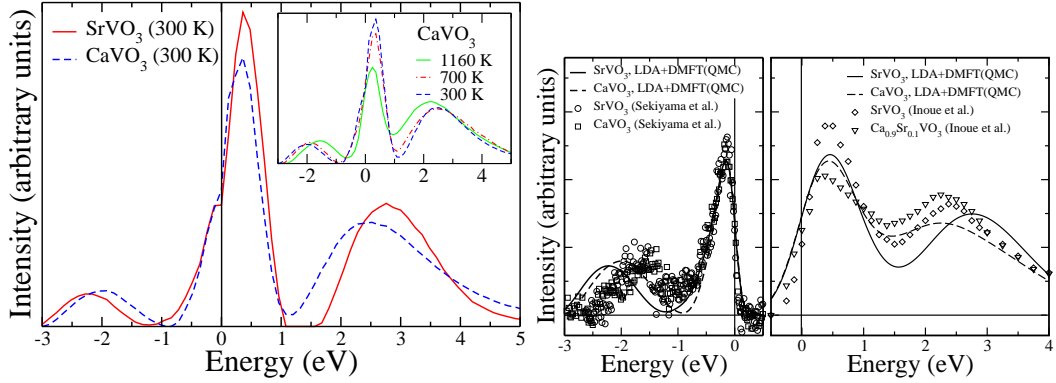

\centering \epsfig{file=lda_dmft.eps,width=0.43\textwidth,clip}
\centering \epsfig{file=suga_exp_spline.eps,width=0.43\textwidth,clip}
\caption{On the left side: LDA+DMFT(QMC) spectra for SrVO$_{3}$ (solid line) and 
CaVO$_{3}$ (dashed line) at T=300K (inset: temperature influence on CaVO$_{3}$ 
spectrum lineshape).
On the right side: comparison of calculated (without adjustable parameters)
LDA+DMFT(QMC) spectra for SrVO$_{3}$ (solid line) and CaVO$_{3}$ (dashed line) 
with high resolution bulk sensitive photoemission data 
(SrVO$_{3}$ -- circles; CaVO$_{3}$ -- rectangles)~\protect\cite{Sekiyama02} 
(left picture) and 1$s$- XAS-spectra: (SrVO$_{3}$ -- diamonds, 
Ca$_{0.9}$Sr$_{0.1}$VO$_{3}$ -- triangles)~\protect\cite{Inoue94}
(right figure). 
Horizontal line -- experimental background.}
\label{fig_ldadmft}
\end{figure}

Further calculations with Hamiltonian (\ref{H}) were performed in the framework
of DMFT with quantum Monte-Carlo method (QMC)~\cite{QMC} as an impurity solver.
In QMC the Green's function was obtained for the imaginary time and then 
continued on the real time (frequency) by maximum entropy method \cite{MEM}.
In LDA+DMFT(QMC) calculated spectra for SrVO$_{3}$ and CaVO$_{3}$ shown in
Fig.~\ref{fig_ldadmft} (on the left side) we observe manifestations of 
correlation effects, such as formation of lower Hubbard bands near
$-1.5$~eV and upper Hubbard bands at about $2.5$~eV with well developed
quasiparticle peaks on the Fermi level.
Thus both SrVO$_{3}$ and CaVO$_{3}$ are strongly correlated metals.
Difference of bare bandwidth (about 4\%) only leads to a small
additional spectral weight transfer from quasiparticle peak to Hubbard bands
and slight changes of Hubbard bands positions.
Obviously, both systems are close to Mott-Hubbard metal insulator transition.
Manybody densities of states for both systems (Fig.~\ref{fig_ldadmft}) are 
similar but not identical.
Indeed, SrVO$_{3}$ is a bit less correlated than CaVO$_{3}$ in agreement with
difference of the LDA bandwidths. The inset in Fig.~\ref{fig_ldadmft}
shows that temperature influence on the spectrum is small for $T \lesssim 700$~K. 

In the middle part of Fig.~\ref{fig_ldadmft} LDA+DMFT(QMC) densities of states 
(obtained at $T$=300K, multiplied with the Fermi function at 20$\,$K and
broadened with Gaussian of the width 0.1 eV to mimic experimental 
resolution~\cite{Sekiyama02})
are compared with experimental photoemission data after subtraction of oxygen 
and surface contributions. In particular, the height and the width of these 
spectra are almost the same in SrVO$_{3}$ and CaVO$_{3}$ (with a bit of 
difference above the Fermi level). On the other hand, positions of lower Hubbard 
band differs quite noticeably. This distinctions might occur because of 
subtraction of (estimated) oxygen contribution, which can delete part of 3$d$ 
spectral weight below $-2$~eV, as well as with uncertainties of $\bar{U}$ 
calculations. 

On the right side of Fig.~\ref{fig_ldadmft} we show comparison with XAS data.
Finite lifetime effects for holes are taken into account by broadening
of theoretical spectra with Lorentzian of the width of 0.2~eV~\cite{Krause79},
multiplication by inverse Fermi function (at $T$=80 K) and further broadening 
with Gaussian for experimental resolution $0.36$ eV.
General agreement of weights and positions of quasiparticle and upper Hubbard 
bands for $t_{2g}$-band is good, including tendencies associated with transition 
from .SrVO$_{3}$ to CaVO$_{3}$ (in the experiment Ca$_{0.9}$Sr$_{0.1}$VO$_{3}$). 
For CaVO$_{3}$ quasiparticle spectral peak weight is a bit smaller than in the 
experiment. In contrast to single band Hubbard model calculations LDA+DMFT
accounts for peculiarities of the systems and reproduce strong
asymmetry of the spectra close to the Fermi energy, including
relative weights and bandwidths. These results give different
interpretation of XAS as compared with Ref.~\cite{Inoue94}, where maximum at
$2.5\,$ eV was associated with $e_g$ band and not with upper Hubbard band
of $t_{2g}$ band. Small differences of qusiparticle peaks
(see Fig.~\ref{fig_ldadmft}) lead to different values of effective masses:
$m^*/m_0\!=\!2.1$ for SrVO$_{3}$ and $m^*/m_0\!=\!2.4$ for CaVO$_{3}$.
These theoretical values agree with  $m^{\ast}/m_{0}\!=\!2-3$ for SrVO$_{3}$ and 
CaVO$_{3}$ obtained from de Haas-van Alphen experiments and thermodynamic data
~\cite{old_experiments1,old_experiments2,old_experiments4,Inoue02}. 
Note that the effective mass for CaVO$_{3}$ determined from optical experiments
is slightly larger: $m^{\ast }/m_{0}\!=\!3.9$~\cite{Makino98}. 

\subsubsection{Kinks in spectral function of SrVO$_3$.}
\label{ARPES}

Let us consider in detail LDA+DMFT(QMC) results for spectral function
$A({\BK},\omega)$ for SrVO$_3$ obtained in Ref. \cite{Nekrasov06}.
Owing to ideal cubic lattice symmetry self-energy matrix $\Sigma(\omega)$
is diagonal and all diagonal elements are the same for all $t_{2g}$ orbitals.
Spectral function is defined by imaginary part of Green function 
${\rm Im}G({\BK},\omega)$, i.e. in fact by self-energy $\Sigma(\omega)$
on real axis. This self-energy was calculated by numerical solution of
Dyson's equation for the known interacting and bare Green functions as
described in the Appendix of Ref. \cite{Anis}.

In the Fig.~\ref{sigma_qmc} this self-energy $\Sigma(\omega)$
is plotted as a function of real frequencies. It is essentially
asymmetric with respect to the Fermi level, as could be assumed
from the asymmetry of LDA density of states and band filling 1/6.
At energies $\omega\sim\pm$1.5 eV the real part of self-energy has extrema,
corresponding to transition region from quasiparticle peak to
lower and upper Hubbard bands. Two extrema in the imaginary part
of the self-energy coinciding with Re$\Sigma$ zeros
\footnote{Here we remind that real part of self-energy
is connected with its imaginary part via Kramers-Kronig relation \cite{AGD}.}
determine energy positions of lower and upper Hubbard bands
(see Fig. \ref{fig_ldadmft}).

Asymmetric quasiparticle peak in the density of states (DOS) is situated in
the energy range -0.8--1.4 eV Fig.~\ref{fig_ldadmft}.
We see that imaginary part of self-energy ${\rm Im} \Sigma(\omega)$
is sufficiently small for these energies, while the real part can be roughly 
approximated with the dashed straight line shown in Fig.\ \ref{sigma_qmc}). 
The slope of this line defines mass renormalization value 
$Z=m^\star/m=1-\frac{\partial {\rm{Re}} \Sigma(\omega)}{\partial \omega}|_{\omega=0}=1.9$.
Such $Z$ value agrees with one obtained from QMC data in Matsubara frequencies:
$m^\star/m=1-\frac{{\rm{Im}}\Sigma(\omega_0)}{\omega_0}\approx 2$, where  
$\omega_0$ is ``zeroth'' Matsubara frequency.
This value of renormalization is in accord with the value  $m^*/m=2.2$
from the works \cite{Pavarini03,Pavarini05} as well as with experimental
estimate from ARPES data \cite{Yoshida05}.

In the inset in Fig.~\ref{sigma_qmc} it is visible that Fermi liquid behavior
of self-energy ${\rm Im}\Sigma(\omega) \sim -\omega^2$, together with 
${\rm Re} \Sigma(\omega) \sim -\omega$,
is fulfilled only in the interval from -0.2$\,$ to 0.15$\,$ eV.
The slope of  ${\rm{Re}} \Sigma(\omega)$  in immediate proximity to
the Fermi level is steeper than in wider energy interval
(dashed line in Fig.\ \ref{sigma_qmc}).
Thus Fermi liquid mass renormalization value is larger than
$m^*/m=1.9$ and is equal to $m^*_{{\rm low}E}/m=3$ 
(dashed line on the inset of Fig.\ \ref{sigma_qmc}).
On the edges of Fermi liquid regime, sharp bends of
${\rm Re} \Sigma(\omega)$ at energies $\omega = \pm 0.25$~eV are seen. 
As the border of Fermi liquid regime we can consider energies,
where the self-energy behavior starts to differ from  ${\rm Im}\Sigma
(\omega) \sim -\omega^2$, which because of Kramers-Kronig relation
corresponds to ${\rm Re} \Sigma(\omega) \sim-\omega$.
Deviation from the square behavior of ${\rm Im}\Sigma$ at  energies
of the order of $\omega = \pm 0.25$~eV immediately leads to
cusps in the ${\rm Re} \Sigma(\omega)$.

\begin{figure}
\begin{center}
\includegraphics[clip=true,width=0.5\textwidth]{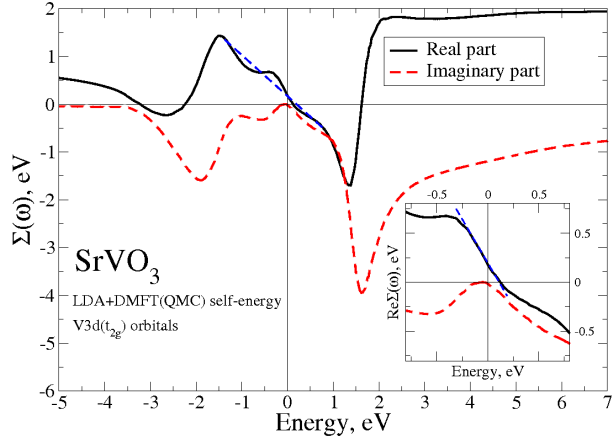}
\end{center}
\caption{
Real (black solid line) and imaginary (grey line) parts of
self-energy $\Sigma(\omega)$ obtained from LDA+DMFT(QMC) calculation
for V3d-$t_{2g}$ orbitals of SrVO$_3$ \cite{Nekrasov06}.
In the inset self-energy in the vicinity of the Fermi level is shown
in more details.
Dashed lines give the slope of ${\rm Re} \Sigma(\omega)$
far away and close to the Fermi level.}
\label{sigma_qmc}
\end{figure}

\begin{figure}
\begin{center}\vspace*{-2mm}\includegraphics[height=85mm]{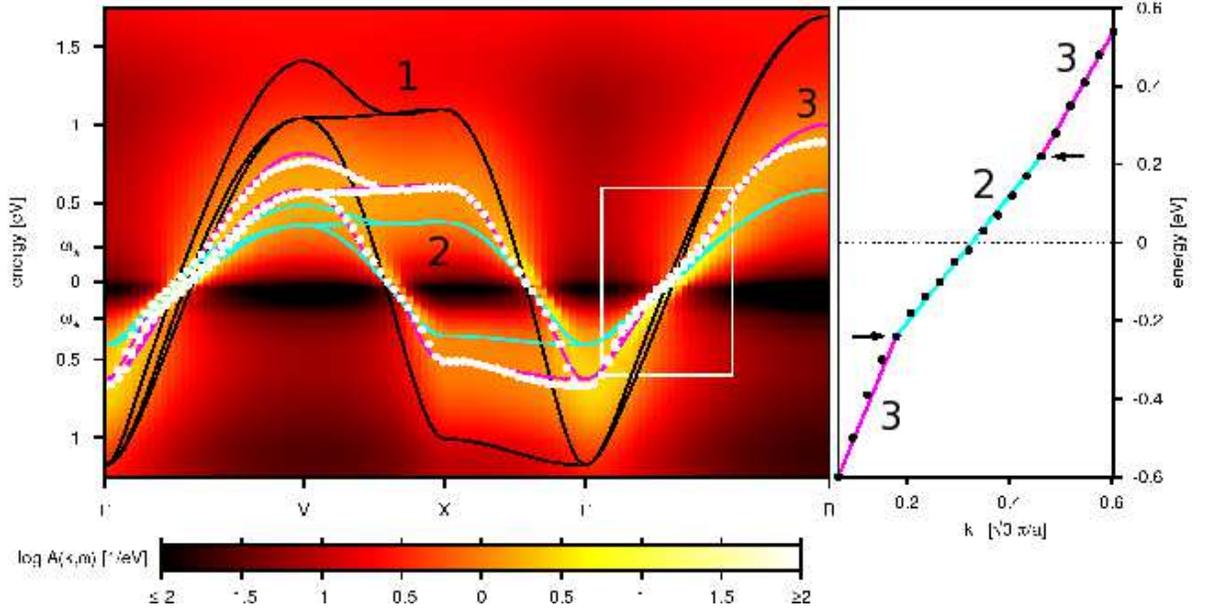}\vspace*{-4mm}\end{center}
\caption{%
Kinks in the dispersion relation $E_{n\bf{k}}$ (white dots),
for SrVO$_3$ obtained from LDA+DMFT calculation. 
Line 1 -- LDA band structure $\epsilon_{n\bf{k}}$;
line 2 -- LDA bands renormalized with Fermi liquid mass renormalization factor.
lines 3 -- band structure within the intermediate regime.
Kinks are marked with arrows on the right panel, which
corresponds to the area restricted with white rectangle on the main part
of the figure.
\vspace*{-3mm}%
}
\label{fig:srvo3}
\end{figure}

If the self-energy on the real axis is known one can compute
spectral function $A({\BK},\omega)$ and also the quasiparticle dispersion
determined by momentum dependence of its maxima. In Fig.~\ref{fig:srvo3}
we show the map of the spectral density for SrVO$_3$ obtained in 
Ref.\cite{Nekrasov06}. In this multiband system (with degenerate bands)
further analysis is similar to that for the single band case
of section \ref{ekink}. White dots denote dispersion curves
$E_{n\bf{k}}$ obtained from LDA+DMFT calculation for SrVO$_3$.
In the narrow vicinity of the Fermi level they coincide with
LDA band structure $\epsilon_{n\bf{k}}$ (line 1)
renormalized by Fermi liquid factor $Z_{\text{FL}}=0.35$,
so that $E_{n\bf{k}}=Z_{\text{FL}}\epsilon_{n\bf{k}}$ (line 2).
Outside the Fermi liquid region dispersion curves correspond to LDA
band structure with different renormalization factor:
$E_{n\bf{k}}=Z_{\text{CP}}\epsilon_{n\bf{k}}+ c_\pm$ (line 3),
where $Z_{\text{CP}}=0.64$, $c_+$=0.086 eV, $c_-=$0.13 eV.
Along high-symmetry $\Gamma$-M and $\Gamma$-R directions
in the Brillouin zone, transition between these two regimes
leads to formation of kinks in the effective dispersions
at energies $\omega_{\star,+}$=0.22 eV and $\omega_{\star,-}$=-0.24 eV.
These kinks are marked with arrows on the right side of Fig. \ref{fig:srvo3},
which corresponds to the area surrounded with the white rectangle on the
main part of the figure. On the contour plot of spectral function 
$A(\bf{k},\omega)$ it is seen that in the energy region sufficiently far away 
from the Fermi level, spectral function keeps explicit ${\bf k}$-dependence, 
despite pretty large damping value, replacing traditional band 
structure picture for systems with strong electron-electron correlations.

Kinks of electronic nature were discovered in this system in ARPES experiments
\cite{Yoshida05} in $\Gamma$-M direction at energies of the order of 0.25 eV, 
which agrees quite well with results of LDA+DMFT(QMC) calculations.

\subsection{Electronic structure of copper oxides in the pseudogap state:\\ 
LDA+DMFT+$\Sigma$.}
\label{lda_dmfs_s_res}

Pseudogap state as was already pointed above is one of main anomalies of the 
normal state of high-T$_c$ cuprates and it is thought that clarification of its 
physical nature is the key point to understand high-temperature 
superconductivity mechanism \cite{Tim,MS,Tam}.
Most powerful tool to investigate this state in recent years became angular
resolved photoemission (ARPES). During last ten years in this area there was
a remarkable progress related to significant growth of ARPES resolution
both in energy and in momentum space \cite{Shen03,ARPES_Bi}.
From ARPES data the Fermi surface (FS) shape, quasiparticle dispersion and
damping, even self-energy can be directly restored \cite{Shen03,ARPES_Bi}.
This allowed to study in detail formation of the pseudogap,
``shadow'' bands, quite unusual phenomena of Fermi arcs formation,
interlayer hybridization effects (bilayer splitting) in double layer systems,
\cite{Shen03,ARPES_Bi}, to determine qualitative distinctions
between electron and hole doped cuprates \cite{Shen03,ARPES_Bi}. 
The purpose of theory is an explanation of all these peculiarities and this
problem is much complicated by rather strong electronic correlations, typical
for these systems and making doubtful the standard band theory
and Fermi liquid approach.

In this section we shall demonstrate that an account of AFM short range order 
fluctuations is in principle enough to describe a number of ARPES experiments 
in real systems. To this purpose we use LDA+DMFT+$\Sigma$ hybrid computational 
scheme \cite{Real1,Real5,Real2,Real3,Real4,Real6}. 
On the one hand this scheme inherits all advantages of
LDA+DMFT \cite{poter97,LDADMFT1,Nekrasov00, psik, LDADMFT}, i.e. the
combination of single electron first principle density functional theory
within the local density approximation (DFT/LDA) \cite{DFT_LDA,DFT_LDA_1} with
dynamical mean-field theory (DMFT) for strongly correlated electrons
\cite{MetzVoll89,vollha93,pruschke,georges96,PT}. 
On another hand this scheme allows to consider nonlocal correlations
by introduction of momentum dependent self-energy, while the usual 
self-consistent set of DMFT equations is preserved \cite{cm05,JTL05,FNT06}. 
To solve effective single impurity problem of DMFT in the works described below
we employed numerical renormalization group (NRG) \cite{NRG,BPH}. 

Such computational scheme fits very well to describe electronic
properties of high-T$_c$ cuprates in normal (underdoped) state.
Firstly, all material specific model parameters of physically relevant
Cu-3d $x^2-y^2$ orbital can be obtained from LDA calculations.
Secondly, stoichiometric cuprates are antiferromagnetic Mott insulators
with $U\gg W$ ($U$ -- local Coulomb interaction 
$W$ -- conduction band width), so that correlation effects there are very 
important.
At finite doping (at least up to optimal doping) cuprates are typical
strongly correlated metals and DMFT stage of the computational scheme 
allows one to account for strong electronic correlations.
Finally, to study ``antiferromagnetic scenario'' of pseudogap formation
we introduce into the standard LDA+DMFT scheme {\bf k}- dependent self-energy 
$\Sigma_{\bf k}$, describing nonlocal correlations induced by (quasi) static 
Heisenberg spin fluctuations of short range AFM order \cite{Sch,KS99}.

In the framework of LDA+DMFT+$\Sigma$ approach we performed calculations
for a series of high temperature superconductors: hole doped
Bi$_2$Sr$_2$CaCu$_2$O$_{8-\delta}$ (Bi2212)~\cite{Real1} and
La$_{2-x}$Sr$_{x}$CuO$_4$ (LSCO)~\cite{Real5}, and also for electron doped 
Nd$_{2-x}$Ce$_x$CuO$_4$ (NCCO)~\cite{Real2,Real3} and
Pr$_{2-x}$Ce$_x$CuO$_4$ (PCCO)~\cite{Real4}.
LDA+DMFT+$\Sigma$ calculation results for Fermi surfaces and spectral functions
can be compared with ARPES data for quasiparticles bands and experimental
Fermi surface maps.

Crystal structure of Bi2212~\cite{Real1}, NCCO~\cite{Real2,Real3} and PCCO~\cite{Real4}
have tetragonal symmetry with the space group I4/mmm, while LSCO has
distorted orthorhombic structure Bmab~\cite{Real5}.
In more details crystallographic data used in LDA+DMFT+$\Sigma$
calculations are presented in Refs.~\cite{Real1,Real5,Real2,Real3,Real4,Real6}. 

It is well known that physical properties of cuprates in many respects are 
determined by quasi-two-dimensionality of their electronic properties. 
From this point of view, the main interest is attracted to electronic states of 
CuO$_2$ plane, where we are dealing with partially filled antibonding 
Cu-3$d$($x^2\!-\!y^2$) orbitals with dispersion crossing the Fermi level.
In tight-binding approximation this dispersion has the following form:
\begin{eqnarray}
\label{disp}
\varepsilon({\bf k})=&-2t&(\cos k_xa+\cos k_ya)
-4t'\cos k_xa\cos k_ya\\ \nonumber
&-2t''&(\cos 2k_xa+\cos 2k_ya)
-2t'''(\cos k_xa\cos 2k_ya +\cos 2k_ya \cos k_ya).
\end{eqnarray}
Here $t$, $t^{\prime}$, $t^{\prime\prime}$, $t^{\prime\prime\prime}$ are
Cu-Cu transfer integrals in first four coordination spheres in the
CuO$_2$ plane, $a$ is the lattice constant. Values of these effective transfer 
integrals calculated with the use of Wannier functions obtained
within the N-th order muffin-tin orbitals method (NMTO) of Ref.~\cite{NMTO}
are listed in the Table~\ref{tab_12}. In the following we shall exploit LDA 
calculated effective antibonding Cu-3$d$($x^2\!-\!y^2$) band as a ``bare'' one 
in  LDA+DMFT+$\Sigma$ calculations.

\begin{table}
\caption {Calculated model energy parameters (eV) and
experimental correlation length $\xi$.
First four Cu-Cu transfer integrals in the CuO$_2$ plane  
$t$, $t^{\prime}$, $t^{\prime\prime}$, $t^{\prime\prime\prime}$; 
effective interlayer transfer integral~$t_\perp$,
local Coulomb interaction~$U$ and pseudogap amplitude~$\Delta$.}
\centering
\begin{tabular}{|l|c|c|c|c|c|c|c|c|}
\hline
&$t$&$t^{\prime}$&$t^{\prime\prime}$&$t^{\prime\prime\prime}$&$t_\perp$&$U$&$\Delta$&$\xi$ \\
\hline
Bi2212&-0.627& 0.133&0.061& -0.015&0.083&1.51&0.21&10a\\
\hline
NCCO& -0.44&0.153&0.063&-0.01&---&1.1&0.36&50a\\
\hline
PCCO&-0.438&0.156&0.098&---&---&1.1&0.275&50a\\
\hline
LSCO&-0.476&0.077&-0.025&-0.015&---&1.1&0.21&10a\\
\hline
\end{tabular}
\label{tab_12}
\end{table}

In double layer systems, e.g. in Bi2212, hopping between two
neighboring planes is also important.
In tight-binding approximation an expression for corresponding
interlayer dispersion derived in Ref.~\cite{Andersen95} has the form: 
\begin{equation}
t_{\perp}({\bf k}) = \frac{t_{\perp}}{4}(\cos k_xa-\cos k_ya)^2
\label{tperp}
\end{equation}
The value of $t_\perp$ is given in Table~\ref{tab_12}.
Consideration of interlayer hopping and ``bilayer splitting''
effects requires certain generalization of LDA+DMFT+$\Sigma$ computational 
scheme \cite{Real1}. 

To perform DMFT calculations one should also calculate a value of onsite
Coulomb interaction. The value of this interaction for effective
Cu-3$d$($x^2\!-\!y^2$) orbital obtained within the constrained
LDA method~\cite{Gunnarsson88} is also given in Table~\ref{tab_12}. 

To account for AFM spin fluctuations we employed two-dimensional model of the
pseudogap state \cite{Sch,KS99},
generalized for DMFT+$\Sigma$ calculations \cite{cm05,FNT06}.
Additional ``external'' \textbf{k}-dependent self-energy
$\Sigma_{\bf k}$~\cite{cm05,FNT06} describes nonlocal correlations caused by
(quasi)static\footnote{Quasistatic approximation for AFM fluctuations 
necessarily restricts this approach to rather high temperatures 
(and energies not very close to the Fermi level)\cite{Sch,KS99}.
Thus we can not judge about the nature of low temperature (low energy)
damping which is determined by dynamical (inelastic) scattering processes.}
AFM  spin fluctuations.

To specify $\Sigma_{\textbf{k}}$ it is necessary to know two important 
parameters -- the pseudogap amplitude $\Delta$, giving energy scale of 
fluctuating SDW and correlation length $\xi$.
The value of $\Delta$ were calculated as described in Refs.~\cite{cm05,FNT06,Real1}.
The values of correlation length were taken in accordance with values obtained
in neutron scattering experiments for NCCO~\cite{NCCOxi} and LSCO~\cite{LSCOxi}.
The values of $\Delta$ and $\xi$ used for all systems under consideration
are also listed in Table~\ref{tab_12}.
To solve effective Anderson single impurity problem in DMFT we used
numerical renormalization group (NRG~\cite{NRG,BPH}).
Temperature in DMFT(NRG) calculations  was chosen to be 0.011~eV
and electron or hole concentration (doping level) was taken to be 15\%.  

LDA+DMFT+$\Sigma$ calculations produce a clear picture of ``hot spots'' behavior 
in the spectral function and on maps of the Fermi surfaces for electron doped 
systems \cite{Real2,Real3,Real4}, while for hole doped systems
only Fermi arcs arise~\cite{Real1,Real5}.

\begin{figure}
\includegraphics[clip=true,angle=270,width=0.4\columnwidth]{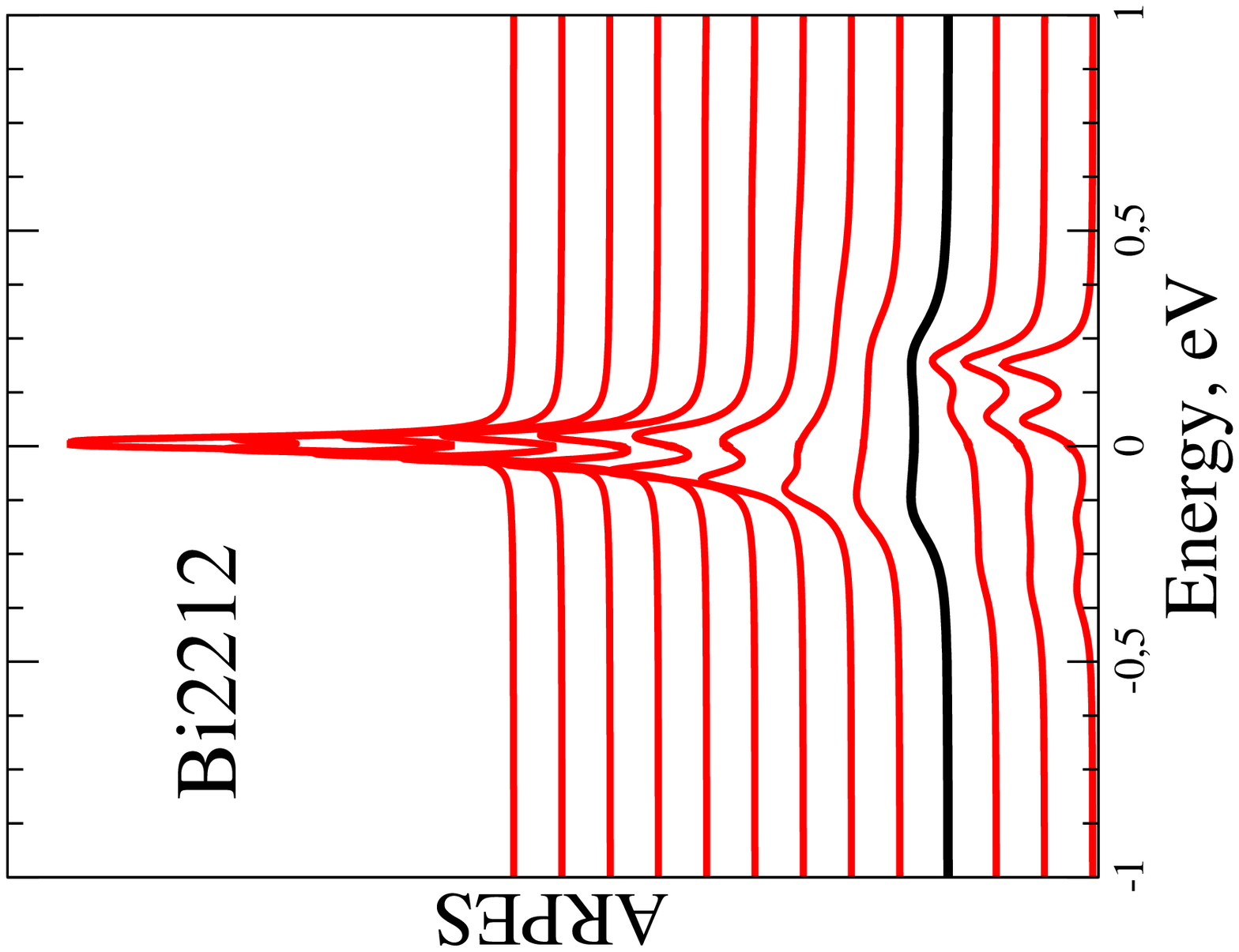}
\includegraphics[clip=true,angle=270,width=0.4\columnwidth]{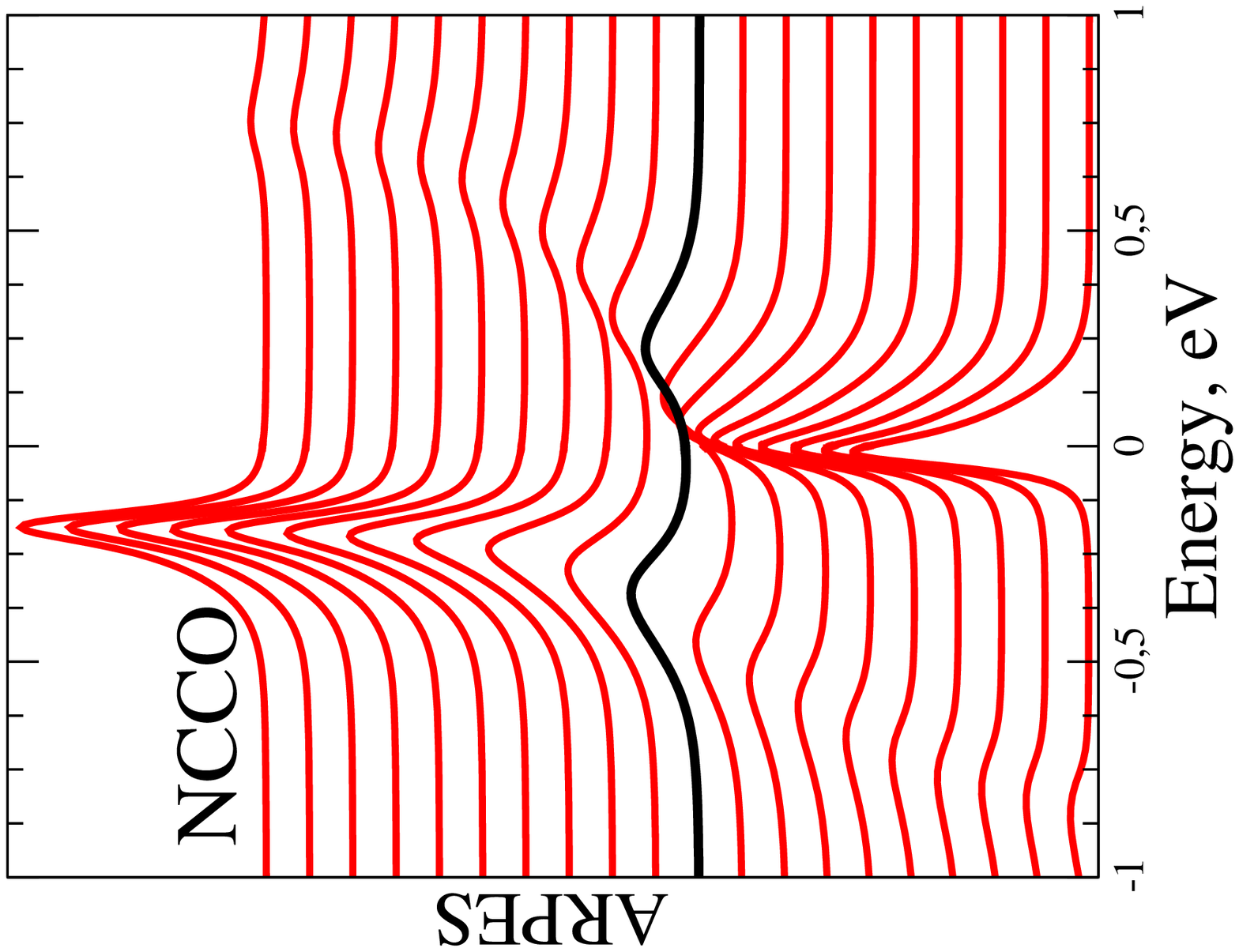}
\caption{\small LDA+DMFT+$\Sigma$ spectral functions for Bi2212 
(leftpanel) and NCCO (right panel) along the ``bare'' Fermi surface 
in the 1/8th of the Brillouin zone.
Black line corresponds to the ``hot spot''~\cite{Real2,Real3}}.
\label{sdfs}
\end{figure}

In Fig.~\ref{sdfs} LDA+DMFT+$\Sigma$ we show spectral functions along 1/8 part
of bare Fermi surface from nodal point on the diagonal of the Brillouin zone
(upper curve) downto antinodal point at the boundary of the zone (lower curve).
Results for Bi2212 are shown on the left panel and for NCCO on the right panel 
of Fig.~\ref{sdfs}. For both systems in nodal direction quasiparticles are well 
defined --- sharp peak of spectral function situated practically on the Fermi 
level is clearly seen.
As one moves to antinodal point quasiparticle damping grows reaching
the maximum at the ``hot-spot'' and the peak of spectral density moves
away from the Fermi level.
This behavior is in complete agreement with results of Refs. 
\cite{Armitage02,Kaminski05} 
(comparison with experiment see in~\cite{Real2,Real3}).
From LDA+DMFT+$\Sigma$ results shown in Fig.~\ref{sdfs} it is directly
seen, that for Bi2212 antinodal states are formed by
low energy edge of the pseudogap\footnote{Especially clear it is visible for 
the case of smaller correlation length $\xi =5a$ considered in Ref. \cite{Real1}.},
while for NCCO by high energy edge.
For Bi2212 we also observe bilayer splitting of quasiparticle peak
which is related to the presence of two CuO$_2$ planes in the elementary cell.

The ``hot spots'' for NCCO are located closer to the Brillouin zone
diagonal~\cite{Real2,Real3}. This can be seen from black lines on Fig.~\ref{sdfs},
which correspond to the ``hot spots''.
Moreover, correlation length in NCCO is much larger than in Bi2212.
Thus for NCCO (in contrast to Bi2212) in the antinodal direction
quasiparticles again are rather well defined. For Bi2212 scattering
near by Brillouin zone boundaries is strong everywhere and instead of
``hot spots'' picture we observe quite strong ``destruction''
of the Fermi surface close to these boundaries.
Qualitatively the same picture is observed also in LSCO.

\begin{figure}%[htb]
\includegraphics[clip=true,width=0.6\textwidth]{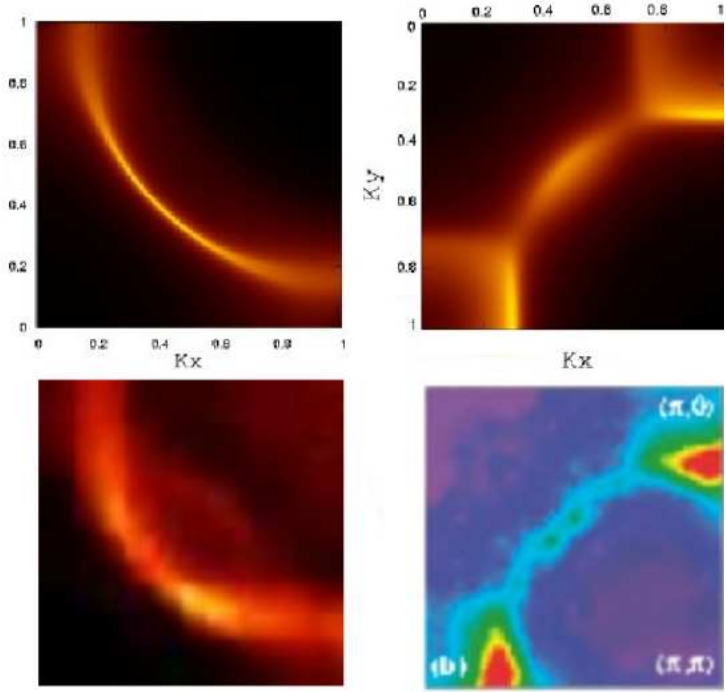}
\caption{\small LDA+DMFT+$\Sigma$ Fermi surface maps obtained in
Refs. \cite{Real2,Real3} for Bi2212 (upper left panel) 
and NCCO (upper right panel) in the quarter of the Brillouin zone, ($k_x,k_y$ are given
in units of $\pi/a$). Experimental Fermi surface for Bi2212 (lower left panel~\cite{Borisenko00}) and 
NCCO (lower right panel~\cite{Armitage02}).}
\label{sk_fs}
\end{figure}

In Fig.~\ref{sk_fs}, on the upper panel, LDA+DMFT+$\Sigma$ Fermi surface maps
in the quarter of the Brilloun zone for Bi2212 (on the left) and NCCO (on the 
right) are presented. In Bi2212 we observe strong ``destruction'' of Fermi surface by 
scattering by pseudogap fluctuations close to Brillouin zone boundaries\footnote{Analogous 
behavior in entire accordance with ARPES results is realized also in another 
hole doped system -- LSCO. The LDA+DMFT+$\Sigma$
calculations for this system were performed in~\cite{Real5}}.
On the contrary, in NCCO the Fermi surface is almost recovered close to the
Brillouin zone boundaries. On the other hand Fermi arc around nodal direction
in Bi2212 is clearly pronounced, while for NCCO it is noticeably smeared.
This is another consequence of the fact that in NCCO
``hot spots'' are located closer to the Brillouin zone diagonal.
A bit larger value of the pseudogap amplitude $\Delta$
also favours the stronger damping of Fermi arcs in NCCO.
One should note the appearance of the ``shadow'' Fermi surface,
which is much more intensive for NCCO.

Qualitatively the same Fermi surfaces were observed experimentally
in real Bi ~\cite{Borisenko00} and  Nd~\cite{Armitage02} systems
(lower panel of Fig.~\ref{sk_fs}). Thus, the distinction of 
Fermi surface maps for these systems is related mainly to the
distinction of band structure parameters of these materials.
In particular, LDA Fermi surfaces of NCCO are more bended and ``hot spots''
appear to be rather far from the Brillouin zone boundaries, consequently
the Fermi surface in the vicinity of these boundaries almost does not
feel scattering by AFM fluctuations.
In Bi2212 LDA Fermi surface is rather close to the Brillouin zone boundaries 
and ($\pi/a$,0) point, so that ``hot spots'' are also close to this point.
Thus in Bi2212 they are more ``washed away'' by strong  pseudogap
scattering close to ($\pi/a$,0) point and are not observed.
``Hot spots'' in NCCO are more vivid also because of much larger
value of correlation length of fluctuations.

\begin{figure}
\includegraphics[clip=true,width=0.6\textwidth]{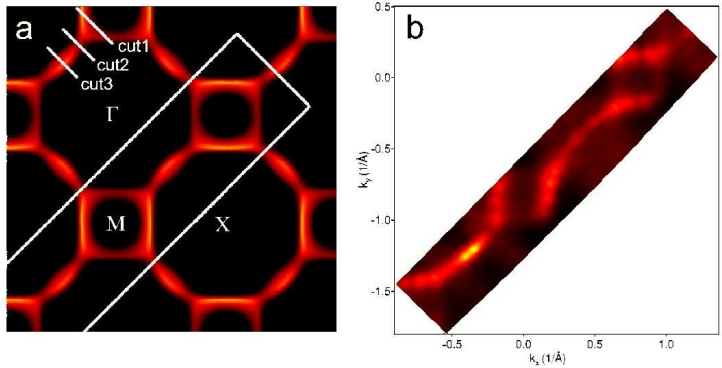}
\caption{\small  Fermi surface map for PCCO. 
(a) --- results of LDA+DMFT+$\Sigma$ calculation.
White rectangle on the panel (a) shows the part of reciprocal space
where ARPES measurements (panel b) were done. 
Lower left corner coincides with X-point ($\pi/a,0$) of the Brillouin zone~\cite{Real4}.}
\label{FermiSurface}
\end{figure}

Not less graphic results were obtained in LDA+DMFT+$\Sigma$ calculations
and ARPES experiments for ~\cite{Real4}. In Fig.~\ref{FermiSurface} we show 
PCCO Fermi surface map (panel (a) -- LDA+DMFT+$\Sigma$ results,
panel (b) -- experimental ARPES data).
Fermi surface here is clearly distinguishable only near 
Brillouin zone boundaries and around $(\pi/a/2,\pi/a/2)$ point (Fermi arc). 
Again, as in NCCO we observe ``destruction'' of the Fermi surface in 
``hot spots'', located at the intersection of Fermi surface and its AFM shadow
``replica'' is detected. This `` destruction'' of the Fermi surface is due to 
the strong electron scattering by AFM spin (pseudogap) fluctuations.
The ``shadow'' Fermi surface is observed, as it happens in the case of AFM
doubling of the lattice period.
However, since there is no long-range order in the underdoped region,
in which we are interested, this ``shadow'' Fermi surface is strongly eroded.
Fermi surface of PCCO is very similar to that observed in
Nd$_{2-x}$Ce$_{x}$CuO$_4$ (NCCO), which belongs to the same family
of superconductors~\cite{Real2,Real3,Armitage02}.

\begin{figure}%[hb]
\includegraphics[clip=true,width=0.5\textwidth]{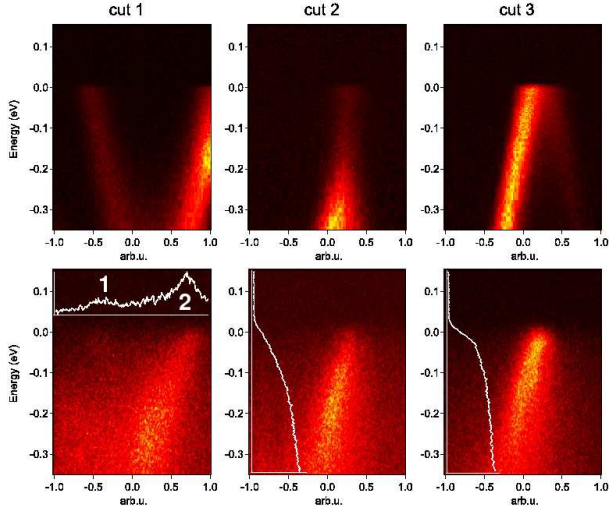}
\caption{\small Momentum -- energy distribution curves for 
characteristic cuts of the Brillouin zone plotted in Fig.~\ref{FermiSurface}
(upper panels --- theoretical data, lower panels --- experimental
photoemission intensity).
For the cut 1 we show momentum distribution  curve (MDC) integrated over energy 
window of the width of 60 meV around the Fermi level. Analogous MDC for the 
cut 2 (through the ``hot spot'') shows ARPES intensity suppression as
compared with MDC for the cut 3,
which is located quite far away from the ``hot spot''~\cite{Real2,Real3}.}
\label{Cuts}
\end{figure}

Let us compare (see Fig.~\ref{Cuts}) theoretical (upper panel) and 
experimental (lower panel) quasiparticle dispersions along
most characteristic cuts of the Fermi surface shown in Fig.~\ref{FermiSurface}.
Theoretical data are multiplied by the Fermi function with a temperature of 30K and 
convoluted (in energy) with a Gaussian distribution to simulate the experimental 
resolution. Cut 1 crosses the quasiparticle and the ``shadow'' Fermi surface 
near the Brillouin zone boundary.
Correspondingly, here it is possible to detect ``fork''-like structure, 
formed by suppressed ``shadow'' band and much better defined quasiparticle band.
This structure corresponds to the beginning of formation of the
Fermi surface cylinder around ($ \ pi / a $, 0) point.
Cut 2 passes exactly through the `` hot spot''. Here we see a strong
suppression of the quasiparticle band near the Fermi level.
Cut 3 crosses the Fermi arc and we can see fairly well-defined
quasiparticle band. However, the `` shadow'' band of low intensity is
also present. In the case of long-range AFM order and a full doubling
of the period, the Fermi surface and its ``shadow'' form a closed ``pocket'',
of the Fermi surface around the ($\pi/2a$, $\pi/2a$) point.
while in the present case a part of the pocket formed by ``shadow'' band
is strongly blurred. One can see that there is a good agreement
between calculated and experimental data.

As was already noted, within LDA+DMFT+$\Sigma$ scheme two-particle properties
can also be calculated ~\cite{PRB07}, which allowed to
investigate optical conductivity of Bi and Nd cuprates~\cite{Real2,Real3},
also demonstrating significant differences in the effects of pseudogap
fluctuations. In particular, in optical conductivity of NCCO, in qualitative 
agreement with experiment~\cite{Onose}, we observe a characteristic pseudogap 
dip and a smooth maximum due absorption through the pseudogap at frequencies
$\sim 2\Delta $. However, in optical conductivity of Bi2212 characteristic 
pseudogap structure practically does not occur neither in theory or in 
experiments~\cite{Quijada}, which is related to sufficiently small values of 
$\Delta$ and fluctuation correlation length in this system.

Let us summarize some of our conclusions. For all the systems studied,
LDA+DMFT+$\Sigma$ calculations show, that Fermi-liquid behavior persists only 
rather far away from the ``hot spots'' (nodal direction), and ``destruction''
of the Fermi surface is observed near the ``hot spots''.
This destruction is due to strong scattering of correlated electrons on
short-range order AFM (pseudogap) fluctuations.
Comparison of ARPES data and LDA+DMFT+$\Sigma$ calculations shows the
existence of quite distinct ``hot spots'' in the behavior of the spectral 
density and maps of the Fermi surface in electron doped systems, in contrast to
hole doped systems, where we only observe only a strong ``destruction'' of the
Fermi surface near Brillouin zone boundaries and Fermi arc around its
diagonal. There are several reasons for this difference:
(1) ``hot spots'' in electron doped systems are
located closer to the center of the Brillouin zone, (2) correlation length of 
AFM fluctuations in electron doped systems is bigger;
(3) the width of the pseudogap in the electron doped systems is also larger 
than in hole doped ones. Experimental and theoretical results discussed here 
clearly confirm the AFM scenario of the formation of the pseudogap in both 
hole-doped \cite{Real1,Real5}, and electron doped cuprates~\cite{Real2,Real3,Real4}.

\section{Conclusion}
\label{concl}

In this review we discussed DMFT + $\Sigma$ generalization of the standard 
dynamical mean-field theory (DMFT), which allows to include  non-local
correlations or additional (relative to Hubbard one) interactions
(in principle of any type), while remaining within the single-impurity picture 
of DMFT and retaining the same set of self-consistent DMFT equations. The basic 
approximation of this method is the neglect of interference contributions of 
DMFT diagrams and additional interactions included into the analysis. Precisely 
this (strictly speaking not completely controllable) approximation 
allows to preserve the overall structure of DMFT equations, which permits to
solve DMFT+$\Sigma$ equations with well developed methods used in the standard 
DMFT. It must be emphasized that the self-consistent account of additional
interactions at every step of DMFT loop leads to a rather complicated procedure,
equivalent to the summation of infinite classes of diagrams.

The proposed approach proved to be versatile enough to be applied to a
number of problems in systems with strong electron correlations -- from
semi-phenomenological account of non-local short range order pseudogap 
fluctuations to the self-consistent scheme for metal-insulator transition
in the disordered Hubbard-Anderson model and account of the effects of 
electron-phonon interaction in electronic spectra of strongly correlated systems.
A remarkable feature of DMFT+$\Sigma $ approach is the possibility to
study, along with one-particle characteristics, also the two-particle properties,
i.e., in principle, any kind of response functions (optical conductivity, 
magnetic susceptibility, charge screening, etc.). The universality of the method 
allows one to hope for its successful application in a number of future problems.

Discussing all problems under consideration, one should keep in mind
that in many respects similar physical results can be obtained with more
sophisticated approaches, using these or other methods of direct numerical 
simulation. For example, similar results for the formation of the pseudogap in 
the single-particle characteristics of the two-dimensional Hubbard model were 
obtained in the cluster generalizations of DMFT \cite{TMrmp,KSPB}. However, 
these methods have specific limitations
(e.g. of cluster size) and are still not widely used to calculate the 
two-particle properties, such as the general response functions, in particular, 
the optical conductivity. DMFT + $\Sigma$ approach has obvious advantages,
associated with savings of computational resources. It requires a significantly 
lower cost of computational time, and its advantage in calculating the 
two-particle response functions is quite obvious. This opens up additional 
opportunities for the systematic comparison of various types of non-local 
fluctuations or additional interactions and their influence on the electronic 
properties of strongly correlated systems,
providing intuitively clear path to the analysis of experiments and theoretical 
results obtained via more complicated schemes.

Rather simple generalization of our computational scheme
enabled us to formulate also the generalized LDA+DMFT+$\Sigma $ approach, 
that allows to perform calculations of all of the effects discussed above for
real compounds of transition elements with strong electronic correlations.
One can expect that these calculations will be useful in analyzing and 
explaining the new experimental data.

\section{Acknowledgements}
\label{acknow}

The authors thank Th. Pruschke for his significant contribution to the 
development of DMFT+$\Sigma $ approach in its initial stages, as well as for 
providing us with very efficient NRG code that was used in most calculations.

We are also grateful to S. Borisenko and other members of Dresden ARPES group
with whom ARPES studies of PCCO were performed.

Investigations of Sr(Ca)VO$_3$, LSCO and NCCO were done in the framework
of joint project together with University of Osaka (group of
Prof. S. Suga and Prof. A. Sekiyama).

This work was partially supported by RFBR grant 11-02-00147 and performed
within the framework of programs of Presidium of RAS ``Quantum physics of
condensed matter'' (UB RAS 09-$\Pi$-2-1009) and Physics Division of RAS 
``Strongly correlated electrons in solids''(UB RAS 09-T-2-1011 .)

\end{document}